\documentclass[a4paper,11pt]{article}

\usepackage{xcolor}
\usepackage{graphicx}
\usepackage{amsmath}
\usepackage{mathrsfs}
\usepackage{arydshln}
\usepackage{color}
\usepackage{amssymb}
\usepackage{ulem}
\usepackage{tikz}
\usetikzlibrary{decorations.pathmorphing}
\usepackage{bm}
\usepackage{cite}
\usepackage{xfrac}
\usepackage{nicefrac}
\usepackage{braketmod}

\usepackage{geometry}
\numberwithin{equation}{section}
\allowdisplaybreaks

\begin{document}

\newcommand{\hiduke}[1]{\hspace{\fill}{\small [{#1}]}}
\newcommand{\aff}[1]{${}^{#1}$}
\renewcommand{\thefootnote}{\fnsymbol{footnote}}

%------Naotaka------
\global\long\def\cs{\text{CS}}%

\global\long\def\Iside{\mathrm{vec}}%

\global\long\def\Icross{\mathrm{mat}}%

\global\long\def\bra#1{\Bra{#1}}%

\global\long\def\bbra#1{\Bbra{#1}}%

\global\long\def\ket#1{\Ket{#1}}%

\global\long\def\kket#1{\Kket{#1}}%

\global\long\def\braket#1{\Braket{#1}}%

\global\long\def\bbraket#1{\Bbraket{#1}}%

\global\long\def\brakket#1{\Brakket{#1}}%

\global\long\def\bbrakket#1{\Bbrakket{#1}}%

\newcommand{\fmmn}{
\left[ \text{FMMN} \right]}
%------Naotaka------

\begin{titlepage}
% \begin{flushright}
% %{\footnotesize preprint SISSA xx/202y/FISI},\,
% {\footnotesize YITP-22-11}
% \end{flushright}
\begin{center}
{\Large\bf
Fermi gas formalism for D-type quiver Chern-Simons theory\\
with non-uniform ranks
}\\
\bigskip\bigskip
\bigskip\bigskip
{\large Naotaka Kubo\footnote{\tt naotaka.kubo(at)yukawa.kyoto-u.ac.jp}}\aff{1}
{\large and Tomoki Nosaka\footnote{\tt nosaka(at)yukawa.kyoto-u.ac.jp}}\aff{2}\\
\bigskip\bigskip
\aff{1} {\small
\it 
Center for Joint Quantum Studies and Department of Physics, School of Science, Tianjin University,\\
135 Yaguan Road, Tianjin 300350, China
}
\\
\aff{2} {\small
\it Kavli Institute for Theoretical Sciences and CAS Center for Excellence in Topological Quantum Computation, University of Chinese Academy of Sciences, Beijing, 100190, China
}\\

\bigskip
\end{center}
\bigskip
\bigskip
\begin{abstract}
We construct the Fermi gas formalism for the partition function of supersymmetric Chern-Simons theories with affine $D$-type quiver diagrams with non-uniform ranks of the gauge groups and Fayet-Illiopoulos parameters by two different approaches: the open string formalism and the closed string formalism.
In the closed string formalism approach, we find a novel connection between the partition function of this theory and the partition function of a four-nodes circular quiver supersymmetric Chern-Simons theory.
We also studied a symmetry of a density matrix appeared in the closed string formalism.
We further calculate the exact values of the partition function for finite $N$, with which we identified the exponent of the leading non-perturbative effect in $1/N$ corresponding to the worldsheet instantons in the circular quiver supersymmetric Chern-Simons theories.
\end{abstract}

\bigskip\bigskip\bigskip

\end{titlepage}

\renewcommand{\thefootnote}{$\dagger$\arabic{footnote}}
\setcounter{footnote}{0}

\tableofcontents

\section{Introduction and Summary}
\label{sec_intro}

Connections between matrix models and integrable systems such as Painlev\'e equations and their discrete generalizations have been discussed for a long time (see e.g.~\cite{Marino:2008ya,Morozov:1994hh}).
A famous example is the connection between the Hermitian matrix models and Painlev\'e equations, which arises through the method of orthogonal polynomial \cite{Bessis:1979aaa,1980Bessis}.
The Painlev\'e equations are known to be transcendental in the sense that their general solutions, called Painlev\'e transcendentals, cannot be written in closed form expressions in terms of well-known special functions such as hypergeometric functions or elliptic theta functions.
The connection between matrix models and integrable systems would provide a powerful tool to study various expansions of the Painlev\'e transcendentals through the standard techniques to evaluate the matrix models.
Conversely, the connection can also be viewed as a powerful tool to study the large $N$ expansion of the matrix models.
This aspect has also been utilized in various contexts in theoretical high energy physics, such as two-dimensional gravity \cite{Gross:1989vs,Okuyama:2019xbv,Eynard:2023qdr} and Gross-Witten matrix model \cite{Marino:2008ya}.

Recently, a new example of the connection between matrix models and integrable systems was discovered in theoretical high energy physics \cite{Bonelli:2017gdk,Nosaka:2020tyv,Bonelli:2022dse,Moriyama:2023mjx,Moriyama:2023pxd,Bonelli:2016idi,Bonelli:2017ptp,Grassi:2018spf,Francois:2023trm}.
In this connection, the matrix models are the partition functions of three-dimensional circular quiver superconformal Chern-Simons theories describing M2-branes in M-theory on $S^3$, and the corresponding integrable systems are $\mathfrak{q}$-deformed Painlev\'e equations \cite{Bonelli:2017gdk,Bonelli:2022dse,Moriyama:2023mjx,Moriyama:2023pxd} or $\mathfrak{q}$-deformed Toda equations \cite{Nosaka:2020tyv}.
In this paper we shall call this connection ``$\mathfrak{q}$-integrable equation/M2-matrix model ($\mathfrak{q}$-IE/M2-MM) correspondence'' for short.
More concretely, the parameters in the integrable equations such as the dynamical time or other Painlev\'e parameters are mapped under the $\mathfrak{q}$-IE/M2-MM correspondence to the deformation parameters of the three-dimensional theory such as FI parameters and the relative ranks of the gauge groups which do not modify the orbifold background probed by the M2-branes.
On the other hand, the overall rank of the three-dimensional theory corresponds to a parameter which does not appear explicitly in the integrable equations, namely to one of the initial conditions.
The fact that the integrable equation holds for arbitrary initial conditions implies that there is an infinite set of relations among the partition function at different values of the overall rank and the deformation parameters.

In all examples of the $\mathfrak{q}$-IE/M2-MM correspondence discovered so far, the partition function of the theories of M2-branes with overall rank $N$ can be written as the partition function of an $N$ particle ideal Fermi gas in one dimensional quantum mechanics \cite{Marino:2011eh}, which played a crucial role in the discovery and the subsequent tests of the correspondence in the following senses.
First, the one-particle density matrix for each theory of M2-branes enjoys the symmetry of a discrete group.
When the integrable equation is one of the $\mathfrak{q}$-Painlev\'e equations, which are characterized by the discrete groups under the Sakai's classification \cite{Sakai:2001aaa,Kajiwara:2015aaa}, the corresponding M2-brane matrix model and the dictionary between the parameters can be identified by matching the discrete group associated with the $\mathfrak{q}$-Painlev\'e equation with the symmetry of the density matrix of the Fermi gas formalism.
Second, the Fermi gas formalism allows us to calculate the exact values of the partition function for finite $N$ efficiently, with which we can check the integrable equation for the grand partition function at leading orders in the fugacity dual to $N$ \cite{Bonelli:2017gdk,Nosaka:2020tyv,Bonelli:2022dse,Moriyama:2023mjx}.
Third, the Fermi gas formalism relates $\mathfrak{q}$-IE/M2-MM correspondence to the Painlev\'e/gauge correspondence \cite{Bershtein:2014yia,Bershtein:2016uov,Gavrylenko:2020gjb,Bonelli:2019boe,Gavrylenko:2017lqz,Gavrylenko:2016zlf,Gavrylenko:2015wla,Nagoya:2015,Bonelli:2016qwg,Gamayun:2012ma,Iorgov:2014vla} or its $\mathfrak{q}$-uplift \cite{Bershtein:2016aef,Bershtein:2017swf,Bershtein:2018srt,Bershtein:2018zcz,Jimbo:2017ael,Matsuhira:2018qtx}.
Namely, the inverse of the density matrix of the Fermi gas formalism takes the form of a one-dimensional algebraic curve, which can be regarded as the Seiberg-Witten curve of the five-dimensional ${\cal N}=1$ super Yang-Mills theory.
Indeed, through topological string/spectral theory correspondence \cite{Hatsuda:2013oxa,Honda:2014npa,Grassi:2014zfa,Codesido:2015dia} and the geometric engineering \cite{Katz:1996fh,Leung:1997tw,Gopakumar:1998jq}, the grand partition function of the theory of M2-branes is connected to the discrete Fourier transformation of the Nekrasov partition function of the five-dimensional Yang-Mills theory, which is precisely what the Painlev\'e $\tau$-functions correspond to under the ($\mathfrak{q}$-)Painlev\'e/gauge correspondence.

The simplest example of the $\mathfrak{q}$-Painlev\'e/gauge correspondence, which is the correspondence between $\mathfrak{q}\text{PIII}_3$ and five-dimensional pure $\text{SU}(2)$ Yang-Mills theory, can be derived \cite{Bershtein:2018zcz,Bershtein:2018srt} by the non-linear self-consistency equations satisfied by the Nekrasov-partition function called Nakajima-Yoshioka blowup equations \cite{Nakajima:2003pg,Nakajima:2005fg,Gottsche:2006bm,Shchechkin:2020ryb}.
Since the blowup equations exist for a large class of five-dimensional ${\cal N}=1$ Yang-Mills theories with various gauge groups and matter fields \cite{Keller:2012da,Kim:2019uqw}, one would expect that the other examples of the $\mathfrak{q}$-Painlev\'e/gauge correspondence can be derived in the same way.
On the other hand, although the connection between the circular quiver Chern-Simons theories for M2-branes and $\mathfrak{q}$-integrable systems has been established through several examples, it is still not clear how such a connection arises from the viewpoint of the physics of M2-branes or matrix models.\footnote{
In some special cases the $\mathfrak{q}$-IE/M2-MM correspondence may be viewed as a $\mathfrak{q}$-uplift of the connection between Painlev\'e equation and critical Ising model found in \cite{Barouch:1976vt,McCoy:1976cd,Zamolodchikov:1994uw,1996CMaPh.179....1T}.
However, it is not clear how to extend the way to understand/prove this connection to the case with $\mathfrak{q}\neq 1$ as well as to more general setups of M2-branes.
}

To break through this situation, it would be useful to investigate the $\mathfrak{q}$-IE/M2-MM correspondence for a more general class of the theories of M2-branes.
Besides the circular quiver Chern-Simons theories \cite{Hosomichi:2008jd,Imamura:2008nn} realized on the $S^1$-compactified Hanany-Witten brane setup \cite{Hanany:1996ie,Kitao:1998mf,Bergman:1999na}, there are various different three-dimensional gauge theories which are believed to describe the M2-branes on different backgrounds (see e.g.~\cite{Amariti:2019pky} and references therein).
In particular, in this paper we consider the supersymmetric Chern-Simons matter theories with affine $D_\ell$ (or ${\hat D}_\ell$) quivers \cite{Gulotta:2011vp,Gulotta:2012yd,Crichigno:2012sk,Jain:2015kzj,Crichigno:2017rqg}.
The partition function of this model was found to be written in the Fermi gas formalism when all the deformation parameters are turned off\footnote{
The Fermi gas formalism for a ${\hat D}_3$ quiver theory with the rank and FI deformations is performed in \cite{Kubo:2024zug}.
Note that the assignment of the Chern-Simons levels in \cite{Kubo:2024zug} is different from those in our current setup, which causes a technical difference in the detail of the Fermi gas approaches.
}
\cite{Moriyama:2015jsa} (see also \cite{Assel:2015hsa}), while it is not clear whether the partition function enjoys the correspondence to some topological string partition function or five-dimensional Nekrasov partition function.
Nevertheless, it would be worth investigating the $\mathfrak{q}$-IE/M2-MM correspondence also for this class of theories.
Indeed, when $\ell=3$, ${\hat D}_3$ quiver Chern-Simons theory is equivalent to a four-node circular quiver Chern-Simons theory where the inverse of the density matrix of the Fermi gas formalism is clearly identified with that of a five-dimensional linear quiver Yang-Mills theory.
Hence we expect the $\mathfrak{q}$-IE/M2-MM correspondence to hold at least for this special case.
On the other hand, the Fermi gas formalism of the ${\hat D}_3$ quiver theory (as a ${\hat D}$-type quiver theory \cite{Moriyama:2015jsa,Honda:2015rbb,Kubo:2024zug}) has the same structure as the Fermi gas formalism of the other ${\hat D}_\ell$ ($\ell>3$) quiver theories, with which we expect that the $\mathfrak{q}$-IE/M2-MM correspondence holds also for the ${\hat D}_\ell$ quiver theories with $\ell>3$.

If the $\mathfrak{q}$-IE/M2-MM correspondence holds for ${\hat D}_\ell$ quiver theories, the integrable equation would appear as a relation among the grand partition function with different relative ranks and FI parameters, as is the case for the circular quiver Chern-Simons theories.
Therefore, as a preparation to find such a relation in a future study, in this paper we extend the Fermi gas formalism of ${\hat D}_\ell$ quiver theory \cite{Moriyama:2015jsa} to the case when these deformation parameters are turned on.
In the circular quiver Chern-Simons theories, there are two different ways to extend the Fermi gas formalism for the theory with non-zero relative ranks, namely the open string formalism \cite{Matsumoto:2013nya} and the closed string formalism \cite{Awata:2012jb,Honda:2013pea}.
The open string formalism is more straightforward to construct than the closed string formalism and also more convenient in some setups for the calculation of the exact values of the partition function at finite $N$.
On the other hand, the closed string formalism is more suitable when we study the discrete symmetry of the partition function and the connection to the five-dimensional theory.
In this paper we first construct the open string formalism for general ${\hat D}_\ell$ quiver theory with a two-parameter rank deformation.
Then, focusing on $\ell=4$, which is the simplest non-trivial case in the ${\hat D}_\ell$ quiver theories, we also construct the closed string formalism with a three-parameter rank deformation.
In the closed string formalism, we find a curious connection between the density matrix of ${\hat D}_4$ quiver theory and the density matrix of a circular quiver theory, with which we also comment on the partial inverse of the density matrix.
We also study the symmetry of the density matrix and reinterpret it as dualities of the ${\hat D}_4$ quiver theory.
Note that both of the constructions are worked out with completely generic FI parameters.
Lastly, we further establish the algorithm to calculate the exact values of the partition function without rank/FI deformations, which was not investigated in \cite{Moriyama:2015jsa}.
The exact values thus obtained show good agreement with the formula for the all order $1/N$ expansion obtained by the semiclassical (i.e.~small $k$) analysis of the Fermi gas formalism, and also allow us to propose the exponent of the $1/N$ non-perturbative effect analogous to the worldsheet instanton in ABJM theory which cannot be accessed in the semiclassical expansion.

The rest of this paper is organized as follows.
In section \ref{sec:Model} we introduce the partition function of ${\hat D}$-type quiver super Chern-Simons theory given by the supersymmetric localization formula.               
In section \ref{sec_open} we construct an open string formalism for the partition function with two-parameter rank deformation where all the rank deformations on the non-affine nodes are turned off.
In section \ref{sec_closed} we construct a closed string formalism for ${\hat D}_4$ quiver super Chern-Simons theory, with the rank deformation on the non-affine node also turned on and discuss two symmetries in terms of the density matrix obtained there.
In section \ref{sec_TWPY} we calculate the exact values of the partition function by generalizing the technique developed in \cite{Putrov:2012zi} (see also \cite{Moriyama:2014nca}) for ${\hat D}_4$ quiver super Chern-Simons theory without rank/FI deformations.
In section \ref{sec:Discussion} we conclude and list several future directions of the research.                        
By using these exact values we identify the exponent of the leading worldsheet instanton.                              
In appendix \ref{sec_1dqmnotation} we provide our notation for one-dimensional quantum mechanics used in the Fermi gas formalism and formulas for quantum mechanics.
In appendix \ref{app_detformulas} we list the formulas which we use in sections \ref{sec_open} and \ref{sec_closed} to handle various determinant/Pfaffian arising in the Fermi gas formalism.
In appendix \ref{sec:YkQC} we provide a conjecture and evidence which is crucial for applying the closed string formalism.
In appendix \ref{app_exactvalues} we list the exact values of the partition function of ${\hat D}_4$ quiver super Chern-Simons theory without rank/FI deformations obtained in section \ref{sec_TWPY}.

\section{${\hat D}$-type quiver Chern-Simons theories\label{sec:Model}}
\begin{figure}
\begin{center}
\includegraphics[width=16cm]{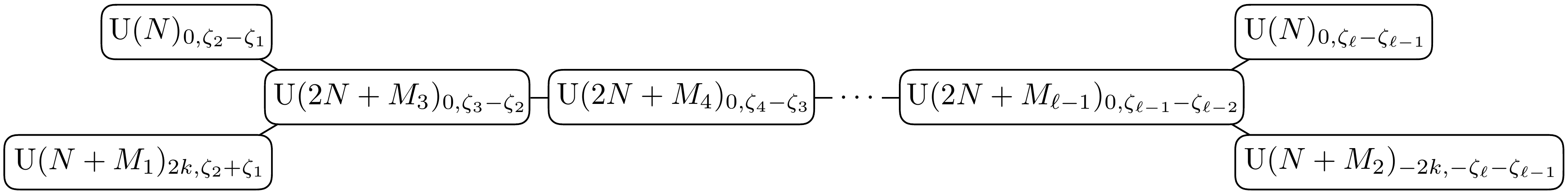}
\caption{Quiver diagram of the ${\hat D}_\ell$ quiver superconformal Chern-Simons theory \eqref{Dr}.
Each rounded rectangle with ${\rm U}\left( n \right)_{k,\eta}$ represents ${\rm U}\left( n \right)$ vector multiplet with Chern-Simons level $k$ and FI parameter $\eta$, and each edge represents bi-fundamental hypermultiplet.
}
\label{Drquiver}
\end{center}
\end{figure}

Let us consider the supersymmetric Chern-Simons matter theory on affine $D_\ell$ quiver with two rank deformations $M_1,M_2$ on the affine nodes with levels $\pm 2k$, rank deformations $M_3,M_4,\cdots,M_{\ell-1}$ on the non-affine nodes and FI deformations (see figure \ref{Drquiver} for the assignment of the Chern-Simons levels and rank/FI deformations).
We assume $k>0$ without loss of generality.
The partition function of this theory on the three sphere is given by the supersymmetric localization formula \cite{Kapustin:2009kz} as
\begin{align}
&Z=\frac{(-1)^{N(M_1-M_2)}i^{-\frac{M_1^2}{2}+\frac{M_2^2}{2}}}{(N!)^2(N+M_1)!(N+M_2)!\prod_{a=1}^{\ell-3}(2N+M_{a+2})!}\nonumber \\
&\quad \times \int_{-\infty}^\infty
\frac{d^N\mu}{(2\pi k)^N}
\frac{d^{N+M_1}\mu'}{(2\pi k)^{N+M_1}}
\left(\prod_{a=1}^{\ell-3}\frac{d^{2N+M_{a+2}}\lambda^{(a)}}{(2\pi k)^{2N+M_{a+2}}}\right)
\frac{d^N\nu}{(2\pi k)^N}
\frac{d^{N+M_2}\nu'}{(2\pi k)^{N+M_2}}\nonumber \\
&\quad \times e^{\frac{i}{2\pi k}(\sum_{m=1}^{N+M_1}(\mu_m')^2-\sum_{m=1}^{N+M_2}(\nu_m')^2)}
e^{-\frac{i}{2\pi k}((\zeta_2-\zeta_1)\sum_{m=1}^N\mu_m+(\zeta_2+\zeta_1)\sum_{m=1}^{N+M_1}\mu_m')}\nonumber \\
&\quad \times e^{-\frac{i}{2\pi k}\sum_{a=1}^{\ell-3}(\zeta_{a+2}-\zeta_{a+1})\sum_{m=1}^{2N+M_{a+2}}\lambda_m^{(a)})}
e^{-\frac{i}{2\pi k}((\zeta_{\ell}-\zeta_{\ell-1})\sum_{m=1}^N\nu_m+(-\zeta_{\ell}-\zeta_{\ell-1})\sum_{m=1}^{N+M_2}\nu_m')}\nonumber \\
&\quad \times \frac{
\prod_{m<m'}^N(2\sinh\frac{\mu_m-\mu_{m'}}{2k})^2
\prod_{m<m'}^{N+M_1}(2\sinh\frac{\mu'_m-\mu'_{m'}}{2k})^2
}{
\prod_{m=1}^N\prod_{n=1}^{2N+M_3}2\cosh\frac{\mu_m-\lambda_n^{(1)}}{2k}
\prod_{m=1}^{N+M_1}\prod_{n=1}^{2N+M_3}2\cosh\frac{\mu'_m-\lambda_n^{(1)}}{2k}
}
\frac{\prod_{a=1}^{\ell-3}\prod_{m<m'}^{2N+M_{a+2}}(2\sinh\frac{\lambda_m^{(a)}-\lambda_{m'}^{(a)}}{2k})^2}{\prod_{a=1}^{\ell-4}\prod_{m=1}^{2N+M_{a+2}}\prod_{n=1}^{2N+M_{a+3}}2\cosh\frac{\lambda_m^{(a)}-\lambda_n^{(a+1)}}{2k}}\nonumber \\
&\quad \times \frac{
\prod_{m<m'}^N(2\sinh\frac{\nu_m-\nu_{m'}}{2k})^2
\prod_{m<m'}^{N+M_2}(2\sinh\frac{\nu'_m-\nu'_{m'}}{2k})^2
}{
\prod_{m=1}^{2N+M_{\ell-1}}\prod_{n=1}^N2\cosh\frac{\lambda_m^{(\ell-3)}-\nu_n}{2k}
\prod_{m=1}^{2N+M_{\ell-1}}\prod_{n=1}^{N+M_2}2\cosh\frac{\lambda_m^{(\ell-3)}-\nu'_n}{2k}
}.
\label{Dr}
\end{align}
Here we have chosen the overall phase to simplify the final results of the open/closed string Fermi gas formalism.
We have also rescaled all of the integration variables in the original expression \cite{Kapustin:2009kz} from
$\mu_m,\mu'_m,\nu_m,\nu'_m,\lambda^{(a)}_m$
to
$\mu_m/(2\pi k),\mu'_m/(2\pi k),\nu_m/(2\pi k),\nu'_m/(2\pi k),\lambda^{(a)}_m/(2\pi k)$ for later convenience.

Note that the integration \eqref{Dr} is convergent only when the following condition is satisfied for all nodes with vanishing Chern-Simons levels:\footnote{
Here we have assumed that $\zeta_{a+1}-\zeta_a$ are real numbers.
The conditions for the convergence becomes more strict if some of $\zeta_{a+1}-\zeta_a$ have non-zero imaginary part.
}
\begin{align}
2((\text{rank of the node under concern})-1)<(\text{sum of ranks over adjacent nodes}),
\end{align}
that is, only when the theory is not ``bad'' for any node with vanishing Chern-Simons level \cite{Gaiotto:2008ak}.
This can be seen by studying the behavior of the integrand in the limit $\mu_1\rightarrow\pm \infty$, $\lambda_1^{(1)}\rightarrow\pm \infty$, and so on.
The conditions for ${\hat D}_\ell$ quiver with $\ell\ge 5$ are expressed in terms of $M_1,M_2,\cdots,M_{\ell-1}$ as
\begin{align}
&-2<M_3,\quad -2<M_{\ell-1},\nonumber \\
&2(M_3-1)<M_1+M_4,\nonumber \\
&2(M_a-1)<M_{a-1}+M_{a+1},\quad (a=4,5,\cdots,\ell-2)\nonumber \\
&2(M_{\ell-1}-1)<M_2+M_{\ell-2}.
\end{align}
For $\ell=4$, the condition reduces to
\begin{align}
-2<M_3<\frac{M_1+M_2}{2}+1.
\end{align}

Also note that although we have introduced $\ell$ FI parameters, the partition function \eqref{Dr} is invariant under the uniform shift of $\zeta_1,\zeta_2,\cdots,\zeta_\ell$
\begin{align}
\zeta_a\rightarrow \zeta_a+c, \label{eq:FIshift}   
\end{align}
for any $c$, up to a trivial overall factor independent of $N$:
\begin{align}
Z_k(N,M_a,\zeta_a+c)=
e^{\frac{i(M_1-M_2)c^2}{2\pi k}}
Z_k(N,M_a,\zeta_a).
\end{align}

In the remaining part of this section, we rewrite the partition function into a form which is the common starting point for both the open string formalism and the closed string formalism.
First, we rewrite the one-loop determinant factors involving $\mu_m,\mu'_m,\nu_m,\nu'_m$ as
\begin{align}
&\frac{
\prod_{m<m'}^N(2\sinh\frac{\mu_m-\mu_{m'}}{2k})^2
\prod_{m<m'}^{N+M_1}(2\sinh\frac{\mu'_m-\mu'_{m'}}{2k})^2
}{
\prod_{m=1}^N\prod_{n=1}^{2N}2\cosh\frac{\mu_m-\lambda_n^{(1)}}{2k}
\prod_{m=1}^{N+M_1}\prod_{n=1}^{2N}2\cosh\frac{\mu'_m-\lambda_n^{(1)}}{2k}
}\nonumber \\
&=
\frac{\prod_{m<m'}^N2\sinh\frac{\mu_m-\mu_{m'}}{2k}
\prod_{m<m'}^{N+M_1}2\sinh\frac{\mu'_m-\mu'_{m'}}{2k}
}{
\prod_{m=1}^N\prod_{n=1}^{N+M_1}2\sinh\frac{\mu_m-\mu'_n}{2k}
}
\frac{
\prod_{m<m'}^{2N+M_1}2\sinh\frac{\bar{\mu}_m-\bar{\mu}_{m'}}{2k}
}{
\prod_{m=1}^{2N+M_1}\prod_{n=1}^{2N}2\cosh\frac{\bar{\mu}_m-\lambda_n^{(1)}}{2k}
},\nonumber \\
&\frac{
\prod_{m<m'}^N(2\sinh\frac{\nu_m-\nu_{m'}}{2k})^2
\prod_{m<m'}^{N+M_2}(2\sinh\frac{\nu'_m-\nu'_{m'}}{2k})^2
}{
\prod_{m=1}^{2N}\prod_{n=1}^N2\cosh\frac{\lambda_m^{(\ell-3)}-\nu_n}{2k}
\prod_{m=1}^{2N}\prod_{n=1}^{N+M_2}2\cosh\frac{\lambda_m^{(\ell-3)}-\nu'_n}{2k}
}\nonumber \\
&=
\frac{\prod_{m<m'}^N2\sinh\frac{\nu_m-\nu_{m'}}{2k}
\prod_{m<m'}^{N+M_2}2\sinh\frac{\nu'_m-\nu'_{m'}}{2k}
}{
\prod_{m=1}^N\prod_{n=1}^{N+M_2}2\sinh\frac{\nu_m-\nu'_n}{2k}
}
\frac{
\prod_{m<m'}^{2N+M_2}2\sinh\frac{\bar{\nu}_m-\bar{\nu}_{m'}}{2k}
}{
\prod_{m=1}^{2N}\prod_{n=1}^{2N+M_2}2\cosh\frac{\lambda_m^{(\ell-3)}-\bar{\nu}_n}{2k}
},
\end{align}
where
\begin{align}
\bar{\mu}_m=(\mu_1,\cdots,\mu_N,\mu'_1,\cdots,\mu'_{N+M_1}),\quad
\bar{\nu}_m=(\nu_1,\cdots,\nu_N,\nu'_1,\cdots,\nu'_{N+M_2}).
\end{align}
Then, by using Cauchy-Vandermonde determinant formulas \eqref{CauchyVdm3}-\eqref{CauchyVdm4} we obtain
\begin{align}
&Z=\frac{(-1)^{N(M_1-M_2)}i^{-\frac{M_1^2}{2}+\frac{M_2^2}{2}}}{(N!)^2(N+M_1)!(N+M_2)!}\int_{-\infty}^\infty
\frac{d^{2N+M_1}\bar{\mu}}{(2\pi)^{2N+M_1}}
\frac{d^{2N+M_2}\bar{\nu}}{(2\pi)^{2N+M_2}}
e^{\frac{i}{2\pi k}(\sum_{m=1}^{N+M_1}(\mu_m')^2-\sum_{m=1}^{N+M_2}(\nu_m')^2)}\nonumber \\
&\quad \times e^{-\frac{i}{2\pi k}((\zeta_2-\zeta_1)\sum_{m=1}^N\mu_m+(\zeta_2+\zeta_1)\sum_{m=1}^{N+M_1}\mu_m')}
e^{-\frac{i}{2\pi k}((\zeta_{\ell}-\zeta_{\ell-1})\sum_{m=1}^N\nu_m+(-\zeta_{\ell}-\zeta_{\ell-1})\sum_{m=1}^{N+M_2}\nu_m')}\nonumber \\
&\quad \times
\det\begin{pmatrix}
\left[\braket{\mu_m|\frac{\tanh\frac{{\hat p}-\pi iM_1}{2}}{2}|\mu'_n}\right]_{m,n}^{N\times (N+M_1)}\\
\left[\bbraket{t_{M_1,r}|\mu'_n}\right]_{r,n}^{M_1\times (N+M_1)}
\end{pmatrix}
{\cal Z}(\bar{\mu}_m,\bar{\nu}_n)
\nonumber \\
&\quad \times \det
\left(
\begin{array}{cc}
\left[\braket{\nu'_m|\frac{\tanh\frac{{\hat p}+\pi iM_2}{2}}{2}|\nu_n}\right]_{m,n}^{(N+M_2)\times N}&
\left[\brakket{\nu'_m|-t_{M_2,s}}\right]_{m,s}^{(N+M_2)\times M_2}
\end{array}
\right),
\label{Dr_centergeneral}
\end{align}
where $\ket{\cdot}$ and $\kket{\cdot}$ are position/momentum eigenstates (see appendix \ref{sec_1dqmnotation} for the detail of notation for 1d quantum mechanics), and
\begin{align}
{\cal Z}(\bar{\mu}_m,\bar{\nu}_n)&=
\frac{k^{-2N-\frac{M_1+M_2}{2}}}{\prod_{a=1}^{\ell-3}(2N+M_{a+2})!}\int_{-\infty}^\infty \Bigl(\prod_{a=1}^{\ell-3}\frac{d^{2N+M_{a+2}}\lambda^{(a)}}{(2\pi k)^{2N+M_{a+2}}}\Bigr)
e^{-\frac{i}{2\pi k}\sum_{a=1}^{\ell-3}(\zeta_{a+2}-\zeta_{a+1})\sum_{m=1}^{2N+M_{a+2}}\lambda_m^{(a)}}\nonumber \\
&\quad \times \frac{
\prod_{m<n}^{2N+M_1}2\sinh\frac{\bar{\mu}_m-\bar{\mu}_n}{2k}
}{\prod_{m=1}^{2N+M_1}\prod_{n=1}^{2N+M_3}2\cosh\frac{\bar{\mu}_m-\lambda^{(1)}_n}{2k}}
\frac{\prod_{a=1}^{\ell-3}\prod_{m<n}^{2N+M_{a+2}}(2\sinh\frac{\lambda_m^{(a)}-\lambda_n^{(a)}}{2k})^2}{\prod_{a=1}^{\ell-4}
\prod_{m=1}^{2N+M_{a+2}}\prod_{n=1}^{2N+M_{a+3}}2\cosh\frac{\lambda^{(a)}_m-\lambda^{(a+1)}_n}{2k}}\nonumber \\
&\quad \times \frac{
\prod_{m<n}^{2N+M_2}2\sinh\frac{\bar{\nu}_m-\bar{\nu}_n}{2k}
}{\prod_{m=1}^{2N+M_{\ell-1}}\prod_{n=1}^{2N+M_2}2\cosh\frac{\lambda^{(\ell-3)}_m-\bar{\nu}_n}{2k}}.
\label{calZ}
\end{align}
% In the following two sections we apply the open string formalism and closed string formalism starting from this matrix model.
Let us define the grand partition function $\Xi(\kappa)$ as
\begin{align}
\Xi(\kappa)=\sum_{N=0}^\infty \kappa^NZ(N).\label{eq:GPFdef}
\end{align}
When there are no rank/FI deformations, $M_1=\cdots=M_{\ell-1}=\zeta_1=\cdots=\zeta_\ell=0$, $\Xi(\kappa)$ can be rewritten in the following form \cite{Moriyama:2015jsa}
\begin{align}
\Xi(\kappa)=\sqrt{\text{Det}(1+\kappa{\hat\rho})},
\label{Xiundeformed}
\end{align}
with a one-dimensional quantum mechanical operator ${\hat\rho}$ whose explicit expression was obtained in \cite{Moriyama:2015jsa} as
\begin{align}
{\hat \rho}
=
\frac{1}{(2\cosh\frac{{\hat p}}{2})^{\ell-2}}
\Bigl(
e^{-\frac{i{\hat x}^2}{2\pi k}}\frac{\tanh\frac{{\hat p}}{2}}{2}
+\frac{\tanh\frac{{\hat p}}{2}}{2}e^{-\frac{i{\hat x}^2}{2\pi k}}
\Bigr)
\frac{1}{(2\cosh\frac{{\hat p}}{2})^{\ell-2}}
\Bigl(
\frac{\tanh\frac{{\hat p}}{2}}{2}e^{\frac{i{\hat x}^2}{2\pi k}}
+e^{\frac{i{\hat x}^2}{2\pi k}}\frac{\tanh\frac{{\hat p}}{2}}{2}
\Bigr).\label{rhowithoutdeformation}
\end{align}

In the following two sections we apply the open string formalism and closed string formalism starting from \eqref{Dr_centergeneral}, and see how the grand partition function \eqref{eq:GPFdef} is generalized in the two different formalism.

\section{Open string formalism for $\hat{D}_\ell$ model}
\begin{figure}
\begin{center}
\includegraphics[width=16cm]{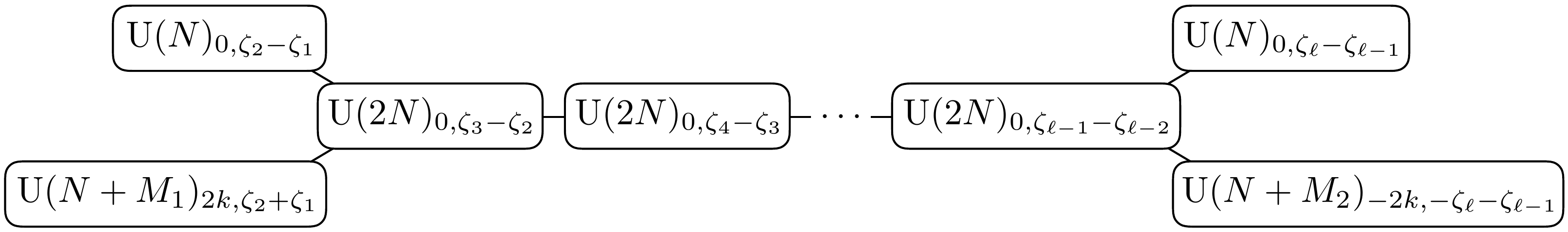}
\caption{Quiver diagram of the models considered in section \ref{sec_open}.
}
\label{Drquiver_Laall0}
\end{center}
\end{figure}

In this section we apply the open string formalism to the partition function for $M_3=M_4=\cdots=M_{\ell-1}=0$ case as depicted in figure \ref{Drquiver_Laall0}.
\label{sec_open}
When there are no rank deformations, $M_1=M_2=0$, the grand partition function of ${\hat D}_\ell$ theory, $\Xi(\kappa)$, is written as \eqref{Xiundeformed}.
In the following, we show that when there are rank deformations $M_1,M_2\ge 0$ this result is modified as
\begin{align}
\Xi(\kappa)=
e^{\pi i(-\frac{M_1^2}{4}+\frac{M_2^2}{4}+M_1M_2)}
\sqrt{\text{Det}(1+\kappa{\hat\rho}^{\text{(open)}})}\sqrt{\det\left(\left[H_{ab}(\kappa)\right]_{a,b}^{(2M_1+2M_2)\times (2M_1+2M_2)}\right)},
\label{Xiopen}
\end{align}
with ${\hat\rho}^{\text{(open)}}$ and $H_{ab}(\kappa)$ given below as \eqref{rhoopen} and \eqref{Hab}.
Note that this is analogous to the open string formalism for the circular quiver Chern-Simons matter theories obtained in \cite{Matsumoto:2013nya,Moriyama:2017gye}.

When $M_3=M_4=\cdots=M_{\ell-1}=0$, ${\cal Z}(\bar{\mu}_m,\bar{\nu}_n)$ defined in \eqref{calZ} can be written, by using Cauchy-Vandermonde determinant formulas \eqref{CauchyVdm1}-\eqref{CauchyVdm2} and further using Cauchy-Binet formula \eqref{CauchyBinet}, as
\begin{align}
{\cal Z}(\bar{\mu}_m,\bar{\nu}_n)=(-1)^{M_1M_2}
\det\begin{pmatrix}
\left[\braket{\bar{\mu}_m|{\hat S}|\bar{\nu}_n}\right]_{m,n}^{(2N+M_1)\times (2N+M_2)}
&\left[\brakket{\bar{\mu}_m|-t_{M_1,s}}\right]_{m,s}^{(2N+M_1)\times M_1}\\
\left[\bbraket{t_{M_2,r}|\bar{\nu}_n}\right]_{r,n}^{M_2\times (2N+M_2)}&[0]^{M_2\times M_1}
\end{pmatrix},
\end{align}
with
\begin{align}
{\hat S}=
\frac{1}{2\cosh\frac{{\hat p}+\pi iM_1}{2}}
e^{-\frac{i(\zeta_3-\zeta_2)}{2\pi k}{\hat x}}
\frac{1}{2\cosh\frac{{\hat p}}{2}}
e^{-\frac{i(\zeta_4-\zeta_3)}{2\pi k}{\hat x}}
\frac{1}{2\cosh\frac{{\hat p}}{2}}
\cdots
e^{-\frac{i(\zeta_{\ell-1}-\zeta_{\ell-2})}{2\pi k}{\hat x}}
\frac{1}{2\cosh\frac{{\hat p}-\pi iM_2}{2}}.
\end{align}
For example, for ${\hat D}_4$ quiver with $M_3=0$, we have ${\hat S}=\frac{1}{2\cosh\frac{{\hat p}+\pi iM_1}{2}}e^{-\frac{i(\zeta_3-\zeta_2)}{2\pi k}{\hat x}}\frac{1}{2\cosh\frac{{\hat p}-\pi iM_2}{2}}$.
To obtain \eqref{Xiopen}, first we apply the Cauchy-Binet formula \eqref{CauchyBinet} for the integrations over $\mu_m'$ and $\nu_m'$, to obtain
\begin{align}
&Z(N)=\frac{(-1)^{M_1M_2-\frac{M_1^2}{4}+\frac{M_2^2}{4}}}{(N!)^2}\int_{-\infty}^\infty
\frac{d^N\mu}{(2\pi)^N}
\frac{d^N\nu}{(2\pi)^N}\nonumber \\
&\quad
\times \det\begin{pmatrix}
\hspace{-4pt}\left[\braket{\mu_m|{\hat S}|\nu_n}\right]_{m,n}^{\substack{N\\ \times N}}
&\left[\braket{\mu_m|{\hat S}{\hat U}|\nu_n}\right]_{m,n}^{\substack{N\\ \times N}}
&\left[\braket{\mu_m|{\hat S}|V_{2,s}}\right]_{m,s}^{\substack{N\\ \times M_2}}
&\left[\brakket{\mu_m|-t_{M_1,s}}\right]_{m,s}^{\substack{N\\ \times M_1}}\hspace{-4pt}\\
\hspace{-4pt}\left[\braket{\mu_m|{\hat T}{\hat S}|\nu_n}\right]_{m,n}^{\substack{N\\ \times N}}
&\left[\braket{\mu_m|{\hat T}{\hat S}{\hat U}|\nu_n}\right]_{m,n}^{\substack{N\\ \times N}}
&\left[\braket{\mu_m|{\hat T}{\hat S}|V_{2,s}}\right]_{m,s}^{\substack{N\\ \times M_2}}
&\hspace{-6pt}\left[\brakket{\mu_m|{\hat T}|-t_{M_1,s}}\right]_{m,s}^{\substack{N\\ \times M_1}}\hspace{-4pt}\\
\hspace{-4pt}\left[\braket{V_{1,r}|{\hat S}|\nu_n}\right]_{r,n}^{\substack{M_1\\ \times N}}
&\left[\braket{V_{1,r}|{\hat S}{\hat U}|\nu_n}\right]_{r,n}^{\substack{M_1\\ \times N}}
&\left[\braket{V_{1,r}|{\hat S}|V_{2,s}}\right]_{r,s}^{\substack{M_1\\ \times M_2}}
&\left[\brakket{V_{1,r}|-t_{M_1,s}}\right]_{r,s}^{\substack{M_1\\ \times M_1}}\hspace{-4pt}\\
\hspace{-4pt}\left[\bbraket{t_{M_2,r}|\nu_n}\right]_{r,n}^{\substack{M_2\\ \times N}}
&\hspace{-6pt}\left[\bbraket{t_{M_2,r}|{\hat U}|\nu_n}\right]_{r,n}^{\substack{M_2\\ \times N}}
&\hspace{-6pt}\left[\bbraket{t_{M_2,r}|V_{2,s}}\right]_{r,s}^{\substack{M_2\\ \times M_2}}
&\left[0\right]^{\substack{M_2\\ \times M_1}}\hspace{-4pt}
\end{pmatrix},\label{openmunu}
\end{align}
where
\begin{align}
&{\hat T}=(-1)^{M_1}e^{-\frac{i(\zeta_2-\zeta_1){\hat x}}{2\pi k}}\frac{\tanh\frac{{\hat p}-\pi iM_1}{2}}{2}e^{\frac{i{\hat x}^2}{2\pi k}}e^{-\frac{i(\zeta_2+\zeta_1){\hat x}}{2\pi k}},\quad
{\hat U}=(-1)^{M_2}e^{-\frac{i{\hat x}^2}{2\pi k}}e^{-\frac{i(-\zeta_{\ell}-\zeta_{\ell-1}){\hat x}}{2\pi k}}\frac{\tanh\frac{{\hat p}+\pi iM_2}{2}}{2}e^{-\frac{i(\zeta_{\ell}-\zeta_{\ell-1}){\hat x}}{2\pi k}},\nonumber \\
&\ket{V_{2,s}}=e^{-\frac{i{\hat x}^2}{2\pi k}}e^{-\frac{i(-\zeta_{\ell}-\zeta_{\ell-1}){\hat x}}{2\pi k}}\kket{-t_{M_2,s}},\quad
\bra{V_{1,r}}=\bbra{t_{M_1,r}}
e^{\frac{i{\hat x}^2}{2\pi k}}
e^{-\frac{i(\zeta_2+\zeta_1){\hat x}}{2\pi k}}.
\end{align}
Next we perform the $\nu$-integrations in \eqref{openmunu} by using the Cauchy-Binet-like formula for Pfaffian \eqref{CBPfaffian}.
As a result we obtain
\begin{align}
&Z(N)\nonumber \\
&=\frac{(-1)^{\frac{N(N-1)}{2}+\frac{M_1(M_1-2)}{4}-\frac{M_2(M_2-2)}{4}}}{N!}\int_{-\infty}^\infty \frac{d^N\mu}{(2\pi)^N}\nonumber \\
&\quad
{\fontsize{10pt}{20pt}\selectfont
\times \text{pf}
\begin{pmatrix}
\hspace{-4pt}\left[\braket{\mu_m|{\hat *}^{11}|\nu_n}\right]_{m,n}^{\substack{N\\ \times N}}
&\hspace{-6pt}\left[\braket{\mu_m|{\hat *}^{12}|\nu_n}\right]_{m,n}^{\substack{N\\ \times N}}
&\hspace{-6pt}\left[\braket{\mu_m|*^{13}_s}\right]_{m,s}^{\substack{N\\ \times M_1}}
&\hspace{-6pt}\left[\braket{\mu_m|*^{14}_s}\right]_{m,s}^{\substack{N\\ \times M_2}}
&\hspace{-6pt}\left[\braket{\mu_m|*^{15}_s}\right]_{m,s}^{\substack{N\\ \times M_2}}
&\hspace{-6pt}\left[\braket{\mu_m|*^{16}_s}\right]_{m,s}^{\substack{N\\ \times M_1}}\hspace{-4pt}\\
\hspace{-4pt}\left[\braket{\mu_m|{\hat *}^{21}|\nu_n}\right]_{m,n}^{\substack{N\\ \times N}}
&\hspace{-6pt}\left[\braket{\mu_m|{\hat *}^{22}|\nu_n}\right]_{m,n}^{\substack{N\\ \times N}}
&\hspace{-6pt}\left[\braket{\mu_m|*^{23}_s}\right]_{m,s}^{\substack{N\\ \times M_1}}
&\hspace{-6pt}\left[\braket{\mu_m|*^{24}_s}\right]_{m,s}^{\substack{N\\ \times M_2}}
&\hspace{-6pt}\left[\braket{\mu_m|*^{25}_s}\right]_{m,s}^{\substack{N\\ \times M_2}}
&\hspace{-6pt}\left[\braket{\mu_m|*^{26}_s}\right]_{m,s}^{\substack{N\\ \times M_1}}\hspace{-4pt}\\
\hspace{-4pt}\left[\braket{*^{31}_r|\nu_n}\right]_{r,n}^{\substack{M_1\\ \times N}}
&\hspace{-6pt}\left[\braket{*^{32}_r|\nu_n}\right]_{r,n}^{\substack{M_1\\ \times N}}
&\hspace{-6pt}\left[*^{33}_{rs}\right]_{r,s}^{\substack{M_1\\ \times M_1}}
&\hspace{-6pt}\left[*^{34}_{rs}\right]_{r,s}^{\substack{M_1\\ \times M_2}}
&\hspace{-6pt}\left[*^{35}_{rs}\right]_{r,s}^{\substack{M_1\\ \times M_2}}
&\hspace{-6pt}\left[*^{36}_{rs}\right]_{r,s}^{\substack{M_1\\ \times M_1}}\hspace{-4pt}\\
\hspace{-4pt}\left[\braket{*^{41}_r|\nu_n}\right]_{r,n}^{\substack{M_2\\ \times N}}
&\hspace{-6pt}\left[\braket{*^{42}_r|\nu_n}\right]_{r,n}^{\substack{M_2\\ \times N}}
&\hspace{-6pt}\left[*^{43}_{rs}\right]_{r,s}^{\substack{M_2\\ \times M_1}}
&\hspace{-6pt}\left[*^{44}_{rs}\right]_{r,s}^{\substack{M_2\\ \times M_2}}
&\hspace{-6pt}\left[*^{45}_{rs}\right]_{r,s}^{\substack{M_2\\ \times M_2}}
&\hspace{-6pt}\left[0\right]^{\substack{M_2\\ \times M_1}}\hspace{-4pt}\\
\hspace{-4pt}\left[\braket{*^{51}_r|\nu_n}\right]_{r,n}^{\substack{M_2\\ \times N}}
&\hspace{-6pt}\left[\braket{*^{52}_r|\nu_n}\right]_{r,n}^{\substack{M_2\\ \times N}}
&\hspace{-6pt}\left[*^{53}_{rs}\right]_{r,s}^{\substack{M_2\\ \times M_1}}
&\hspace{-6pt}\left[*^{54}_{rs}\right]_{r,s}^{\substack{M_2\\ \times M_2}}
&\hspace{-6pt}\left[0\right]^{\substack{M_2\\ \times M_2}}
&\hspace{-6pt}\left[0\right]^{\substack{M_2\\ \times M_1}}\hspace{-4pt}\\
\hspace{-4pt}\left[\braket{*^{61}_r|\nu_n}\right]_{r,n}^{\substack{M_1\\ \times N}}
&\hspace{-6pt}\left[\braket{*^{62}_r|\nu_n}\right]_{r,n}^{\substack{M_1\\ \times N}}
&\hspace{-6pt}\left[*^{63}_{rs}\right]_{r,s}^{\substack{M_1\\ \times M_1}}
&\hspace{-6pt}\left[0\right]^{\substack{M_1\\ \times M_2}}
&\hspace{-6pt}\left[0\right]^{\substack{M_1\\ \times M_2}}
&\hspace{-6pt}\left[0\right]^{\substack{M_1\\ \times M_1}}\hspace{-4pt}
\end{pmatrix}
},
\end{align}
with
\begin{align}
&{\hat *}^{11}=
-{\hat S}{\hat {\tilde U}}{\hat S}^t,\quad
{\hat *}^{12}=-{\hat S}{\hat {\tilde U}}{\hat S}^t{\hat T}^t,\quad
\ket{*^{13}_s}=
-{\hat S}{\hat {\tilde U}}{\hat S}^t\ket{{\tilde V}_{1,s}},\quad
\ket{*^{14}_s}=
-{\hat S}{\hat {\tilde U}}\kket{-t_{M_2,s}},\nonumber \\
&\ket{*^{15}_s}=
{\hat S}\ket{V_{2,s}},\quad
\ket{*^{16}_s}=
\kket{-t_{M_1,s}},\nonumber \\
&{\hat *}^{21}=-{\hat T}{\hat S}{\hat {\tilde U}}{\hat S}^t,\quad
{\hat *}^{22}=-{\hat T}{\hat S}{\hat {\tilde U}}{\hat S}^t{\hat T}^t,\quad
\ket{*^{23}_s}=-{\hat T}{\hat S}{\hat {\tilde U}}{\hat S}^t\ket{{\tilde V}_{1,s}},\quad
\ket{*^{24}_s}=-{\hat T}{\hat S}{\hat {\tilde U}}\kket{-t_{M_2,s}},\nonumber \\
&\ket{*^{25}_s}={\hat T}{\hat S}\ket{V_{2,s}},\quad
\ket{*^{26}_s}={\hat T}\kket{-t_{M_1,s}},\nonumber \\
&\bra{*^{31}_r}=
-\bra{V_{1,r}}{\hat S}{\hat {\tilde U}}{\hat S}^t,\quad
\bra{*^{32}_r}=
-\bra{V_{1,r}}{\hat S}{\hat {\tilde U}}{\hat S}^t{\hat T}^t,\quad
*^{33}_{rs}=
-\braket{V_{1,r}|{\hat S}{\hat {\tilde U}}{\hat S}^t|{\tilde V}_{1,s}},\nonumber \\
&*^{34}_{rs}=
-\brakket{V_{1,r}|{\hat S}{\hat {\tilde U}}|-t_{M_2,s}},\quad
*^{35}_{rs}=
\braket{V_{1,r}|{\hat S}|V_{2,s}},\quad
*^{36}_{rs}=
\brakket{V_{1,r}|-t_{M_1,s}},\nonumber \\
&\bra{*^{41}_r}=
-\bbra{t_{M_2,r}}{\hat {\tilde U}}{\hat S}^t,\quad
\bra{*^{42}_r}=
-\bbra{t_{M_2,r}}{\hat {\tilde U}}{\hat S}^t{\hat T}^t,\quad
*^{43}_{rs}=
-\bbraket{t_{M_2,r}|{\hat {\tilde U}}{\hat S}^t|{\tilde V}_{1,s}},\nonumber \\
&*^{44}_{rs}=
-\bbrakket{t_{M_2,r}|{\hat {\tilde U}}|-t_{M_2,s}},\quad
*^{45}_{rs}=
\bbraket{t_{M_2,r}|V_{2,s}},\nonumber \\
&\bra{*^{51}_r}=
-\bra{{\tilde V}_{2,r}}{\hat S}^t,\quad
\bra{*^{52}_r}=
-\bra{{\tilde V}_{2,r}}{\hat S}^t{\hat T}^t,\quad
*^{53}_{rs}=
-\braket{{\tilde V}_{2,r}|{\hat S}^t|{\tilde V}_{1,s}},\quad
*^{54}_{rs}=
-\brakket{{\tilde V}_{2,r}|-t_{M_2,s}},\nonumber \\
&\bra{*^{61}_r}=
-\bbra{t_{M_1,r}},\quad
\bra{*^{62}_r}=
-\bbra{t_{M_1,r}}{\hat T}^t,\quad
*^{63}_{rs}=
-\bbraket{t_{M_1,r}|{\tilde V}_{1,s}}.
\end{align}
Here ${\hat {\cal O}}^t$ is the transpose of an operator in position basis which can be calculated according to the rules \eqref{transposerules}.
We have also defined
\begin{align}
\ket{{\tilde V}_{1,r}}=(\bra{V_{1,r}})^t=e^{\frac{i{\hat x}^2}{2\pi k}}e^{-\frac{i(\zeta_2+\zeta_1){\hat x}}{2\pi k}}\kket{-t_{M_1,r}},\quad
\bra{{\tilde V}_{2,r}}=(\ket{V_{2,r}})^t=\bbra{t_{M_2,r}}e^{-\frac{i{\hat x}^2}{2\pi k}}e^{-\frac{i(-\zeta_{\ell}-\zeta_{\ell-1}){\hat x}}{2\pi k}},\quad
{\hat {\tilde U}}={\hat U}-{\hat U}^t.
\end{align}

By using the Fredholm Pfaffian formula \eqref{FredholmPfaffian2} we obtain
\begin{align}
\Xi(\kappa)=
e^{\pi i(-\frac{M_1^2}{4}+\frac{M_2^2}{4}+M_1M_2)}
\sqrt{\text{Det}
\begin{pmatrix}
\hspace{-4pt}\kappa{\hat *}^{11}
&\hspace{-6pt}1+\kappa{\hat *}^{12}
&\hspace{-6pt}\left[\kappa\ket{*^{13}_s}\right]_s^{M_1}
&\hspace{-6pt}\left[\kappa\ket{*^{14}_s}\right]_s^{M_2}
&\hspace{-6pt}\left[\kappa\ket{*^{15}_s}\right]_s^{M_2}
&\hspace{-6pt}\left[\kappa\ket{*^{16}_s}\right]_s^{M_1}\hspace{-4pt}\\
\hspace{-4pt}-1+\kappa{\hat *}^{21}
&\hspace{-6pt}\kappa{\hat *}^{22}
&\hspace{-6pt}\left[\kappa\ket{*^{23}_s}\right]_s^{M_1}
&\hspace{-6pt}\left[\kappa\ket{*^{24}_s}\right]_s^{M_2}
&\hspace{-6pt}\left[\kappa\ket{*^{25}_s}\right]_s^{M_2}
&\hspace{-6pt}\left[\kappa\ket{*^{26}_s}\right]_s^{M_1}\hspace{-4pt}\\
\hspace{-4pt}\left[\bra{*^{31}_r}\right]_r^{M_1}
&\hspace{-6pt}\left[\bra{*^{32}_r}\right]_r^{M_1}
&\hspace{-6pt}\left[*^{33}_{rs}\right]_{r,s}^{\substack{M_1\\ \times M_1}}
&\hspace{-6pt}\left[*^{34}_{rs}\right]_{r,s}^{\substack{M_1\\ \times M_2}}
&\hspace{-6pt}\left[*^{35}_{rs}\right]_{r,s}^{\substack{M_1\\ \times M_2}}
&\hspace{-6pt}\left[*^{36}_{rs}\right]_{r,s}^{\substack{M_1\\ \times M_1}}\hspace{-4pt}\\
\hspace{-4pt}\left[\bra{*^{41}_r}\right]_r^{M_2}
&\hspace{-6pt}\left[\bra{*^{42}_r}\right]_r^{M_2}
&\hspace{-6pt}\left[*^{43}_{rs}\right]_{r,s}^{\substack{M_2\\ \times M_1}}
&\hspace{-6pt}\left[*^{44}_{rs}\right]_{r,s}^{\substack{M_2\\ \times M_2}}
&\hspace{-6pt}\left[*^{45}_{rs}\right]_{r,s}^{\substack{M_2\\ \times M_2}}
&\hspace{-6pt}\left[0\right]^{\substack{M_2\\ \times M_1}}\hspace{-4pt}\\
\hspace{-4pt}\left[\bra{*^{51}_r}\right]_r^{M_2}
&\hspace{-6pt}\left[\bra{*^{52}_r}\right]_r^{M_2}
&\hspace{-6pt}\left[*^{53}_{rs}\right]_{r,s}^{\substack{M_2\\ \times M_1}}
&\hspace{-6pt}\left[*^{54}_{rs}\right]_{r,s}^{\substack{M_2\\ \times M_2}}
&\hspace{-6pt}\left[0\right]^{\substack{M_2\\ \times M_2}}
&\hspace{-6pt}\left[0\right]^{\substack{M_2\\ \times M_1}}\hspace{-4pt}\\
\hspace{-4pt}\left[\bra{*^{61}_r}\right]_r^{M_1}
&\hspace{-6pt}\left[\bra{*^{62}_r}\right]_r^{M_1}
&\hspace{-6pt}\left[*^{63}_{rs}\right]_{r,s}^{\substack{M_1\\ \times M_1}}
&\hspace{-6pt}\left[0\right]^{\substack{M_1\\ \times M_2}}
&\hspace{-6pt}\left[0\right]^{\substack{M_1\\ \times M_2}}
&\hspace{-6pt}\left[0\right]^{\substack{M_1\\ \times M_1}}\hspace{-4pt}\\
\end{pmatrix}}.
\end{align}
The semi-functional matrix under the determinant can be transformed into a block diagonal matrix by the elementary row/column operations in the following way.
For the purpose of explanations let us write the matrix components explicitly
\begin{align}
&\frac{\Xi(\kappa)}{e^{\pi i(-\frac{M_1^2}{4}+\frac{M_2^2}{4}+M_1M_2)}}\nonumber \\
&=
{\fontsize{7pt}{1pt}\selectfont
\sqrt{\text{Det}\begin{pmatrix}
-\kappa{\hat S}{\hat {\tilde U}}{\hat S}^t&
1-\kappa{\hat S}{\hat {\tilde U}}{\hat S}^t{\hat T}^t&
-\kappa{\hat S}{\hat {\tilde U}}{\hat S}^t\ket{{\tilde V}_{1,s}}&
-\kappa{\hat S}{\hat {\tilde U}}\kket{-t_{M_2,s}}&
\kappa{\hat S}\ket{V_{2,s}}&
\kappa\kket{-t_{M_1,s}}\\
-1-\kappa{\hat T}{\hat S}{\hat {\tilde U}}{\hat S}^t&
-\kappa{\hat T}{\hat S}{\hat {\tilde U}}{\hat S}^t{\hat T}^t&
-\kappa{\hat T}{\hat S}{\hat {\tilde U}}{\hat S}^t\ket{{\tilde V}_{1,2}}&
-\kappa{\hat T}{\hat S}{\hat {\tilde U}}\kket{-t_{M_2,s}}&
\kappa{\hat T}{\hat S}\ket{V_{2,s}}&
\kappa{\hat T}\kket{-t_{M_1,s}}\\
\bra{V_{1,r}}(-{\hat S}{\hat {\tilde U}}{\hat S}^t)&
\bra{V_{1,r}}(-{\hat S}{\hat {\tilde U}}{\hat S}^t{\hat T}^t)&
\braket{V_{1,r}|(-{\hat S}{\hat {\tilde U}}{\hat S}^t)|{\tilde V}_{1,s}}&
\brakket{V_{1,r}|(-{\hat S}{\hat {\tilde U}})|-t_{M_2,s}}&
\braket{V_{1,r}|{\hat S}|V_{2,s}}&
\brakket{V_{1,r}|-t_{M_1,s}}\\
\bbra{t_{M_2,r}}(-{\hat {\tilde U}}{\hat S}^t)&
\bbra{t_{M_2,r}}(-{\hat {\tilde U}}{\hat S}^t{\hat T}^t)&
\bbraket{t_{M_2,r}|(-{\hat {\tilde U}}{\hat S}^t)|{\tilde V}_{1,s}}&
\bbrakket{t_{M_2,r}|(-{\hat {\tilde U}})|-t_{M_2,s}}&
\bbraket{t_{M_2,r}|V_{2,s}}&
0\\
\bra{{\tilde V}_{2,r}}(-{\hat S}^t)&
\bra{{\tilde V}_{2,r}}(-{\hat S}^t{\hat T}^t)&
\braket{{\tilde V}_{2,r}|(-{\hat S}^t)|{\tilde V}_{1,s}}&
-\brakket{{\tilde V}_{2,r}|-t_{M_2,s}}&
0&
0\\
-\bbra{t_{M_1,r}}&
\bbra{t_{M_1,r}}(-{\hat T}^t)&
-\bbraket{t_{M_1,r}|{\tilde V}_{1,s}}&
0&
0&
0
\end{pmatrix}
}
}.
\end{align}
To simplify the component first we add $(\text{first column})\times(-\ket{{\tilde V}_{1,s}})$ to the third column, add $(-{\hat T})\times(\text{first row})$ to the second row and then add $(\text{first column})\times(-{\hat T}^t)$ to the second column:
\begin{align}
&\frac{\Xi(\kappa)}{e^{\pi i(-\frac{M_1^2}{4}+\frac{M_2^2}{4}+M_1M_2)}}\nonumber \\
&=
{\fontsize{7pt}{1pt}\selectfont
\sqrt{\text{Det}\begin{pmatrix}
-\kappa{\hat S}{\hat {\tilde U}}{\hat S}^t&
1-\kappa{\hat S}{\hat {\tilde U}}{\hat S}^t{\hat T}^t&
0&
-\kappa{\hat S}{\hat {\tilde U}}\kket{-t_{M_2,s}}&
\kappa{\hat S}\ket{V_{2,s}}&
\kappa\kket{-t_{M_1,s}}\\
-1-\kappa{\hat T}{\hat S}{\hat {\tilde U}}{\hat S}^t&
-\kappa{\hat T}{\hat S}{\hat {\tilde U}}{\hat S}^t{\hat T}^t&
\ket{{\tilde V}_{1,s}}&
-\kappa{\hat T}{\hat S}{\hat {\tilde U}}\kket{-t_{M_2,s}}&
\kappa{\hat T}{\hat S}\ket{V_{2,s}}&
\kappa{\hat T}\kket{-t_{M_1,s}}\\
\bra{V_{1,r}}(-{\hat S}{\hat {\tilde U}}{\hat S}^t)&
\bra{V_{1,r}}(-{\hat S}{\hat {\tilde U}}{\hat S}^t{\hat T}^t)&
0&
\brakket{V_{1,r}|(-{\hat S}{\hat {\tilde U}})|-t_{M_2,s}}&
\braket{V_{1,r}|{\hat S}|V_{2,s}}&
\brakket{V_{1,r}|-t_{M_1,s}}\\
\bbra{t_{M_2,r}}(-{\hat {\tilde U}}{\hat S}^t)&
\bbra{t_{M_2,r}}(-{\hat {\tilde U}}{\hat S}^t{\hat T}^t)&
0&
\bbrakket{t_{M_2,r}|(-{\hat {\tilde U}})|-t_{M_2,s}}&
\bbraket{t_{M_2,r}|V_{2,s}}&
0\\
\bra{{\tilde V}_{2,r}}(-{\hat S}^t)&
\bra{{\tilde V}_{2,r}}(-{\hat S}^t{\hat T}^t)&
0&
-\brakket{{\tilde V}_{2,r}|-t_{M_2,s}}&
0&
0\\
-\bbra{t_{M_1,r}}&
\bbra{t_{M_1,r}}(-{\hat T}^t)&
0&
0&
0&
0
\end{pmatrix}
}
}\nonumber \\
&=
{\fontsize{7pt}{1pt}\selectfont
\sqrt{\text{Det}\begin{pmatrix}
-\kappa{\hat S}{\hat {\tilde U}}{\hat S}^t&
1-\kappa{\hat S}{\hat {\tilde U}}{\hat S}^t{\hat T}^t&
0&
-\kappa{\hat S}{\hat {\tilde U}}\kket{-t_{M_2,s}}&
\kappa{\hat S}\ket{V_{2,s}}&
\kappa\kket{-t_{M_1,s}}\\
-1&
-{\hat T}&
\ket{{\tilde V}_{1,s}}&
0&
0&
0\\
\bra{V_{1,r}}(-{\hat S}{\hat {\tilde U}}{\hat S}^t)&
\bra{V_{1,r}}(-{\hat S}{\hat {\tilde U}}{\hat S}^t{\hat T}^t)&
0&
\brakket{V_{1,r}|(-{\hat S}{\hat {\tilde U}})|-t_{M_2,s}}&
\braket{V_{1,r}|{\hat S}|V_{2,s}}&
\brakket{V_{1,r}|-t_{M_1,s}}\\
\bbra{t_{M_2,r}}(-{\hat {\tilde U}}{\hat S}^t)&
\bbra{t_{M_2,r}}(-{\hat {\tilde U}}{\hat S}^t{\hat T}^t)&
0&
\bbrakket{t_{M_2,r}|(-{\hat {\tilde U}})|-t_{M_2,s}}&
\bbraket{t_{M_2,r}|V_{2,s}}&
0\\
\bra{{\tilde V}_{2,r}}(-{\hat S}^t)&
\bra{{\tilde V}_{2,r}}(-{\hat S}^t{\hat T}^t)&
0&
-\brakket{{\tilde V}_{2,r}|-t_{M_2,s}}&
0&
0\\
-\bbra{t_{M_1,r}}&
\bbra{t_{M_1,r}}(-{\hat T}^t)&
0&
0&
0&
0
\end{pmatrix}
}
}\nonumber \\
&=
{\fontsize{7pt}{1pt}\selectfont
\sqrt{\text{Det}\begin{pmatrix}
-\kappa{\hat S}{\hat {\tilde U}}{\hat S}^t&
1&
0&
-\kappa{\hat S}{\hat {\tilde U}}\kket{-t_{M_2,s}}&
\kappa{\hat S}\ket{V_{2,s}}&
\kappa\kket{-t_{M_1,s}}\\
-1&
-{\hat T}&
\ket{{\tilde V}_{1,s}}&
0&
0&
0\\
\bra{V_{1,r}}(-{\hat S}{\hat {\tilde U}}{\hat S}^t)&
0&
0&
\brakket{V_{1,r}|(-{\hat S}{\hat {\tilde U}})|-t_{M_2,s}}&
\braket{V_{1,r}|{\hat S}|V_{2,s}}&
\brakket{V_{1,r}|-t_{M_1,s}}\\
\bbra{t_{M_2,r}}(-{\hat {\tilde U}}{\hat S}^t)&
0&
0&
\bbrakket{t_{M_2,r}|(-{\hat {\tilde U}})|-t_{M_2,s}}&
\bbraket{t_{M_2,r}|V_{2,s}}&
0\\
\bra{{\tilde V}_{2,r}}(-{\hat S}^t)&
0&
0&
-\brakket{{\tilde V}_{2,r}|-t_{M_2,s}}&
0&
0\\
-\bbra{t_{M_1,r}}&
0&
0&
0&
0&
0
\end{pmatrix}
}
}.
\end{align}
Then we add $(-\kappa{\hat S}{\hat {\tilde U}}{\hat S}^t)\times(\text{second row})$ to the first row to find
\begin{align}
&\frac{\Xi(\kappa)}{e^{\pi i(-\frac{M_1^2}{4}+\frac{M_2^2}{4}+M_1M_2)}}\nonumber \\
&=
{\fontsize{9.5pt}{1pt}\selectfont
\sqrt{\text{Det}\begin{pmatrix}
0&
1+\kappa{\hat S}{\hat {\tilde U}}{\hat S}^t{\hat T}&
-\kappa{\hat S}{\hat {\tilde U}}{\hat S}^t\ket{{\tilde V}_{1,s}}&
-\kappa{\hat S}{\hat {\tilde U}}\kket{-t_{M_2,s}}&
\kappa{\hat S}\ket{V_{2,s}}&
\kappa\kket{-t_{M_1,s}}\\
-1&
-{\hat T}&
\ket{{\tilde V}_{1,s}}&
0&
0&
0\\
\bra{V_{1,r}}(-{\hat S}{\hat {\tilde U}}{\hat S}^t)&
0&
0&
\brakket{V_{1,r}|(-{\hat S}{\hat {\tilde U}})|-t_{M_2,s}}&
\braket{V_{1,r}|{\hat S}|V_{2,s}}&
\brakket{V_{1,r}|-t_{M_1,s}}\\
\bbra{t_{M_2,r}}(-{\hat {\tilde U}}{\hat S}^t)&
0&
0&
\bbrakket{t_{M_2,r}|(-{\hat {\tilde U}})|-t_{M_2,s}}&
\bbraket{t_{M_2,r}|V_{2,s}}&
0\\
\bra{{\tilde V}_{2,r}}(-{\hat S}^t)&
0&
0&
-\brakket{{\tilde V}_{2,r}|-t_{M_2,s}}&
0&
0\\
-\bbra{t_{M_1,r}}&
0&
0&
0&
0&
0
\end{pmatrix}
}
}.
\end{align}
Finally we perform the column operations with the second column to turn the 13,14,15,16 components to zero, and then perform the column operations with the first column to turn the 23,24,25,26 components to zero.
As a result we obtain
\begin{align}
\Xi(\kappa)
=
e^{\pi i(-\frac{M_1^2}{4}+\frac{M_2^2}{4}+M_1M_2)}
\sqrt{\text{Det}
\begin{pmatrix}
\hspace{-2pt}0
&\hspace{-6pt}1+\kappa{\hat \rho}^{\text{(open)}}
&\hspace{-6pt}\left[0\right]^{2M_2}
\hspace{-2pt}\\
\hspace{-2pt}-1
&\hspace{-6pt}-{\hat {\tilde T}}
&\hspace{-6pt}\left[0\right]^{2M_2}\hspace{-2pt}\\
\hspace{-2pt}\left[0\right]^{2M_1}
&\hspace{-6pt}\left[0\right]^{2M_1}
&\left[H_{ab}(\kappa)\right]_{a,b}^{(2M_1+2M_2)\times (2M_1+2M_2)}\hspace{-2pt}
\end{pmatrix}
},
\end{align}
where
\begin{align}
{\hat {\tilde T}}={\hat T}-{\hat T}^t,
\end{align}
and
\begin{align}
&{\hat\rho}^{\text{(open)}}={\hat S}{\hat {\tilde U}}{\hat S}^t{\hat {\tilde T}},\label{rhoopen} \\
&\left[H_{ab}(\kappa)\right]_{a,b}^{(2M_1+2M_2)\times (2M_1+2M_2)}
=\begin{pmatrix}
\hspace{-2pt}\left[H^{(11)}_{rs}(\kappa)\right]_{r,s}^{M_1\times M_1}
&\hspace{-6pt}\left[H^{(12)}_{rs}(\kappa)\right]_{r,s}^{M_1\times M_2}
&\hspace{-6pt}\left[H^{(13)}_{rs}(\kappa)\right]_{r,s}^{M_1\times M_2}
&\hspace{-6pt}\left[H^{(14)}_{rs}(\kappa)\right]_{r,s}^{M_1\times M_1}\hspace{-2pt}\\
\hspace{-2pt}\left[H^{(21)}_{rs}(\kappa)\right]_{r,s}^{M_2\times M_1}
&\hspace{-6pt}\left[H^{(22)}_{rs}(\kappa)\right]_{r,s}^{M_2\times M_2}
&\hspace{-6pt}\left[H^{(23)}_{rs}(\kappa)\right]_{r,s}^{M_2\times M_2}
&\hspace{-6pt}\left[H^{(24)}_{rs}(\kappa)\right]_{r,s}^{M_2\times M_1}\hspace{-2pt}\\
\hspace{-2pt}\left[H^{(31)}_{rs}(\kappa)\right]_{r,s}^{M_2\times M_1}
&\hspace{-6pt}\left[H^{(32)}_{rs}(\kappa)\right]_{r,s}^{M_2\times M_2}
&\hspace{-6pt}\left[H^{(33)}_{rs}(\kappa)\right]_{r,s}^{M_2\times M_2}
&\hspace{-6pt}\left[H^{(34)}_{rs}(\kappa)\right]_{r,s}^{M_2\times M_1}\hspace{-2pt}\\
\hspace{-2pt}\left[H^{(41)}_{rs}(\kappa)\right]_{r,s}^{M_1\times M_1}
&\hspace{-6pt}\left[H^{(42)}_{rs}(\kappa)\right]_{r,s}^{M_1\times M_2}
&\hspace{-6pt}\left[H^{(43)}_{rs}(\kappa)\right]_{r,s}^{M_1\times M_2}
&\hspace{-6pt}\left[H^{(44)}_{rs}(\kappa)\right]_{r,s}^{M_1\times M_1}\hspace{-2pt}
\end{pmatrix},
\label{Hab}
\end{align}
with
\begin{align}
&H^{(11)}_{rs}(\kappa)=-\braket{V_{1,r}|\frac{1}{1+\kappa{\hat S}{\hat {\tilde U}}{\hat S}^t{\hat {\tilde T}}}{\hat S}{\hat {\tilde U}}{\hat S}^t|{\tilde V}_{1,s}},\quad
H^{(12)}_{rs}(\kappa)=-\brakket{V_{1,r}|\frac{1}{1+\kappa{\hat S}{\hat {\tilde U}}{\hat S}^t{\hat {\tilde T}}}{\hat S}{\hat {\tilde U}}|-t_{M_2,s}},\nonumber \\
&H^{(13)}_{rs}(\kappa)=\braket{V_{1,r}|\frac{1}{1+\kappa{\hat S}{\hat {\tilde U}}{\hat S}^t{\hat {\tilde T}}}{\hat S}|V_{2,s}},\quad
H^{(14)}_{rs}(\kappa)=\brakket{V_{1,r}|\frac{1}{1+\kappa{\hat S}{\hat {\tilde U}}{\hat S}^t{\hat {\tilde T}}}|-t_{M_1,s}},\nonumber \\
&H^{(21)}_{rs}(\kappa)=-\bbraket{t_{M_2,r}|\frac{1}{1+\kappa{\hat {\tilde U}}{\hat S}^t{\hat {\tilde T}}{\hat S}}{\hat {\tilde U}}{\hat S}^t|{\tilde V}_{1,s}},\quad
H^{(22)}_{rs}(\kappa)=-\bbrakket{t_{M_2,r}|\frac{1}{1+\kappa{\hat {\tilde U}}{\hat S}^t{\hat {\tilde T}}{\hat S}}{\hat {\tilde U}}|-t_{M_2,s}},\nonumber \\
&H^{(23)}_{rs}(\kappa)=\bbraket{t_{M_2,r}|\frac{1}{1+\kappa{\hat {\tilde U}}{\hat S}^t{\hat {\tilde T}}{\hat S}}|V_{2,s}},\quad
H^{(24)}_{rs}(\kappa)=-\bbrakket{t_{M_2,r}|\frac{\kappa}{1+\kappa{\hat {\tilde U}}{\hat S}^t{\hat {\tilde T}}{\hat S}}{\hat {\tilde U}}{\hat S}^t{\hat {\tilde T}}|-t_{M_1,s}},\nonumber \\
&H^{(31)}_{rs}(\kappa)=-\braket{{\tilde V}_{2,r}|\frac{1}{1+\kappa{\hat S}^t{\hat {\tilde T}}{\hat S}{\hat {\tilde U}}}{\hat S}^t|{\tilde V}_{1,s}},\quad
H^{(32)}_{rs}(\kappa)=-\brakket{{\tilde V}_{2,r}|\frac{1}{1+\kappa{\hat S}^t{\hat {\tilde T}}{\hat S}{\hat {\tilde U}}}|-t_{M_2,s}},\nonumber \\
&H^{(33)}_{rs}(\kappa)=-\braket{{\tilde V}_{2,r}|\frac{\kappa}{1+\kappa {\hat S}^t{\hat {\tilde T}}{\hat S}{\hat {\tilde U}}}{\hat S}^t{\hat {\tilde T}}{\hat S}|V_{2,s}},\quad
H^{(34)}_{rs}(\kappa)=-\brakket{{\tilde V}_{2,r}|\frac{\kappa}{1+\kappa{\hat S}^t{\hat {\tilde T}}{\hat S}{\hat {\tilde U}}}{\hat S}^t{\hat {\tilde T}}|-t_{M_1,s}},\nonumber \\
&H^{(41)}_{rs}(\kappa)=-\bbraket{t_{M_1,r}|\frac{1}{1+\kappa{\hat {\tilde T}}{\hat S}{\hat {\tilde U}}{\hat S}^t}|{\tilde V}_{1,s}},\quad
H^{(42)}_{rs}(\kappa)=\bbrakket{t_{M_1,r}|\frac{\kappa}{1+\kappa{\hat {\tilde T}}{\hat S}{\hat {\tilde U}}{\hat S}^t}{\hat {\tilde T}}{\hat S}{\hat {\tilde U}}|-t_{M_2,s}},\nonumber \\
&H^{(43)}_{rs}(\kappa)=-\bbraket{t_{M_1,r}|\frac{\kappa}{1+\kappa{\hat {\tilde T}}{\hat S}{\hat {\tilde U}}{\hat S}^t}{\hat {\tilde T}}{\hat S}|V_{2,s}},\quad
H^{(44)}_{rs}(\kappa)=-\bbrakket{t_{M_1,r}|\frac{\kappa}{1+\kappa{\hat {\tilde T}}{\hat S}{\hat {\tilde U}}{\hat S}^t}{\hat {\tilde T}}|-t_{M_1,s}}.
\end{align}
Note that $H_{ab}(\kappa)$ is an anti-symmetric matrix, namely
\begin{align}
H^{(\alpha\beta)}_{rs}(\kappa)=-H^{(\beta\alpha)}_{sr}(\kappa).
\end{align}
Hence we finally obtain \eqref{Xiopen} with ${\hat \rho}^{\text{(open)}}$ given in \eqref{rhoopen} and $H_{ab}(\kappa)$ given in \eqref{Hab}.

Let us comment that the grand partition function \eqref{Xiopen} reduces drastically in the limit $\kappa\rightarrow 0$.
Besides that the Fredholm determinant becomes $1$, the determinant of the $(2M_1+2M_2)\times (2M_1+2M_2)$ matrix $H$ also simplifies since
$H_{rs}^{(24)}(0)=H_{rs}^{(33)}(0)=H_{rs}^{(34)}(0)=H_{rs}^{(42)}(0)=H_{rs}^{(43)}(0)=H_{rs}^{(44)}(0)=0$, as
\begin{align}
\det\left(\left[H_{ab}(0)\right]_{a,b}^{(2M_1+2M_2)\times (2M_1+2M_2)}\right)
&=\left[(-1)^{M_1M_2}
\det\left(\left[H_{rs}^{(14)}(0)\right]_{r,s}^{M_1\times M_1}\right)
\det\left(\left[H_{rs}^{(23)}(0)\right]_{r,s}^{M_2\times M_2}\right)
\right]^2,
\end{align}
where
\begin{align}
H^{(14)}_{rs}=\brakket{V_{1,r}|-t_{M_1,s}},\quad
H^{(23)}_{rs}=\bbraket{t_{M_2,r}|V_{2,s}}.
\end{align}
Hence we have
\begin{align}
\Xi(0)=
e^{\pi i(-\frac{M_1^2}{4}+\frac{M_2^2}{4})}
\det\left(\left[\brakket{V_{1,r}|-t_{M_1,s}}\right]^{M_1\times M_1}_{r,s}\right)
\det\left(\left[\bbraket{t_{M_2,r}|V_{2,s}}\right]^{M_2\times M_2}_{r,s}\right).
\label{Xiopenatkappa0}
\end{align}
Indeed, when $N=0$ the model in figure \ref{Drquiver_Laall0} reduces to a pair of decoupled pure Chern-Simons theory with rank $M_1,M_2$, level $\pm 2k$ and the FI parameters $\zeta_1+\zeta_2,-\zeta_3-\zeta_4$, whose partition function is given by \eqref{Xiopenatkappa0}.

\section{Closed string formalism for ${\hat{D}}_{4}$ model}\label{sec_closed}
\begin{figure}
\centering{}\includegraphics[width=12cm]{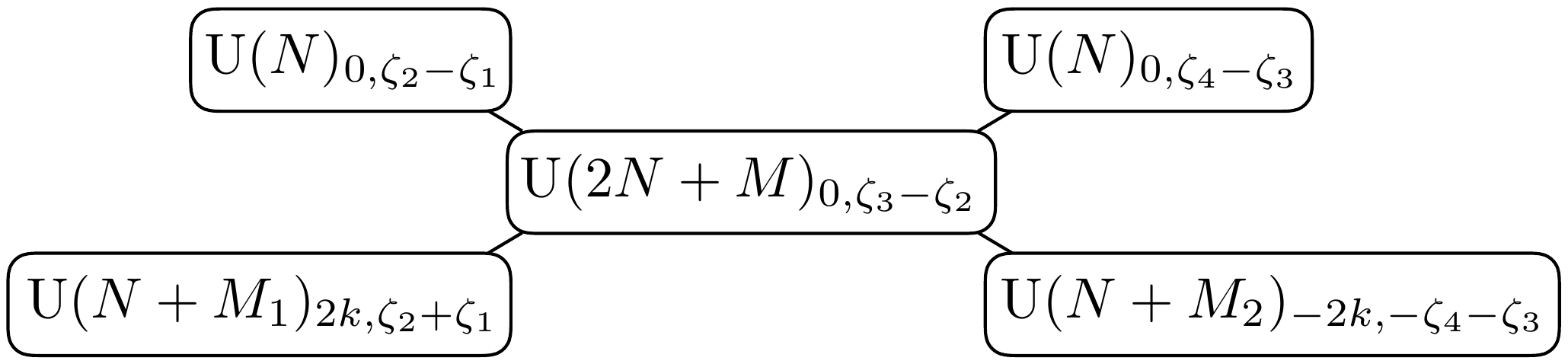}
\caption{Quiver diagram of the ${\hat{D}}_{4}$ quiver superconformal Chern-Simons theory with three rank deformations. Here we have set $M_3=M$.
}
\label{D4quiver} 
\end{figure}

In the previous section we have applied the open string formalism to ${\hat{D}}_{\ell}$ quiver theory where only the ranks of the two affine nodes are deformed (see figure \ref{Drquiver_Laall0}). In this section we apply the closed string formalism to the ${\hat{D}}_{4}$ model where, in addition to the two affine nodes, the rank of the non-affine node is also deformed. Figure \ref{D4quiver} shows the gauge group which we will consider. 

Before going on, we carefully restrict the region of the ranks. First, as discussed in section \ref{sec:Model}, we avoid the ``bad'' region \cite{Gaiotto:2008ak}. This restricts the range of the rank difference of the non-Affine node $M$ to be 
\begin{equation}
2M-1\leq M_{1}+M_{2}.\label{eq:NotBad}
\end{equation}
Second, we find that large $M_{1}$ or $M_{2}$ compared with $2k$ introduces a subtlety when the Fermi gas formalism is applied. Although it may seem rather technical issue, here we restrict the region to
\begin{equation}
M_{1}\leq2k,\quad M_{2}\leq2k.\label{eq:Cascade}
\end{equation}

In section \ref{CSF}
we apply the closed string formalism to the partition function.
This formalism encodes the almost all information of the partition function in the single density matrix except for $N$-independent part.
Based on this result, in section \ref{symmetries} we study symmetries of this density matrix and discuss their physical interpretations.

\subsection{Closed string formalism\label{CSF}}

\subsubsection{Partition function}

In this section we apply the closed string formalism to the partition function. The partition function of the $\hat{D}_{4}$ theory (shown in figure \ref{D4quiver}) is given by \eqref{Dr} with $\ell=4$ and $M_{3}=M$. As explained, this partition function can be rewritten as \eqref{Dr_centergeneral}. By using the Cauchy-Vandermonde determinant formulas \eqref{CauchyVdm1} and \eqref{CauchyVdm2}, the factor ${\cal Z}(\bar{\mu}_m,\bar{\nu}_n)$ defined in
\eqref{calZ} can be rewritten as a product of two determinants. For later convenience we also put the Fresnel factors
$e^{\frac{i}{2\pi k}(\sum_{m=1}^{N+M_1}(\mu_m')^2-\sum_{m=1}^{N+M_2}(\nu_m')^2)}$
and the FI factors properly into the four determinants.
Then we obtain
\begin{equation}
Z_{k}(N;M_{1},M_{2},M;\zeta_{a})=\frac{(-1)^{N(M_1-M_2)}i^{-\frac{M_1^2}{2}+\frac{M_2^2}{2}}}{(2N+M)!}\int_{-\infty}^{\infty}\frac{d^{2N+M}\lambda}{(2\pi)^{2N+M}}{\cal X}^{(\bar{\mu})}(\lambda_{m}){\cal X}^{(\bar{\nu})}(\lambda_{m}),
\end{equation}
with 
\begin{align}
 & {\cal X}^{(\bar{\mu})}(\lambda_{m})\nonumber \\
 & =\frac{1}{N!(N+M_{1})!}\int_{-\infty}^{\infty}\frac{d^{2N+M_{1}}\bar{\mu}}{(2\pi)^{2N+M_{1}}}\det\begin{pmatrix}\left[\braket{\mu_{m}|e^{-\frac{i}{4\pi k}\hat{x}^{2}+\frac{i\zeta_{1}}{2\pi k}\hat{x}}\frac{\tanh\frac{\hat{p}-\pi iM_{1}}{2}}{2}e^{\frac{i}{4\pi k}\hat{x}^{2}-\frac{i\zeta_{1}}{2\pi k}\hat{x}}|\mu_{n}'}\right]_{m,n}^{N\times(N+M_{1})}\\
\left[\bbraket{t_{M_{1},r}|e^{\frac{i}{4\pi k}\hat{x}^{2}-\frac{i\zeta_{1}}{2\pi k}\hat{x}}|\mu_{n}'}\right]_{r,n}^{M_{1}\times(N+M_{1})}
\end{pmatrix}\nonumber \\
 & \times D\left(\bar{\mu}_m,\lambda_n;2N+M_{1},2N+M;e^{\frac{i}{4\pi k}\hat{x}^{2}-\frac{i\zeta_{2}}{2\pi k}\hat{x}}\right),\label{ZmuDef}\\
 & {\cal X}^{(\bar{\nu})}(\lambda_{m})\nonumber \\
 & =\frac{1}{N!(N+M_{2})!}\int_{-\infty}^{\infty}\frac{d^{2N+M_{2}}\bar{\nu}}{(2\pi)^{2N+M_{2}}}D\left(\lambda_m,\bar{\nu}_n;2N+M,2N+M_{2};e^{\frac{i}{4\pi k}\hat{x}^{2}-\frac{i\zeta_{3}}{2\pi k}\hat{x}}\right)\nonumber \\
 & \times\det\left(\hspace{-6pt}\begin{array}{cc}
\left[\braket{\nu'_{m}|e^{-\frac{i}{4\pi k}\hat{x}^{2}+\frac{i\zeta_{4}}{2\pi k}\hat{x}}\frac{\tanh\frac{\hat{p}+\pi iM_{2}}{2}}{2}e^{\frac{i}{4\pi k}\hat{x}^{2}-\frac{i\zeta_{4}}{2\pi k}\hat{x}}|\nu_{n}}\right]_{m,n}^{\substack{(N+M_{2})\\
\times N
}
} & \hspace{-6pt}\left[\brakket{\nu'_{m}|e^{-\frac{i}{4\pi k}\hat{x}^{2}+\frac{i\zeta_{4}}{2\pi k}\hat{x}}|-t_{M_{2},s}}\right]_{m,s}^{\substack{(N+M_{2})\\
\times M_{2}
}
}\end{array}\hspace{-6pt}\right),\label{ZnuDef}
\end{align}
where 
\begin{align}
 & D\left(\alpha_m,\beta_n;N_{1},N_{2};\hat{\mathcal{O}}\right)\nonumber \\
 & =\begin{cases}
\det\left(\begin{array}{cc}
\left[\braket{\alpha_{m}|\hat{\mathcal{O}}\frac{1}{2\cosh\frac{\hat{p}+\pi i\left(N_{1}-N_{2}\right)}{2}}\hat{\mathcal{O}}^{-1}|\beta_{n}}\right]_{m,n}^{N_{1}\times N_{2}} & \left[\brakket{\alpha_{m}|\hat{\mathcal{O}}|-t_{N_{1}-N_{2},s}}\right]_{m,s}^{N_{1}\times\left(N_{1}-N_{2}\right)}\end{array}\right) & \left(N_{1}\geq N_{2}\right)\\
\det\begin{pmatrix}\left[\braket{\alpha_{m}|\hat{\mathcal{O}}\frac{1}{2\cosh\frac{\hat{p}-\pi i\left(N_{2}-N_{1}\right)}{2}}\hat{\mathcal{O}}^{-1}|\beta_{n}}\right]_{m,n}^{N_{1}\times N_{2}}\\
\left[\bbraket{t_{N_{2}-N_{1},r}|\hat{\mathcal{O}}^{-1}|\beta_{n}}\right]_{r,n}^{\left(N_{2}-N_{1}\right)\times N_{2}}
\end{pmatrix} & \left(N_{1}<N_{2}\right)
\end{cases}.
\end{align}
Here all of $k$ factors in the integration measure are canceled by the same factors coming from the determinant formulas.
Next, we perform the integrations over $\mu'_{N+r}$ ($r=1,2,\cdots,M_{1}$) and $\nu'_{N+r}$ ($r=1,2,\cdots,M_{2}$) by generating the delta functions for these variables from the first determinant in \eqref{ZmuDef} and the second determinant in \eqref{ZnuDef}, respectively.
This is achieved by the following replacement of the state vectors 
\begin{align}
 & \bra{\mu_{m}}\rightarrow\bra{\mu_{m}}e^{\frac{i}{4\pi k}{\hat{p}}^{2}},\quad\ket{\mu'_{m}}\rightarrow e^{\frac{i}{4\pi k}{\hat{p}}^{2}}\ket{\mu'_{m}},\quad\bra{\bar{\mu}_{m}}\rightarrow\bra{\bar{\mu}_{m}}e^{-\frac{i}{4\pi k}{\hat{p}}^{2}},\quad\ket{\lambda_{m}}\rightarrow e^{\frac{i}{4\pi k}{\hat{p}}^{2}}\ket{\lambda_{m}},\label{togetdeltafcn1}\\
 & \bra{\lambda_{m}}\rightarrow\bra{\lambda_{m}}e^{-\frac{i}{4\pi k}{\hat{p}}^{2}},\quad\ket{\nu_{m}}\rightarrow e^{-\frac{i}{4\pi k}{\hat{p}}^{2}}\ket{\nu_{m}},\quad\bra{\nu'_{m}}\rightarrow\bra{\nu'_{m}}e^{-\frac{i}{4\pi k}{\hat{p}}^{2}},\quad\ket{\bar{\nu}_{m}}\rightarrow e^{\frac{i}{4\pi k}{\hat{p}}^{2}}\ket{\bar{\nu}_{m}},\label{togetdeltafcn2}
\end{align}
under which the integration is invariant thanks
to the identities \eqref{completebasisid1} and \eqref{completebasisid2}. Under \eqref{togetdeltafcn1}, the $\bar{\mu}_{m}$-dependent part ${\cal X}^{(\bar{\mu})}(\lambda_{m})$ in $Z_{k}(N)$ transforms as 
\begin{align}
 & {\cal X}^{(\bar{\mu})}(\lambda_{m})\nonumber \\
 & =\frac{1}{N!(N+M_{1})!}\int_{-\infty}^{\infty}\frac{d^{2N+M_{1}}\bar{\mu}}{(2\pi)^{2N+M_{1}}}\nonumber \\
 & \times\det\begin{pmatrix}\left[\braket{\mu_{m}|e^{\frac{i}{4\pi k}\hat{p}^{2}}e^{-\frac{i}{4\pi k}\hat{x}^{2}+\frac{i\zeta_{1}}{2\pi k}\hat{x}}\frac{\tanh\frac{\hat{p}-\pi iM_{1}}{2}}{2}e^{\frac{i}{4\pi k}\hat{x}^{2}-\frac{i\zeta_{1}}{2\pi k}\hat{x}}e^{\frac{i}{4\pi k}\hat{p}^{2}}|\mu_{n}'}\right]_{m,n}^{N\times(N+M_{1})}\\
\left[\bbraket{t_{M_{1},r}|e^{\frac{i}{4\pi k}\hat{x}^{2}-\frac{i\zeta_{1}}{2\pi k}\hat{x}}e^{\frac{i}{4\pi k}\hat{p}^{2}}|\mu_{n}'}\right]_{r,n}^{M_{1}\times(N+M_{1})}
\end{pmatrix}\nonumber \\
 & \times D\left(\bar{\mu}_m,\lambda_n;2N+M_{1},2N+M;e^{-\frac{i\hat{p}^{2}}{4\pi k}}e^{\frac{i}{4\pi k}\hat{x}^{2}-\frac{i\zeta_{2}}{2\pi k}\hat{x}}\right)\nonumber \\
 & =\frac{1}{N!(N+M_{1})!}\int_{-\infty}^{\infty}\frac{d^{2N+M_{1}}\bar{\mu}}{(2\pi)^{2N+M_{1}}}\det\begin{pmatrix}\left[\braket{\mu_{m}|e^{\frac{i}{4\pi k}\hat{p}^{2}}\frac{\tanh\frac{\hat{p}+\hat{x}-\zeta_{1}-\pi iM_{1}}{2}}{2}e^{\frac{i}{4\pi k}\hat{p}^{2}}|\mu_{n}'}\right]_{m,n}^{N\times(N+M_{1})}\\
\left[2\pi\sqrt{i}e^{-\frac{i}{4\pi k}(t_{M_{1},r}+\zeta_{1})^{2}}\delta(t_{M_{1},r}+\zeta_{1}-\mu'_{n})\right]_{r,n}^{M_{1}\times(N+M_{1})}
\end{pmatrix}\nonumber \\
 & \times D\left(\bar{\mu}_m,\lambda_n;2N+M_{1},2N+M;e^{-\frac{i\hat{p}^{2}}{4\pi k}}e^{\frac{i}{4\pi k}\hat{x}^{2}-\frac{i\zeta_{2}}{2\pi k}\hat{x}}\right).\label{eq:Zmu1}
\end{align}
Here, to obtain the last expression we have used the formulas of the similarity transformations \eqref{simtrsfformulaop1}, \eqref{simtrsfformulastate1} and \eqref{simtrsfformulastate2} for the first determinant.Now we use the formula \eqref{trivialize} to trivialize the first determinant, which produces the delta functions $(2\pi\sqrt{i})^{M_{1}}\prod_{r=1}^{M_{1}}\delta(t_{M_{1},r}+\zeta_{1}-\mu'_{N+r})$ from the bottom block.

At this stage, we deform the $D$ factor in the last line of \eqref{eq:Zmu1}. By using the formulas of the similarity transformations \eqref{simtrsfformulaPop1} and \eqref{simtrsfformulaPstate1} for this $D$ factor, the momentum operator $\hat{p}$ and its eigenvector become $\hat{P}$ and its eigenvector, respectively. Now we can use the Cauchy-Vandermonde determinant formulas \eqref{CauchyVdm1} and \eqref{CauchyVdm2} inversely to the $D$ factor as 
\begin{align}
 & D\left(\bar{\mu}_m,\lambda_n;2N+M_{1},2N+M;e^{-\frac{i\hat{p}^{2}}{4\pi k}}e^{\frac{i}{4\pi k}\hat{x}^{2}-\frac{i\zeta_{2}}{2\pi k}\hat{x}}\right)\nonumber \\
 & =\frac{1}{\left(2k\right)^{2N+\frac{1}{2}M_{1}+\frac{1}{2}M}}\prod_{r=1}^{M_{1}-M}e^{-\frac{i}{8\pi k}(-t_{M_{1}-M,r}-\zeta_{2})^{2}}\prod_{m=1}^{2N+M_{1}}e^{\frac{i}{8\pi k}\bar{\mu}_{m}^{2}-\frac{i\zeta_{2}}{4\pi k}\bar{\mu}_{m}}\prod_{m=1}^{2N+M}e^{-\frac{i}{8\pi k}\lambda_{m}^{2}+\frac{i\zeta_{2}}{4\pi k}\lambda_{m}}\nonumber \\
 & \times\frac{\prod_{m<m'}^{2N+M_{1}}2\sinh\frac{\bar{\mu}_{m}-\bar{\mu}_{m'}}{4k}\prod_{n<n'}^{2N+M}2\sinh\frac{\lambda_{n}-\lambda_{n'}}{4k}}{\prod_{m=1}^{2N+M_{1}}\prod_{n=1}^{2N+M}2\cosh\frac{\bar{\mu}_{m}-\lambda_{n}}{4k}}.
\end{align}
After performing the integrations over $\mu'_{N+r}$ by using the delta functions,\footnote{
For integrating out the delta functions, we need to shift the integration contours from $\mathbb{R}$ to $\mathbb{R}+t_{M_{1},r}$. The condition \eqref{eq:Cascade} is sufficient for avoiding any poles under the shift.
} we obtain
\begin{align}
 {\cal X}^{(\bar{\mu})}(\lambda_{m})
 & =\frac{i^{\frac{M_{1}^{2}}{2}}e^{i\theta_{2k}(M_{1},-\zeta_{1})+i\theta_{2k}(M_{1}-M,\zeta_{2})-\frac{i}{4\pi k}\zeta_{1}\zeta_{2}M_{1}}Z_{2k}^{\text{(CS)}}(M_{1})}{(2k)^{2N+\frac{M}{2}}N!}\int_{-\infty}^{\infty}\frac{d^{2N}\widetilde{\mu}}{(2\pi)^{2N}}e^{-\frac{i\zeta_{2}}{4\pi k}\sum_{m=1}^{2N}\widetilde{\mu}_{m}}\nonumber \\
 & \times\prod_{m=1}^{N}\braket{\mu_{m}|e^{\frac{i}{8\pi k}\hat{x}^{2}}e^{\frac{i}{4\pi k}\hat{p}^{2}}\frac{\tanh\frac{\hat{p}+\hat{x}-\zeta_{1}-\pi iM_{1}}{2}}{2}e^{\frac{i}{4\pi k}\hat{p}^{2}}e^{\frac{i}{8\pi k}\hat{x}^{2}}|\mu'_{m}}
 \nonumber \\
 & \times
\left(\prod_{m=1}^{2N}\prod_{r=1}^{M_{1}}2\sinh\frac{\widetilde{\mu}_{m}-t_{M_{1},r}-\zeta_{1}}{4k}\right)
\frac{\prod_{m<m'}^{2N}2\sinh\frac{\widetilde{\mu}_{m}-\widetilde{\mu}_{m'}}{4k}\prod_{n<n'}^{2N+M}2\sinh\frac{\lambda_{n}-\lambda_{n'}}{4k}}{\prod_{m=1}^{2N}\prod_{n=1}^{2N+M}2\cosh\frac{\widetilde{\mu}_{m}-\lambda_{n}}{4k}}\nonumber \\
 &\times \left(\prod_{m=1}^{2N+M}\frac{e^{-\frac{i}{8\pi k}\lambda_{m}^{2}+\frac{i\zeta_{2}}{4\pi k}\lambda_{m}}}{\prod_{r=1}^{M_{1}}2\cosh\frac{\lambda_{m}-t_{M_{1},r}-\zeta_{1}}{4k}}\right),\label{eq:Zmu2}
\end{align}
where we have introduced a new abbreviation ${\widetilde{\mu}}_{m}=(\mu_{1},\cdots,\mu_{N},\mu'_{1}\cdots,\mu'_{N})$ and 
\begin{align}
\theta_{k}(L,\zeta)=\frac{1}{4\pi k}\left(\frac{\pi^{2}(L^{3}-L)}{3}-L\zeta^{2}\right),\quad Z_{k}^{\text{(CS)}}(L)=\frac{1}{k^{\frac{L}{2}}}\prod_{r>s}^{L}2\sin\frac{\pi(r-s)}{k}.
\end{align}
$Z_{k}^{\text{(CS)}}(L)$ is the partition function of the pure ${\rm U}\left(L\right)_{k}$ Chern-Simons theory. 
For later convenience, we deform \eqref{eq:Zmu2} by multiplying the second and third lines of \eqref{eq:Zmu2} by a factor
\begin{equation}
\prod_{m=1}^{2N}2\cosh\frac{\widetilde{\mu}_{m}-\zeta_{1}-\pi iM_{1}}{2},
\end{equation}
and its inverse, respectively.
The second line reads
\begin{align}
 & \prod_{m=1}^{2N}2\cosh\frac{\widetilde{\mu}_{m}-\zeta_{1}-\pi iM_{1}}{2} 
 \prod_{m=1}^{N}\braket{\mu_{m}|e^{\frac{i}{8\pi k}\hat{x}^{2}}e^{\frac{i}{4\pi k}\hat{p}^{2}}\frac{\tanh\frac{\hat{p}+\hat{x}-\zeta_{1}-\pi iM_{1}}{2}}{2}e^{\frac{i}{4\pi k}\hat{p}^{2}}e^{\frac{i}{8\pi k}\hat{x}^{2}}|\mu'_{m}}\nonumber \\
 & =\prod_{m=1}^{N}\braket{\mu_{m}|e^{\frac{i}{8\pi k}\hat{x}^{2}}e^{\frac{i}{4\pi k}\hat{p}^{2}}2\cosh\frac{\hat{x}-\hat{p}-\zeta_{1}-\pi iM_{1}}{2}\sinh\frac{\hat{x}+\hat{p}-\zeta_{1}-\pi iM_{1}}{2}e^{\frac{i}{4\pi k}\hat{p}^{2}}e^{\frac{i}{8\pi k}\hat{x}^{2}}|\mu_{m}'},
\end{align}
where we have used the similarity transformations \eqref{simtrsfformulaop1}. Therefore, 
\begin{align}
 & {\cal X}^{(\bar{\mu})}(\lambda_{m})\nonumber \\
 & =\frac{i^{\frac{M_{1}^{2}}{2}}e^{i\theta_{2k}(M_{1},-\zeta_{1})+i\theta_{2k}(M_{1}-M,\zeta_{2})-\frac{i}{4\pi k}\zeta_{1}\zeta_{2}M_{1}}Z_{2k}^{\text{(CS)}}(M_{1})}{(2k)^{2N+\frac{M}{2}}N!}\int_{-\infty}^{\infty}\frac{d^{2N}\widetilde{\mu}}{(2\pi)^{2N}}e^{-\frac{i\zeta_{2}}{4\pi k}\sum_{m=1}^{2N}\widetilde{\mu}_{m}}\nonumber \\
 & \times\prod_{m=1}^{N}\braket{\mu_{m}|e^{\frac{i}{8\pi k}\hat{x}^{2}}e^{\frac{i}{4\pi k}\hat{p}^{2}}2\cosh\frac{\hat{x}-\hat{p}-\zeta_{1}-\pi iM_{1}}{2}\sinh\frac{\hat{x}+\hat{p}-\zeta_{1}-\pi iM_{1}}{2}e^{\frac{i}{4\pi k}\hat{p}^{2}}e^{\frac{i}{8\pi k}\hat{x}^{2}}|\mu_{m}'}\nonumber \\
 & \times\left(\prod_{m=1}^{2N}\frac{\prod_{r=1}^{M_{1}}2\sinh\frac{\widetilde{\mu}_{m}-t_{M_{1},r}-\zeta_{1}}{4k}}{2\cosh\frac{\widetilde{\mu}_{m}-\zeta_{1}-\pi iM_{1}}{2}}\right)\frac{\prod_{m<m'}^{2N}2\sinh\frac{\widetilde{\mu}_{m}-\widetilde{\mu}_{m'}}{4k}\prod_{n<n'}^{2N+M}2\sinh\frac{\lambda_{n}-\lambda_{n'}}{4k}}{\prod_{m=1}^{2N}\prod_{n=1}^{2N+M}2\cosh\frac{\widetilde{\mu}_{m}-\lambda_{n}}{4k}}\left(\prod_{m=1}^{2N+M}\frac{e^{-\frac{i}{8\pi k}\lambda_{m}^{2}+\frac{i\zeta_{2}}{4\pi k}\lambda_{m}}}{\prod_{r=1}^{M_{1}}2\cosh\frac{\lambda_{m}-t_{M_{1},r}-\zeta_{1}}{4k}}\right).\label{eq:Zmu3}
\end{align}

With the replacement \eqref{togetdeltafcn2} and the similar calculations as above, the $\bar{\nu}_{m}$-dependent part ${\cal X}^{(\bar{\nu})}(\lambda_{m})$ in $Z_{k}(N)$ can also be reduced into the following form
\begin{align}
 & {\cal X}^{(\bar{\nu})}(\lambda_{m})\nonumber \\
 & =\frac{i^{-\frac{M_{2}^{2}}{2}}e^{-i\theta_{2k}(M_{2},-\zeta_{4})-i\theta_{2k}(M_{2}-M,\zeta_{3})+\frac{i}{4\pi k}\zeta_{3}\zeta_{4}M_{2}}Z_{2k}^{\text{(CS)}}(M_{2})}{(2k)^{2N+\frac{M}{2}}N!}\int_{\infty}^{\infty}\frac{d^{2N}\widetilde{\nu}}{(2\pi)^{2N}}e^{\frac{i\zeta_{3}}{4\pi k}\sum_{m=1}^{2N}\widetilde{\nu}_{m}}\nonumber \\
 & \times\left(\prod_{m=1}^{2N+M}\frac{e^{\frac{i}{8\pi k}\lambda_{m}^{2}-\frac{i\zeta_{3}}{4\pi k}\lambda_{m}}}{\prod_{r=1}^{M_{2}}2\cosh\frac{\lambda_{m}-t_{M_{2},r}-\zeta_{4}}{4k}}\right)\frac{\prod_{n<n'}^{2N+M}2\sinh\frac{\lambda_{n}-\lambda_{n'}}{4k}\prod_{m<m'}^{2N}2\sinh\frac{\widetilde{\nu}_{m}-\widetilde{\nu}_{m'}}{4k}}{\prod_{m=1}^{2N+M}\prod_{n=1}^{2N}2\cosh\frac{\lambda_{m}-\widetilde{\nu}_{n}}{4k}}\left(\prod_{m=1}^{2N}\frac{\prod_{r=1}^{M_{2}}2\sinh\frac{\widetilde{\nu}_{m}-t_{M_{2},r}-\zeta_{4}}{4k}}{2\cosh\frac{\widetilde{\nu}_{m}-\zeta_{4}-\pi iM_{2}}{2}}\right)\nonumber \\
 & \times\prod_{m=1}^{N}\braket{\nu'_{m}|e^{-\frac{i}{8\pi k}\hat{x}^{2}}e^{-\frac{i}{4\pi k}\hat{p}^{2}}2\sinh\frac{\hat{x}+\hat{p}-\zeta_{4}+\pi iM_{2}}{2}\cosh\frac{\hat{x}-\hat{p}-\zeta_{4}+\pi iM_{2}}{2}e^{-\frac{i}{4\pi k}\hat{p}^{2}}e^{-\frac{i}{8\pi k}\hat{x}^{2}}|\nu_{m}},
\end{align}
where we have introduced a new abbreviation ${\widetilde{\nu}}_{m}=(\nu_{1},\cdots,\nu_{N},\nu'_{1},\cdots,\nu'_{N})$. Putting these results together, we finally obtain
\begin{align}
 & Z_{k}(N;M_{1},M_{2},M;\zeta_{a})\nonumber \\
 & =e^{i\Theta_{2k}(M_{1},M_{2},M;\zeta_{a})}Z_{2k}^{\text{(CS)}}(M_{1})Z_{2k}^{\text{(CS)}}(M_{2})\frac{1}{2^{2N}(N!)^{2}(2k)^{2N}}\int_{-\infty}^{\infty}\frac{d^{2N}\widetilde{\mu}}{(2\pi)^{2N}}\frac{d^{2N}\widetilde{\nu}}{(2\pi)^{2N}}\nonumber \\
 & \times\prod_{m=1}^{N}\braket{\mu_{m}|e^{\frac{i}{8\pi k}\hat{x}^{2}}e^{\frac{i}{4\pi k}\hat{p}^{2}}2\cosh\frac{\hat{x}-\hat{p}-\zeta_{1}-\pi iM_{1}}{2}2\sinh\frac{\hat{x}+\hat{p}-\zeta_{1}-\pi iM_{1}}{2}e^{\frac{i}{4\pi k}\hat{p}^{2}}e^{\frac{i}{8\pi k}\hat{x}^{2}}|\mu_{m}'}\nonumber \\
 & \times{\cal Y}_{2k}(\widetilde{\mu}_{i},\widetilde{\nu}_{j};2N;M_{1},M_{2},M;\zeta_{a})\nonumber \\
 & \times\prod_{m=1}^{N}\braket{\nu_{m}'|e^{-\frac{i}{8\pi k}\hat{x}^{2}}e^{-\frac{i}{4\pi k}\hat{p}^{2}}2\sinh\frac{\hat{x}+\hat{p}-\zeta_{4}+\pi iM_{2}}{2}2\cosh\frac{\hat{x}-\hat{p}-\zeta_{4}+\pi iM_{2}}{2}e^{-\frac{i}{4\pi k}\hat{p}^{2}}e^{-\frac{i}{8\pi k}\hat{x}^{2}}|\nu_{m}},\label{semifinal}
\end{align}
where $\Theta_{k}$ is an $N$-independent phase
\begin{align}
 & \Theta_{k}(L_{1},L_{2},L;\zeta_{a})\nonumber \\
 & =-\frac{1}{2\pi k}(\zeta_{1}\zeta_{2}L_{1}-\zeta_{3}\zeta_{4}L_{2})+\theta_{k}(L_{1},-\zeta_{1})+\theta_{k}(L_{1}-L,\zeta_{2})-\theta_{k}(L_{2},-\zeta_{4})-\theta_{k}(L_{2}-L,\zeta_{3}),\label{Ak}
\end{align}
and ${\cal Y}_{k}$ is 
\begin{align}
 & {\cal Y}_{k}(\alpha_{m},\beta_{n};N;L_{1},L_{2},L;\zeta_{a})\nonumber \\
 & =\frac{i^{-(L_{1}-L_{2})N}}{(N+L)!}\int_{-\infty}^{\infty}\frac{d^{N+L}\gamma}{(2\pi k)^{N+L}}e^{\frac{i}{2\pi k}(-\zeta_{2}\sum_{m=1}^{N}\alpha_{m}+(\zeta_{2}-\zeta_{3})\sum_{m=1}^{N+L}\gamma_{m}+\zeta_{3}\sum_{m=1}^{N}\beta_{m})}\nonumber \\
 & \times\left(\prod_{m=1}^{N}\frac{\prod_{r=1}^{L_{1}}2\sinh\frac{\alpha_{m}-\zeta_{1}-t_{L_{1},r}}{2k}}{2\cosh\frac{\alpha_{m}-\zeta_{1}-\pi iL_{1}}{2}}\right)\frac{\prod_{m<m'}^{N}2\sinh\frac{\alpha_{m}-\alpha_{m'}}{2k}\prod_{m<m'}^{N+L}2\sinh\frac{\gamma_{m}-\gamma_{m'}}{2k}}{\prod_{m=1}^{N}\prod_{n=1}^{N+L}2\cosh\frac{\alpha_{m}-\gamma_{n}}{2k}}\left(\prod_{m=1}^{N+L}\frac{1}{\prod_{r=1}^{L_{1}}2\cosh\frac{\gamma_{m}-\zeta_{1}-t_{L_{1},r}}{2k}}\right)\nonumber \\
 & \times\left(\prod_{m=1}^{N+L}\frac{1}{\prod_{r=1}^{L_{2}}2\cosh\frac{\gamma_{m}-\zeta_{4}-t_{L_{2},r}}{2k}}\right)\frac{\prod_{m<m'}^{N+L}2\sinh\frac{\gamma_{m}-\gamma_{m'}}{2k}\prod_{m<m'}^{N}2\sinh\frac{\beta_{m}-\beta_{m'}}{2k}}{\prod_{m=1}^{N+L}\prod_{n=1}^{N}2\cosh\frac{\gamma_{m}-\beta_{n}}{2k}}\left(\prod_{m=1}^{N}\frac{\prod_{r=1}^{L_{2}}2\sinh\frac{\beta_{m}-\zeta_{4}-t_{L_{2},r}}{2k}}{2\cosh\frac{\beta_{m}-\zeta_{4}+\pi iL_{2}}{2}}\right).\label{eq:Ik}
\end{align}

Although we cannot apply the closed string formalism of the Fermi gas formalism to this factor due to a lack of computation technique, here we conjecture that this factor can be deformed as \eqref{eq:YkConj}.
\begin{figure}
\begin{centering}
\includegraphics[scale=0.6]{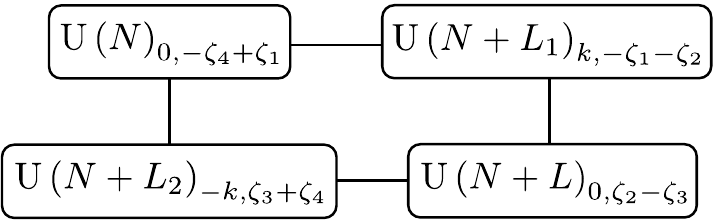}
\par\end{centering}
\caption{The quiver diagram corresponding to the factor ${\cal Y}_{k}$ defined in \eqref{eq:Ik}.
\label{fig:Quiver-22}}
\end{figure}
This comes from an observation in \cite{Kubo:2019ejc,Bonelli:2022dse} that a similar factor appears in the four-node circular quiver Chern-Simons theory (so called the (2,2) model) whose quiver diagram is shown in figure \ref{fig:Quiver-22}.
See appendix \ref{sec:YkQC} for detailed discussion. After plugging in the conjectured relation, we obtain
\begin{align}
 & Z_{k}(N;M_{1},M_{2},M;\zeta_{a})\nonumber \\
 & =e^{i\Theta_{2k}(M_{1},M_{2},M;\zeta_{a})}Z_{2k}^{\text{(CS)}}(M_{1})Z_{2k}^{\text{(CS)}}(M_{2}){\cal Y}_{2k}^{\left(0\right)}(M_{1},M_{2},M;\zeta_{a})\frac{1}{2^{2N}(N!)^{2}}\int_{-\infty}^{\infty}\frac{d^{2N}\widetilde{\mu}}{(2\pi)^{2N}}\frac{d^{2N}\widetilde{\nu}}{(2\pi)^{2N}}\nonumber \\
 & \times\prod_{m=1}^{N}\braket{\mu_{m}|e^{\frac{i}{8\pi k}\hat{x}^{2}}e^{\frac{i}{4\pi k}\hat{p}^{2}}2\cosh\frac{\hat{x}-\hat{p}-\zeta_{1}-\pi iM_{1}}{2}2\sinh\frac{\hat{x}+\hat{p}-\zeta_{1}-\pi iM_{1}}{2}e^{\frac{i}{4\pi k}\hat{p}^{2}}e^{\frac{i}{8\pi k}\hat{x}^{2}}|\mu_{m}'}\nonumber \\
 & \times\det\left(\left[\braket{\widetilde{\mu}_{m}|\frac{1}{\hat{H}_{2k}\left(\hat{x},\hat{P};M_{1},M_{2},M;\zeta_{a}\right)}|\widetilde{\nu}_{n}}\right]_{m,n}^{2N\times2N}\right)\nonumber \\
 & \times\prod_{m=1}^{N}\braket{\nu_{m}'|e^{-\frac{i}{8\pi k}\hat{x}^{2}}e^{-\frac{i}{4\pi k}\hat{p}^{2}}2\sinh\frac{\hat{x}+\hat{p}-\zeta_{4}+\pi iM_{2}}{2}2\cosh\frac{\hat{x}-\hat{p}-\zeta_{4}+\pi iM_{2}}{2}e^{-\frac{i}{4\pi k}\hat{p}^{2}}e^{-\frac{i}{8\pi k}\hat{x}^{2}}|\nu_{m}},\label{ZD4-CSF1}
\end{align}
where the quantum curve $\hat{H}_{k}$ is defined in \eqref{eq:QCConj}.\footnote{The integration in the factor ${\cal Y}_{k}^{\left(0\right)}$ defined in \eqref{eq:Yk0Def} converges if and only if the ranks satisfy the ``good'' or ``ugly'' condition \eqref{eq:NotBad}.}Now we again perform the similarity transformation
\begin{align}
 & \bra{\mu_{m}}\rightarrow\bra{\mu_{m}}e^{-\frac{i}{4\pi k}\hat{p}^{2}}e^{-\frac{i}{8\pi k}\hat{x}^{2}},\quad\ket{\mu'_{m}}\rightarrow e^{-\frac{i}{8\pi k}\hat{x}^{2}}e^{-\frac{i}{4\pi k}\hat{p}^{2}}\ket{\mu'_{m}},\quad\bra{\widetilde{\mu}_{m}}\rightarrow\bra{\widetilde{\mu}_{m}}e^{\frac{i}{4\pi k}\hat{p}^{2}}e^{\frac{i}{8\pi k}\hat{x}^{2}},\label{togetdeltafcn1-1}\\
 & \ket{\nu_{m}}\rightarrow e^{\frac{i}{8\pi k}\hat{x}^{2}}e^{\frac{i}{4\pi k}\hat{p}^{2}}\ket{\nu_{m}},\quad\bra{\nu'_{m}}\rightarrow\bra{\nu'_{m}}e^{\frac{i}{4\pi k}\hat{p}^{2}}e^{\frac{i}{8\pi k}\hat{x}^{2}},\quad\ket{\widetilde{\nu}_{m}}\rightarrow e^{-\frac{i}{8\pi k}\hat{x}^{2}}e^{-\frac{i}{4\pi k}\hat{p}^{2}}\ket{\widetilde{\nu}_{m}},\label{togetdeltafcn2-1}
\end{align}
and use the formula of the similarity transformation \eqref{simtrsfformulaop1}. We also introduce new operators 
\begin{equation}
\hat{u}\equiv\hat{p}+\hat{x},\quad\hat{v}\equiv\hat{p}-\hat{x}.
\end{equation}
We finally arrive at
\begin{align}
 & Z_{k}(N;M_{1},M_{2},M;\zeta_{a})\nonumber \\
 & =e^{i\Theta_{2k}(M_{1},M_{2},M;\zeta_{a})}Z_{2k}^{\text{(CS)}}(M_{1})Z_{2k}^{\text{(CS)}}(M_{2}){\cal Y}_{2k}^{\left(0\right)}(M_{1},M_{2},M;\zeta_{a})\frac{1}{(N!)^{4}}\int_{-\infty}^{\infty}\frac{d^{2N}\widetilde{\mu}}{(2\pi)^{2N}}\frac{d^{2N}\widetilde{\nu}}{(2\pi)^{2N}}\nonumber \\
 & \times\det\left(\left[\braket{\mu_{m}|2\cosh\frac{\hat{v}+\zeta_{1}+\pi iM_{1}}{2}2\sinh\frac{\hat{u}-\zeta_{1}-\pi iM_{1}}{2}|\mu_{m}'}\right]_{m,n}^{N\times N}\right)\nonumber \\
 & \times\det\left(\left[\braket{\widetilde{\mu}_{m}|\frac{1}{2\hat{H}_{2k}\left(\hat{u},\hat{v};M_{1},M_{2},M;\zeta_{a}\right)}|\widetilde{\nu}_{n}}\right]_{m,n}^{2N\times2N}\right)\nonumber \\
 & \times\det\left(\left[\braket{\nu_{m}'|2\sinh\frac{\hat{u}-\zeta_{4}+\pi iM_{2}}{2}2\cosh\frac{\hat{v}+\zeta_{4}-\pi iM_{2}}{2}|\nu_{m}}\right]_{m,n}^{N\times N}\right).\label{ZD4-CSF-Res2}
\end{align}
Here we used the determinant trivialization formula \eqref{trivialize} backward, and we put the $2^{2N}$ factor into the determinant for later convenience.

\subsubsection{Grand partition function}

In the previous section we have applied the closed string formalism to the partition function as \eqref{ZD4-CSF-Res2}. This result can be used for applying the closed string formalism to the grand partition function defined in \eqref{eq:GPFdef}. Since the $N$-independent factors become an overall factor, the grand partition function can be divided as
\begin{equation}
\Xi\left(\kappa\right)=Z_{k}^{\left(0\right)}\sum_{N=0}^{\infty}\kappa^{N}\tilde{Z}_{k}\left(N;M_{1},M_{2},M;\zeta_{a}\right),\label{eq:GPF-Closed0}
\end{equation}
where the overall factor is
\begin{equation}
Z_{k}^{\left(0\right)}=e^{i\Theta_{2k}(M_{1},M_{2},M;\zeta_{a})}Z_{2k}^{\text{(CS)}}(M_{1})Z_{2k}^{\text{(CS)}}(M_{2}){\cal Y}_{k}^{\left(0\right)}(M_{1},M_{2},M;\zeta_{a}),
\end{equation}
and the $N$-dependent part is\footnote{Here $\tilde{Z}_{k}\left(0\right)=1$.}
\begin{align}
 & \tilde{Z}_{k}\left(N;M_{1},M_{2},M;\zeta_{a}\right)
=\frac{Z_{k}\left(N;M_{1},M_{2},M;\zeta_{a}\right)}{Z_k^{\left(0\right)}} \nonumber \\
 & =\frac{1}{(N!)^{4}}\int\frac{d^{N}\mu}{(2\pi)^{N}}\frac{d^{N}\nu}{(2\pi)^{N}}\frac{d^{N}\mu'}{(2\pi)^{N}}\frac{d^{N}\nu'}{(2\pi)^{N}}\nonumber \\
 & \quad\times\det\left(\left[\braket{\mu_{m}|\hat{L}|\mu_{n}'}\right]_{m,n}^{N\times N}\right)\det\left(\left[\braket{\bar{\mu}_{m}|\hat{C}|\bar{\nu}_{n}}\right]_{m,n}^{2N\times2N}\right)\det\left(\left[\braket{\nu_{m}'|\hat{R}|\nu_{n}}\right]_{m,n}^{N\times N}\right),
\end{align}
where
\begin{align}
\hat{C} & =\frac{1}{2\hat{H}_{2k}\left(\hat{u},\hat{v};M_{1},M_{2},M;\zeta_{a}\right)},\nonumber \\
\hat{L} & =2\cosh\frac{\hat{v}+\zeta_{1}+\pi iM_{1}}{2}2\sinh\frac{\hat{u}-\zeta_{1}-\pi iM_{1}}{2},\nonumber \\
\hat{R} & =2\sinh\frac{\hat{u}-\zeta_{4}+\pi iM_{2}}{2}2\cosh\frac{\hat{v}+\zeta_{4}-\pi iM_{2}}{2}.\label{eq:CLR-Def}
\end{align}
For the $N$-dependent part $\sum_{N=0}^{\infty}\kappa^{N}\tilde{Z}\left(N\right)$, we can deform the summation to a square root of a determinant as in section \ref{sec_open}. The way of computation is the same and simpler.

First, we apply the Cauchy-Binet formula \eqref{CauchyBinet} for the integrations over $\mu_{m}'$ and $\nu_{m}'$, which reads
\begin{align}
 & \tilde{Z}_{k}\left(N;M_{1},M_{2},M;\zeta_{a}\right)\nonumber \\
 & =\frac{1}{(N!)^{2}}\int_{-\infty}^{\infty}\frac{d^{N}\mu}{(2\pi)^{N}}\int\frac{d^{N}\nu}{(2\pi)^{N}}\text{det}\left(\begin{array}{cc}
\left[\braket{\mu_{m}|\hat{C}|\mu_{n}}\right]_{m,n}^{N\times N} & \left[\braket{\mu_{m}|\hat{C}\hat{R}|\mu_{n}}\right]_{m,n}^{N\times N}\\
\left[\braket{\mu_{m}|\hat{L}\hat{C}|\mu_{n}}\right]_{m,n}^{N\times N} & \left[\braket{\mu_{m}|\hat{L}\hat{C}\hat{R}|\mu_{n}}\right]_{m,n}^{N\times N}
\end{array}\right).
\end{align}
Second, we perform the integrations over $\nu_{m}$ by using the Cauchy-Binet-like formula for Pfaffian \eqref{CBPfaffian}. As a result we obtain 
\begin{align}
 & \tilde{Z}_{k}\left(N;M_{1},M_{2},M;\zeta_{a}\right)\nonumber \\
 & =\left(-1\right)^{\frac{1}{2}N\left(N-1\right)}\nonumber \\
 &\quad \times\int_{-\infty}^{\infty}\frac{d^{N}\mu}{N!(2\pi)^{N}}\text{pf}\left(\begin{array}{cc}
\left[\braket{\mu_{m}|\left(\hat{C}\left(\hat{R}^{t}-\hat{R}\right)\hat{C}^{t}\right)|\mu_{n}}\right]_{m,n}^{N\times N} & \left[\braket{\mu_{m}|\left(\hat{C}\left(\hat{R}^{t}-\hat{R}\right)\hat{C}^{t}\hat{L}^{t}\right)|\mu_{n}}\right]_{m,n}^{N\times N}\\
\left[\braket{\mu_{m}|\left(\hat{L}\hat{C}\left(\hat{R}^{t}-\hat{R}\right)\hat{C}^{t}\right)|\mu_{n}}\right]_{m,n}^{N\times N} & \left[\braket{\mu_{m}|\left(\hat{L}\hat{C}\left(\hat{R}^{t}-\hat{R}\right)\hat{C}^{t}\hat{L}^{t}\right)|\mu_{n}}\right]_{m,n}^{N\times N}
\end{array}\right).
\end{align}
Third, by using the Fredholm Pfaffian formula \eqref{FredholmPfaffian2}, we obtain 
\begin{align}
 & \sum_{N=0}^{\infty}\kappa^{N}\tilde{Z}_{k}\left(N;M_{1},M_{2},M;\zeta_{a}\right)\nonumber \\
 & =\sqrt{{\rm Det}\left(\begin{array}{cc}
\kappa\left(\hat{C}\left(\hat{R}^{t}-\hat{R}\right)\hat{C}^{t}\right) & 1+\kappa\left(\hat{C}\left(\hat{R}^{t}-\hat{R}\right)\hat{C}^{t}\hat{L}^{t}\right)\\
-1+\kappa\left(\hat{L}\hat{C}\left(\hat{R}^{t}-\hat{R}\right)\hat{C}^{t}\right) & \kappa\left(\hat{L}\hat{C}\left(\hat{R}^{t}-\hat{R}\right)\hat{C}^{t}\hat{L}^{t}\right)
\end{array}\right)}.
\end{align}
This matrix can be transformed into a simpler matrix by a sequence of the elementary row/column operations as
\begin{equation}
\sum_{N=0}^{\infty}\kappa^{N}\tilde{Z}_{k}\left(N;M_{1},M_{2},M;\zeta_{a}\right)=\sqrt{{\rm Det}\left(\begin{array}{cc}
0 & 1+\kappa\hat{C}\left(\hat{R}-\hat{R}^{t}\right)\hat{C}^{t}\left(\hat{L}-\hat{L}^{t}\right)\\
-1 &0
\end{array}\right)}.
\end{equation}
Therefore, we arrive at

\begin{equation}
\sum_{N=0}^{\infty}\kappa^{N}\tilde{Z}_{k}\left(N;M_{1},M_{2},M;\zeta_{a}\right)=\sqrt{{\rm Det}\left(1+\kappa\hat{\rho}_{2k}^{\text{(closed)}}\left(M_{1},M_{2},M;\zeta_{a}\right)\right)},\label{eq:GPFform-Det}
\end{equation}
where 
\begin{equation}
\hat{\rho}_{2k}^{\text{(closed)}}\left(M_{1},M_{2},M;\zeta_{a}\right)=\hat{C}\left(\hat{R}-\hat{R}^{t}\right)\hat{C}^{t}\left(\hat{L}-\hat{L}^{t}\right).\label{eq:rho-Closed}
\end{equation}
The grand partition function \eqref{eq:GPF-Closed0} is now
\begin{align}
\Xi\left(\kappa\right) & =Z_{k}^{\left(0\right)}\sqrt{{\rm Det}\left(1+\kappa\hat{\rho}_{2k}^{\text{(closed)}}\left(M_{1},M_{2},M;\zeta_{a}\right)\right)}.\label{eq:GPF-Closed}
\end{align}

The density matrix for the closed string formalism $\hat{\rho}_{2k}^{\text{(closed)}}$ can be further simplified.
Since $\hat{C}$, $\hat{L}$ and $\hat{R}$ are \eqref{eq:CLR-Def},
\begin{align}
{\hat L}-{\hat L}^t=4e^{-\frac{\pi ik}{2}}\sinh {\hat p},\quad
{\hat R}-{\hat R}^t=4e^{\frac{\pi ik}{2}}\sinh {\hat p}.
\end{align}
Therefore, we have 
\begin{align}
 & \hat{\rho}_{2k}^{\text{(closed)}}\left(M_{1},M_{2},M;\zeta_{a}\right)\nonumber \\
 & =2\sinh\hat{p}\frac{1}{\hat{H}_{2k}\left(\hat{u},\hat{v};M_{1},M_{2},M;\zeta_{a}\right)}2\sinh\hat{p}\frac{1}{\hat{H}_{2k}\left(\hat{u},\hat{v};M_{1},M_{2},M;\zeta_{a}\right)^{t}}.\label{eq:DM-Closed}
\end{align}
In the next section, we study this density matrix.

\subsection{Symmetries of the density matrix\label{symmetries}}

In this section we study the symmetry of the density matrix \eqref{eq:DM-Closed}. We find two types of symmetries, and we explain in each subsection. 

The symmetry of the density matrix is defined in \cite{Kubo:2018cqw} as an equivalence of the density matrices with different parameters up to similarity transformations. Namely, if two or more density matrices are equivalent up to similarity transformations, they are symmetric.
One of the motivation of this definition is that the symmetry defined in this way implies the duality of the corresponding theories.
The closed type density matrix enters the grand partition function via a determinant as \eqref{eq:GPF-Closed}.
Since the determinant of an operator is invariant under similarity transformations, the second factor in \eqref{eq:DM-Closed} is equal for the symmetric density matrices. Note that, on the other hand, the symmetry does not necessarily imply the equivalence of the first factor $Z_{k}^{\left(0\right)}$. We will also comment on this point in the following subsections.
It is known that the similar phenomenon occurs in the theories with affine $A$-type quivers \cite{Kubo:2020qed,Kubo:2021enh}.

\subsubsection{First symmetry and Aharony duality\label{subsec:Aharony-dual}}

From the expression for the density matrix ${\hat\rho}^{\text{(closed)}}_{2k}$ in \eqref{eq:DM-Closed}, one can easily find
that if the (2,2) model quantum curve $\hat{H}_{2k}$ is invariant under the change of the parameters, it is symmetry of the density matrix. We actually find such symmetry under the exchange of the parameters
\begin{align}
\left(M_{1},M_{2},M;\zeta_{1},\zeta_{2},\zeta_{3},\zeta_{4}\right)
  \leftrightarrow\left(M_{1},M_{2},M_{1}+M_{2}-M;\zeta_{1},\zeta_{3},\zeta_{2},\zeta_{4}\right).\label{eq:SymAD}
\end{align}
By using the expression for the quantum curve \eqref{eq:QCConj}, one can easily check that the quantum curve satisfies
\begin{align}
\hat{H}_{2k}\left(\hat{u},\hat{v};M_{1},M_{2},M;\zeta_{1},\zeta_{2},\zeta_{3},\zeta_{4}\right)=\hat{H}_{2k}\left(\hat{u},\hat{v};M_{1},M_{2},M_{1}+M_{2}-M;\zeta_{1},\zeta_{3},\zeta_{2},\zeta_{4}\right).
\end{align}
Here, we have assumed that $E$ is invariant under the change of the variables. Note that this actually holds when $M=0$ \eqref{eq:E0}. Therefore,
\begin{equation}
\hat{\rho}_{2k}^{\text{(closed)}}\left(M_{1},M_{2},M;\zeta_{1},\zeta_{2},\zeta_{3},\zeta_{4}\right)=\hat{\rho}_{2k}^{\text{(closed)}}\left(M_1,M_2,M_1+M_2-M;\zeta_1,\zeta_3,\zeta_2,\zeta_4\right).
\end{equation}

\begin{figure}
\begin{centering}
\includegraphics[width=14cm]{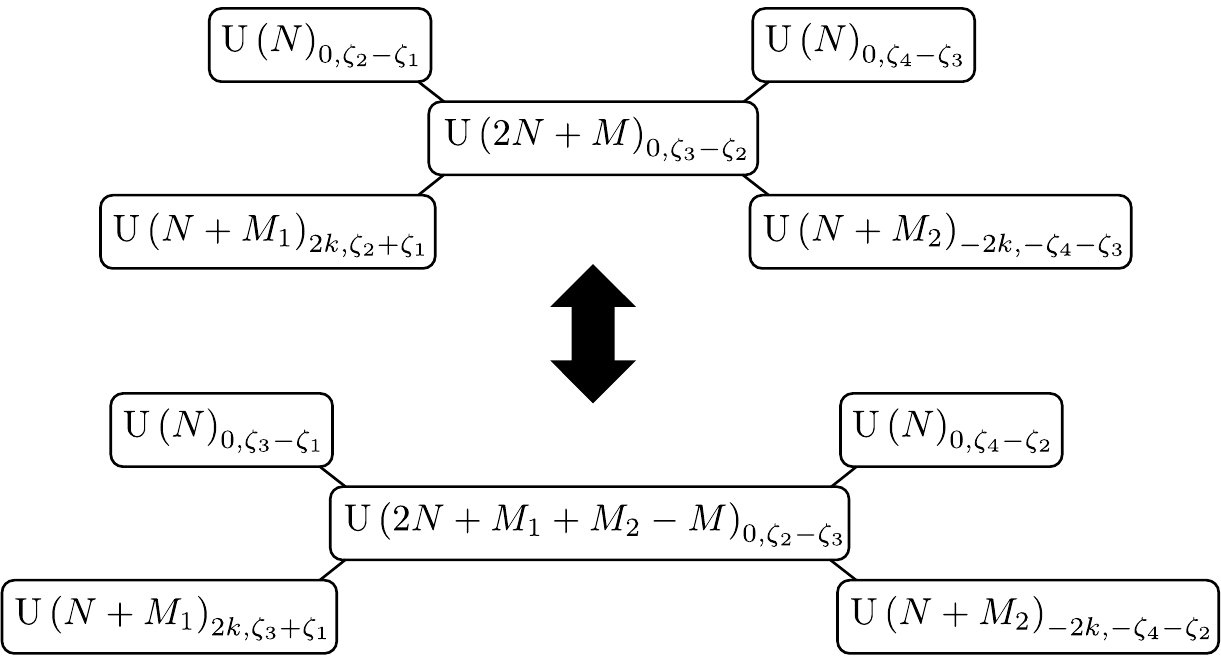}
\par\end{centering}
\caption{Two quiver diagrams corresponding to the symmetry \eqref{eq:SymAD}. This is similar to the Aharony duality.\label{fig:ADual}}
\label{ADual}
\end{figure}

To find physical interpretation, we consider the quiver diagrams corresponding to the both sides, as displayed in figure \ref{fig:ADual}.
This relation is reminiscent of the Aharony duality \cite{Aharony:1997gp}.
The Aharony duality is an IR duality for 3d $\mathcal{N}=2$ theories between ${\rm U}\left(N_{c}\right)$ and ${\rm U}\left(N_{f}-N_{c}\right)$ gauge group where $N_{f}$ is the number of chiral fundamental and anti-fundamental multiplets
Although in our case the theory has $\mathcal{N}=4$, if we naively apply this duality to the center node, it relates ${\rm U}\left(2N+M\right)$ and ${\rm U}\left(2N+M_{1}+M_{2}-M\right)$ group (since there are $4N+M_{1}+M_{2}$ bi-fundamental hypermultiplets). This matches the symmetry \eqref{eq:SymAD}.

One main problem of this naive application is that, in general, it relates a good theory to an ugly/bad theory. In our case, if the original theory satisfies the inequality \eqref{eq:NotBad}, the dual theory does not satisfy (bad) or saturates the inequality (ugly) of the condition. As discussed in section \ref{sec:Model}, the matrix model diverges for a bad theory. Correspondingly, by the same argument, one can check that the factor ${\cal Y}_{2k}^{\left(0\right)}$ diverges when the rank is in the bad region.\footnote{Here we naively apply the closed string formalism also for the region where \eqref{eq:NotBad} is not satisfied.} This divergence is the source of inequality of the matrix models related by the Aharony duality. The symmetry \eqref{eq:SymAD}, nevertheless, means that the Fredholm determinant factor obeys this duality. The similar phenomenon occurs for the A-type quiver theories \cite{Nosaka:2017ohr,Kubo:2020qed,Kubo:2021enh}.

\subsubsection{Second symmetry and Seiberg-like duality\label{subsec:SL-dual}}

In the previous section we considered the symmetry where the quantum curve of (2,2) model $\hat{H}_{2k}$ is invariant.
In this section we consider a symmetry where $\hat{H}_{2k}$ is not invariant. By carefully looking at the form of the density matrix \eqref{eq:DM-Closed}, one would realize that a parameter relation which exchanges $\hat{H}_{2k}$ and $\hat{H}_{2k}^{t}$ is also a symmetry since the symmetry is defined by the equality of the density matrix up to a similarity transformation. We actually find such symmetry under the exchange of the parameters
\begin{align}
\left(M_{1},M_{2},M;\zeta_{1},\zeta_{2},\zeta_{3},\zeta_{4}\right) \leftrightarrow\left(M-M_{2}+2k,M-M_{1}+2k,M;\zeta_{3},\zeta_{4},\zeta_{1},\zeta_{2}\right).\label{eq:SymSLD}
\end{align}
Let us check whether the density matrix \eqref{eq:DM-Closed} is symmetric under this duality. By using the expression for the quantum curve \eqref{eq:QCConj}, one can easily check that the quantum curve satisfies
\begin{align}
\hat{H}_{2k}\left(\hat{u},\hat{v};M_{1},M_{2},M;\zeta_{1},\zeta_{2},\zeta_{3},\zeta_{4}\right) =\hat{H}_{2k}\left(-\hat{v},-\hat{u};M-M_{2}+2k,M-M_{1}+2k,M;\zeta_{3},\zeta_{4},\zeta_{1},\zeta_{2}\right).
\end{align}
Here, we have assumed that $E$ is invariant under the change of the variables. Note this actually holds when $M=0$ \eqref{eq:E0}. It is important to notice that if the quantum curve $\hat{H}_{2k}$ is expressed as \eqref{eq:QCConj}, it satisfies
\begin{equation}
\hat{H}_{2k}\left(\hat{u},\hat{v};M_{1},M_{2},M;\zeta_{a}\right)^{t}=\hat{H}_{2k}\left(-\hat{v},-\hat{u};M_{1},M_{2},M;\zeta_{a}\right).
\end{equation}
Therefore, although the quantum curve is not invariant, the density matrix is equivalent up to an appropriate similarity transformation
\begin{align}
\hat{\rho}_{2k}^{\text{(closed)}}\left(M_{1},M_{2},M;\zeta_{1},\zeta_{2},\zeta_{3},\zeta_{4}\right)=\hat{{\cal O}}\hat{\rho}_{2k}^{\text{(closed)}}\left(M-M_{2}+2k,M-M_{1}+2k,M;\zeta_{3},\zeta_{4},\zeta_{1},\zeta_{2}\right)\hat{{\cal O}}^{-1},
\end{align}
where
\begin{equation}
\hat{{\cal O}}=2\sinh\hat{p}\frac{1}{\hat{H}_{2k}\left(\hat{u},\hat{v};M_{1},M_{2},M;\zeta_{a}\right)}.\label{eq:SLdualSym}
\end{equation}

\begin{figure}
\begin{centering}
\includegraphics[width=14cm]{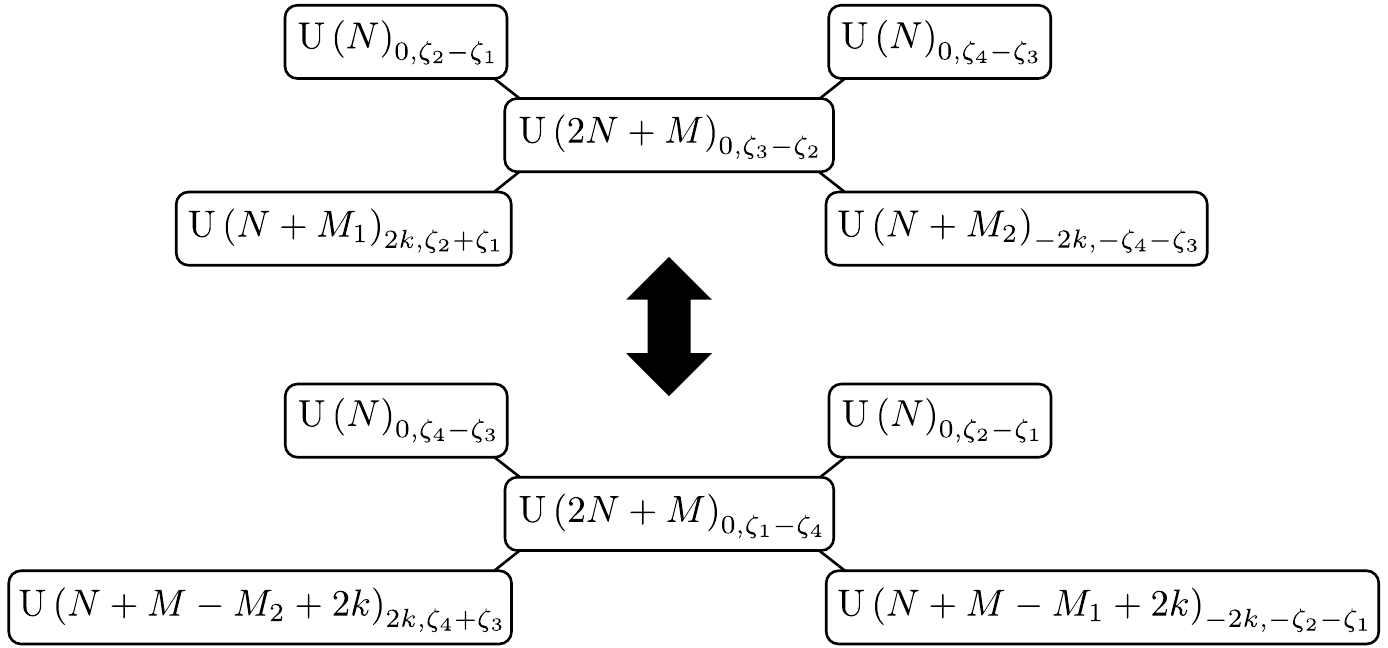}
\par\end{centering}
\caption{Two quiver diagrams corresponding to the symmetry \eqref{eq:SymSLD}. This can be generated by the Seiberg-like dualities.\label{fig:SLDual}}
\label{SLDual}
\end{figure}

To find physical interpretation for the duality, we consider the quiver diagrams corresponding to the both sides of \eqref{eq:SymSLD}, as displayed in figure \ref{fig:SLDual}.
With these quiver diagrams, we find that this symmetry comes from the sequence of the Seiberg-like dualities. The Seiberg-like duality is a duality for the 3d Chern-Simons theories \cite{Giveon:2008zn}. This duality can be applied for any nodes with non-zero Chern-Simons levels.
In that case, if the original node is $\text{U}\left(N\right)_{k,\zeta}$ with level $k$ and the FI parameter $\zeta$ and the sum of the ranks of all the nodes connected to this node is $N_{\text{sum}}$, the Seiberg-like duality changes it to $\text{U}\left(N_{\text{sum}}+|k|-N\right)_{-k,-\zeta}$.
Furthermore, the Chern-Simons levels and the FI parameters of all the connected nodes are added by $k$ and $\zeta$, respectively. Therefore, if we apply the Seiberg-like dualities for the two nodes with non-zero Chern-Simons levels, after exchanging the left side and the right side of the quiver diagram, the quiver diagram returns to the original one with the different ranks and FI parameters. This completely matches with figure \ref{fig:SLDual}, and therefore this is the physical origin of the symmetry \eqref{eq:SymSLD}.
Note that in this case the Seiberg-like duality implies that the absolute values of the prefactors $Z_{k}^{\left(0\right)}$ appeared in \eqref{eq:GPF-Closed} is also invariant under \eqref{eq:SymSLD}. This is indeed satisfied thanks to the symmetry of the partition function of pure Chern-Simons theory $Z^{\text{(CS)}}_k(L)=Z^{\text{(CS)}}_k(k-L)$ and
the known result that the matrix model describing a linear quiver Chern-Simons theory satisfies the Seiberg-like duality \cite{Assel:2014awa}.

\section{Exact values of $Z(N)$ for ${\hat D}_4$ quiver without rank/FI deformations}
\label{sec_TWPY}

In this section we calculate the exact values of the partition of ${\hat D}_\ell$ quiver Chern-Simons theory \eqref{Dr}.
For simplicity let us consider only the ${\hat D}_4$ quiver without FI/rank deformations, $M_1=M_2=M_3=\zeta_1=\zeta_2=\zeta_3=\zeta_4=0$, although the similar calculation should be possible in principle also for more general cases with $M_3=\cdots=M_{\ell-1}=0$ by using the open string formalism \eqref{Xiopen} constructed in section \ref{sec_open}.
In this section we shall denote the partition function as $Z_{k}(N)$.

Without rank/FI deformations, the grand partition function of ${\hat D}_4$ quiver theory is given as \eqref{Xiundeformed}
\begin{align}
\Xi(\kappa)=\sum_{N=0}^\infty \kappa^N
Z_{k}(N)
=\sqrt{\text{Det}(1+\kappa
{\hat \rho}
)},
\label{XiM1M20}
\end{align}
with ${\hat\rho}$ given as \eqref{rhowithoutdeformation} with $\ell=4$.
This implies that we can obtain the partition function $Z_{k}(N)$ by calculating $\text{Tr}{\hat\rho}^n$ with $n\le N$.
After a few manipulations we can rewrite ${\hat \rho}$, up to a similarity transformation which preserves the traces, as
\begin{align}
{\hat \rho}\sim {\hat \rho}'=
\frac{1}{2\cosh\frac{{\hat X}}{2}}
\Bigl(
\frac{\tanh\frac{{\hat X}}{2}}{2}
+\frac{\tanh\frac{{\hat p}}{2}}{2}
\Bigr)
\frac{1}{(2\cosh\frac{{\hat p}}{2})^2}
\Bigl(
\frac{\tanh\frac{{\hat X}}{2}}{2}
+\frac{\tanh\frac{{\hat p}}{2}}{2}
\Bigr)
\frac{1}{2\cosh\frac{{\hat X}}{2}},
\end{align}
where ${\hat X}=2{\hat x}$ which satisfy $[{\hat X},{\hat p}]=4\pi ik$.
We also denote the eigenstates of ${\hat X}$ as $\ket{X}_X$ which differ from the eigenstates of ${\hat x}$ by an overall normalization factor.
See \eqref{ketXketP} and \eqref{ketXketPproperties} for the notation related to ${\hat X}$.
To calculate $\text{Tr}{\hat \rho}^n$, first we write the matrix element of ${\hat\rho}'$ as
\begin{align}
&{}_X\langle x|{\hat \rho}'|y\rangle_X
=\frac{1}{(2\cosh\frac{x}{2})(2\cosh\frac{y}{2})(2\sinh\frac{x-y}{4k})}\frac{1}{2k}\Bigl[
\frac{\tanh\frac{x}{2}}{2}
\frac{\tanh\frac{y}{2}}{2}
\frac{x-y}{4\pi k}
+\Bigl(\frac{\tanh\frac{x}{2}}{2}+\frac{\tanh\frac{y}{2}}{2}\Bigr)\frac{i(x-y)^2}{32\pi^2k^2}\nonumber \\
&\quad +\frac{x-y}{48\pi k}\Bigl(1-\frac{(x-y)^2}{8\pi^2k^2}\Bigr)\Bigr].
\label{rhoprime}
\end{align}
Here we have used the following formulas
\begin{align}
&{}_X\langle x|\frac{1}{(2\cosh\frac{{\hat p}}{2})^2}|y\rangle_X=\frac{x-y}{16\pi k^2\sinh\frac{x-y}{4k}},\quad
{}_X\langle x|\frac{\tanh\frac{{\hat p}}{2}}{2}\frac{1}{(2\cosh\frac{{\hat p}}{2})^2}|y\rangle_X=\frac{i(x-y)^2}{128\pi^2 k^3\sinh\frac{x-y}{4k}},\nonumber \\
&{}_X\langle x|\Bigl(\frac{\tanh\frac{{\hat p}}{2}}{2}\Bigr)^2\frac{1}{(2\cosh\frac{{\hat p}}{2})^2}|y\rangle_X=\frac{x-y}{192\pi k^2\sinh\frac{x-y}{4k}}\Bigl(1-\frac{(x-y)^2}{8\pi^2k^2}\Bigr).
\end{align}

To proceed, we reorganize the terms in $[\cdots]$ in \eqref{rhoprime} into a factorized sum, as
\begin{align}
\frac{\tanh\frac{x}{2}}{2}
\frac{\tanh\frac{y}{2}}{2}
\frac{x-y}{4\pi k}
+\Bigl(\frac{\tanh\frac{x}{2}}{2}+\frac{\tanh\frac{y}{2}}{2}\Bigr)\frac{i(x-y)^2}{32\pi^2k^2}
+\frac{x-y}{48\pi k}\Bigl(1-\frac{(x-y)^2}{8\pi^2k^2}\Bigr)
=\sum_{i=1}^4f_i(x)g_i(y),
\end{align}
with
\begin{align}
&f_1(x)=-\frac{x^3}{384\pi^3k^3}+\frac{ix^2\tanh\frac{x}{2}}{64\pi^2k^2}+\frac{x}{48\pi k},\quad &&g_1(y)=1,\nonumber \\
&f_2(x)=1,\quad &&g_2(y)=\frac{y^3}{384\pi^3k^3}+\frac{iy^2\tanh\frac{y}{2}}{64\pi^2k^2}-\frac{y}{48\pi k},\nonumber \\
&f_3(x)=\frac{ix^2}{32\pi^2k^2}+\frac{x\tanh\frac{x}{2}}{8\pi k},\quad &&g_3(y)=-\frac{iy}{4\pi k}+\frac{\tanh\frac{y}{2}}{2}
,\nonumber \\
&f_4(x)=\frac{ix}{4\pi k}+\frac{\tanh\frac{x}{2}}{2},\quad &&g_4(x)=\frac{iy^2}{32\pi^2k^2}-\frac{y\tanh\frac{y}{2}}{8\pi k}.
\end{align}
Note that $f_i(x)$ and $g_i(x)$ satisfy the following properties
\begin{align}
g_1(x)=f_2(x)|_{i\rightarrow -i},\quad
g_2(x)=-f_1(x)|_{i\rightarrow -i},\quad
g_3(x)=f_4(x)|_{i\rightarrow -i},\quad
g_4(x)=-f_3(x)|_{i\rightarrow -i}.
\label{symmetryoffg}
\end{align}
Now ${\hat\rho}'$ has the following structure:
\begin{align}
e^{\frac{{\hat X}}{2k}}{\hat\rho}'-{\hat\rho}'e^{\frac{{\hat X}}{2k}}
=E({\hat X})\Bigl(\sum_{i=1}^4f_i({\hat X})|0\rangle\!\rangle\langle\!\langle 0|g_i({\hat X})\Bigr)E({\hat X}),
\end{align}
with
\begin{align}
E(x)=\frac{e^{\frac{x}{4k}}}{2\cosh\frac{x}{2}}.
\end{align}
Hence we have
\begin{align}
e^{\frac{{\hat X}}{2k}}({\hat\rho}')^n-({\hat\rho}')^ne^{\frac{{\hat X}}{2k}}
=\sum_{\ell=0}^{n-1}({\hat\rho}')^\ell E({\hat X})\Bigl(\sum_{i=1}^4f_i({\hat X})|0\rangle\!\rangle\langle\!\langle 0|g_i({\hat X})\Bigr)E({\hat X})
({\hat\rho}')^{n-1-\ell},
\end{align}
namely,
\begin{align}
{}_X\langle x|({\hat\rho}')^n|y\rangle_X
=\frac{E(x)E(y)}{e^{\frac{x}{2k}}-e^{\frac{y}{2k}}}\sum_{\ell=0}^{n-1}\sum_{i=1}^4\phi^{(i)}_\ell(x)\psi^{(i)}_{n-1-\ell}(y),
\end{align}
with
\begin{align}
\phi^{(i)}_\ell(x)=\frac{1}{E(x)}{}_X\langle x|({\hat \rho}')^\ell f_i({\hat X})|0\rangle\!\rangle,\quad
\psi^{(i)}_\ell(x)=\langle\!\langle 0|g_i({\hat X})({\hat \rho}')^\ell|x\rangle_X \frac{1}{E(x)}.
\end{align}
Hence $\text{Tr}{\hat\rho}^n=\text{Tr}({\hat\rho}')^n$ is expressed as
\begin{align}
\text{Tr}{\hat\rho}^n
=k\int_{-\infty}^{\infty}\frac{dx}{2\pi}
E(x)^2e^{-\frac{x}{2k}}\sum_{\ell=0}^{n-1}\sum_{i=1}^4\Bigl(
\frac{d\phi^{(i)}_\ell(x)}{dx}\psi^{(i)}_{n-1-\ell}(x)
-\phi^{(i)}_\ell(x)\frac{d\psi^{(i)}_{n-1-\ell}(x)}{dx}
\Bigr).\label{Trrhoprimeton}
\end{align}

The vectors $\phi^{(i)}_\ell(x)$ and $\psi^{(i)}_\ell(x)$ obey the following recursion relations
\begin{align}
\phi_{\ell+1}^{(i)}(x)&=\frac{1}{E(x)}\int_{-\infty}^{\infty}\frac{dy}{2\pi}{}_X\langle x|({\hat\rho}')|y\rangle_X E(y)\phi_\ell^{(i)}(y)\nonumber \\
&=
\frac{1}{2k}\int_{-\infty}^{\infty}\frac{dy}{2\pi}\Bigl[
\frac{\tanh\frac{x}{2}}{2}
\frac{\tanh\frac{y}{2}}{2}
\frac{x-y}{4\pi k}
+\Bigl(
\frac{\tanh\frac{x}{2}}{2}
+\frac{\tanh\frac{y}{2}}{2}
\Bigr)
\frac{i(x-y)^2}{32\pi^2k^2}
+\frac{x-y}{48\pi k}\Bigl(1-\frac{(x-y)^2}{8\pi^2k^2}\Bigr)\Bigr]\nonumber \\
&\quad \times \frac{1}{e^{\frac{x}{2k}}-e^{\frac{y}{2k}}}\frac{e^{\frac{y}{2k}}}{(2\cosh\frac{y}{2})^2}
\phi^{(i)}_\ell(y),\nonumber \\
\psi_{\ell+1}^{(i)}(x)&=
\int_{-\infty}^{\infty}\frac{dy}{2\pi}
\psi_\ell^{(i)}(y)
E(y)
{}_X\langle y|({\hat\rho}')|x\rangle_X
\frac{1}{E(x)}
\nonumber \\
&=
\frac{1}{2k}\int_{-\infty}^{\infty}\frac{dy}{2\pi}\Bigl[
\frac{\tanh\frac{x}{2}}{2}
\frac{\tanh\frac{y}{2}}{2}
\frac{x-y}{4\pi k}
+\Bigl(
\frac{\tanh\frac{x}{2}}{2}
+\frac{\tanh\frac{y}{2}}{2}
\Bigr)
\frac{-i(x-y)^2}{32\pi^2k^2}
+\frac{x-y}{48\pi k}\Bigl(1-\frac{(x-y)^2}{8\pi^2k^2}\Bigr)\Bigr]\nonumber \\
&\quad \times \frac{1}{e^{\frac{x}{2k}}-e^{\frac{y}{2k}}}\frac{e^{\frac{y}{2k}}}{(2\cosh\frac{y}{2})^2}
\psi^{(i)}_\ell(y),\label{recursionphipsi}
\end{align}
together with the initial conditions
\begin{align}
\phi^{(i)}_0(x)=f_i(x),\quad
\psi^{(i)}_0(x)=g_i(x).
\label{M10M20phiic}
\end{align}
Since the recursion relations \eqref{recursionphipsi} preserves the symmetry properties \eqref{symmetryoffg}, it follows that
\begin{align}
\psi^{(1)}_\ell(x)=\phi^{(2)}_\ell(x)|_{i\rightarrow -i},\quad
\psi^{(2)}_\ell(x)=-\phi^{(1)}_\ell(x)|_{i\rightarrow -i},\quad
\psi^{(3)}_\ell(x)=\phi^{(4)}_\ell(x)|_{i\rightarrow -i},\quad
\psi^{(4)}_\ell(x)=-\phi^{(3)}_\ell(x)|_{i\rightarrow -i},
\end{align}
for all $\ell$.
To implement these replacements in the following calculations, it is rather convenient to modify the recursion relation \eqref{recursionphipsi} and the initial conditions \eqref{M10M20phiic} as
\begin{align}
\phi_{\ell+1}^{(c,i)}(x)
&=
\frac{1}{2k}\int_{-\infty}^{\infty}\frac{dy}{2\pi}\Bigl[
\frac{\tanh\frac{x}{2}}{2}
\frac{\tanh\frac{y}{2}}{2}
\frac{x-y}{4\pi k}
+\Bigl(
\frac{\tanh\frac{x}{2}}{2}
+\frac{\tanh\frac{y}{2}}{2}
\Bigr)
\frac{c(x-y)^2}{32\pi^2k^2}
+\frac{x-y}{48\pi k}\Bigl(1-\frac{(x-y)^2}{8\pi^2k^2}\Bigr)\Bigr]\nonumber \\
&\quad \times \frac{1}{e^{\frac{x}{2k}}-e^{\frac{y}{2k}}}\frac{e^{\frac{y}{2k}}}{(2\cosh\frac{y}{2})^2}
\phi^{(c,i)}_\ell(y),\nonumber \\
\phi^{(c,1)}_0(x)&=-\frac{x^3}{384\pi^3k^3}+\frac{cx^2\tanh\frac{x}{2}}{64\pi^2k^2}+\frac{x}{48\pi k},\quad
\phi^{(c,2)}_0(x)=1,\quad
\phi^{(c,3)}_0(x)=\frac{cx^2}{64\pi^2k^2}+\frac{x\tanh\frac{x}{2}}{16\pi k},\nonumber \\
\phi^{(c,4)}_0(x)&=\frac{cx}{2\pi k}+\tanh\frac{x}{2},
\end{align}
calculate $\phi^{(c,i)}_\ell(x)$ recursively in $\ell$, and generate $\phi^{(i)}_\ell(x)$ and $\psi^{(i)}_\ell(x)$ as
\begin{align}
&\phi^{(i)}_\ell(x)=\phi^{(c,i)}_\ell(x)|_{c=i},\nonumber \\
&\psi^{(1)}_\ell(x)=\phi^{(c,2)}_\ell(x)|_{c=-i},\quad
\psi^{(2)}_\ell(x)=-\phi^{(c,1)}_\ell(x)|_{c=-i},\quad
\psi^{(3)}_\ell(x)=\phi^{(c,4)}_\ell(x)|_{c=-i},\quad
\psi^{(4)}_\ell(x)=-\phi^{(c,3)}_\ell(x)|_{c=-i}.
\end{align}

\begin{figure} 
\begin{center} 
\begin{tikzpicture}[scale=0.25]
\draw [thick, ->] (5,0) -- (5,8);
\draw [thick, ->] (2,3) -- (18,3);
\draw [thick] (16,6) -- (16,8);
\draw [thick] (16,6) -- (18,6);
\node [above] at (17,6) {$v$};
\draw [blue, thick, domain=0:360, samples=10] plot({4+(-sin(0.25*\x)+1)},{4+(cos(0.25*\x)-1)});
\draw [blue, thick, domain=0:360, samples=10] plot({4+(-cos(0.25*\x)+1)},{2+(-sin(0.25*\x)+1)});
\draw [blue, thick, ->] (5,4) -- (17,4);
\draw [blue, thick, ->] (17,2) -- (8,2);
\draw [blue, thick] (8,2) -- (5,2);
\node [above] at (17,0) {$\textcolor{blue}{\gamma}$};
\draw [red, thick, domain=0:360, samples=200] plot({5+(12-0)*\x/360},{3-0.2*sin(20*\x)});
\end{tikzpicture}
\end{center} 
\caption{ 
The integration contour $\gamma$ (blue line) used in \eqref{PYrewriting} and the branch cut of $\log^{(+)}(\cdot)$ (wavy red line). 
} 
\label{fig_PYcontour} 
\end{figure}
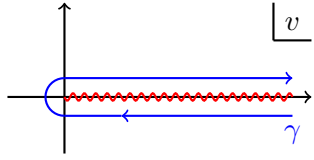

If we introduce new integration variables $u=e^{\frac{x}{2k}},v=e^{\frac{y}{2k}}$, the integrations in the recursion relation for $\phi_n^{(c,i)}(x)$ is written as
\begin{align}
\phi_{\ell+1}^{(c,i)}(u)&=\frac{1}{2\pi}\int_{0}^{\infty} dv\frac{v^{2k}\phi_{\ell}^{(c,i)}(v)}{(u-v)(v^{2k}+1)^2}\biggl[\nonumber \\
&\quad \Bigl[
\frac{1}{8\pi}\frac{u^{2k}-1}{u^{2k}+1}
\frac{v^{2k}-1}{v^{2k}+1}
\log u
+\frac{c}{16\pi^2}\Bigl(\frac{u^{2k}-1}{u^{2k}+1}
+\frac{v^{2k}-1}{v^{2k}+1}
\Bigr)(\log u)^2
+\frac{1}{24\pi}\Bigl(\log u-\frac{1}{2\pi^2}(\log u)^3\Bigr)\Bigr]\nonumber \\
&\quad+\Bigl[-\frac{1}{8\pi}
\frac{u^{2k}-1}{u^{2k}+1}
\frac{v^{2k}-1}{v^{2k}+1}
-\frac{c}{8\pi^2}
\Bigl(\frac{u^{2k}-1}{u^{2k}+1}
+\frac{v^{2k}-1}{v^{2k}+1}
\Bigr)\log u
+\frac{1}{24\pi}\Bigl(-1+\frac{3}{2\pi^2}(\log u)^2\Bigr)\Bigr](\log v)\nonumber \\
&\quad+\Bigl[\frac{c}{16\pi^2}\Bigl(\frac{u^{2k}-1}{u^{2k}+1}
+\frac{v^{2k}-1}{v^{2k}+1}
\Bigr)
-\frac{1}{16\pi^3}\log u\Bigr](\log v)^2\nonumber \\
&\quad+\Bigl[\frac{1}{48\pi^3}\Bigr](\log v)^3
\biggr]\nonumber \\
&=\frac{1}{2\pi}\sum_{j\ge 0}\sum_{w\text{: poles in }\mathbb{C}\backslash \mathbb{R}_{\ge 0}}\text{Res}\Biggl[\frac{v^{2k}\phi_{\ell,j}^{(c,i)}(v)}{(u-v)(v^{2k}+1)^2}\biggl[\nonumber \\
& -\frac{(2\pi i)^{j+1}}{j+1}\Bigl[
\frac{1}{8\pi}\frac{u^{2k}-1}{u^{2k}+1}
\frac{v^{2k}-1}{v^{2k}+1}
\log u
+\frac{c}{16\pi^2}\Bigl(\frac{u^{2k}-1}{u^{2k}+1}
+\frac{v^{2k}-1}{v^{2k}+1}
\Bigr)(\log u)^2\nonumber \\
&\quad\quad\quad \quad +\frac{1}{24\pi}\Bigl(\log u-\frac{1}{2\pi^2}(\log u)^3\Bigr)\Bigr]B_{j+1}\Bigl(\frac{\log^{(+)}v}{2\pi i}\Bigr)\nonumber \\
&-\frac{(2\pi i)^{j+2}}{j+2}\Bigl[-\frac{1}{8\pi}
\frac{u^{2k}-1}{u^{2k}+1}
\frac{v^{2k}-1}{v^{2k}+1}
-\frac{c}{8\pi^2}
\Bigl(\frac{u^{2k}-1}{u^{2k}+1}
+\frac{v^{2k}-1}{v^{2k}+1}
\Bigr)\log u\nonumber \\
&\quad \quad \quad \quad +\frac{1}{24\pi}\Bigl(-1+\frac{3}{2\pi^2}(\log u)^2\Bigr)\Bigr]B_{j+2}\Bigl(\frac{\log^{(+)}v}{2\pi i}\Bigr)\nonumber \\
&-\frac{(2\pi i)^{j+3}}{j+3}\Bigl[\frac{c}{16\pi^2}\Bigl(\frac{u^{2k}-1}{u^{2k}+1}
+\frac{v^{2k}-1}{v^{2k}+1}
\Bigr)
-\frac{1}{16\pi^3}\log u\Bigr]B_{j+3}\Bigl(\frac{\log^{(+)}v}{2\pi i}\Bigr)\nonumber \\
&-\frac{(2\pi i)^{j+4}}{j+4}\Bigl[\frac{1}{48\pi^3}\Bigr]B_{j+4}\Bigl(\frac{\log^{(+)}v}{2\pi i}\Bigr)
\biggr],v\rightarrow w\Biggr],
\label{D4quiver_M0_TWPY_phifinal}
\end{align}
where $\phi_{\ell,j}^{(c,i)}(u)$ are the rational functions of $u$ given by
\begin{align}
\phi_{\ell}^{(c,i)}(u)=\sum_{j\ge 0}\phi_{\ell,j}^{(c,i)}(u)(\log u)^j.
\end{align}
To obtain the second expression in \eqref{D4quiver_M0_TWPY_phifinal} we have used the following formula \cite{Putrov:2012zi}
\begin{align}
&\int_0^\infty dvf(v)(\log v)^j
=-\frac{(2\pi i)^j}{j+1}\int_\gamma dv f(v)B_{j+1}\Bigl(\frac{\log^{(+)}v}{2\pi i}\Bigr)\nonumber \\
&=-\frac{(2\pi i)^{j+1}}{j+1}\sum_{w\text{: poles in }\mathbb{C}\backslash\mathbb{R}_{\ge 0}}\text{Res}\biggl[f(v)B_{j+1}\Bigl(\frac{\log^{(+)}v}{2\pi i}\Bigr),v\rightarrow w\biggr],\quad (j\in\mathbb{Z}_{\ge 0},\,\,f(v)\text{: rational function}),
\label{PYrewriting}
\end{align}
where $\log^{(+)}u$ is the logarithm with branch cut $\mathbb{R}_{\ge 0}$ (i.e.~$\log(re^{i\theta})=\log r+i\theta$ for $0\le\theta<2\pi$), $\gamma$ is the contour depicted in figure \ref{fig_PYcontour} and $B_{j+1}(u)$
are the Bernoulli polynomials (or any polynomials satisfying $B_{j+1}(u+1)-B_{j+1}(u)=(j+1)u^j$).

We can show by induction that at each step of the recursion it is sufficient to pick only the following poles
\begin{align}
v=u,\quad\quad v=e^{\frac{\pi i}{k}(m-\frac{1}{2})},\quad (m=1,2,\cdots,2k).
\end{align}
Similarly, the integration in $\text{Tr}{\hat\rho}^n$ \eqref{Trrhoprimeton} is written as
\begin{align}
\text{Tr}{\widehat\rho}^n=\frac{k}{2\pi}\sum_{j\ge 0}\sum_{w\text{: poles in }\mathbb{C}\backslash\mathbb{R}_{\ge 0}}
\Bigl(-\frac{(2\pi i)^{j+1}}{j+1}\Bigr)
\text{Res}\biggl[\frac{u^{2k}}{(u^{2k}+1)^2}\Phi^{(j)}_n(u)B_{j+1}\Bigl(\frac{\log^{(+)}u}{2\pi i}\Bigr),u\rightarrow w\biggr],
\label{D4quiver_M0_TWPY_trrhofinal}
\end{align}
where $\Phi_{n,j}(u)$ are the rational functions given by
\begin{align}
\sum_{\ell=0}^{n-1}\sum_{i=1}^4\Bigl(
\frac{d\phi_\ell^{(i)}}{du}\psi^{(i)}_{n-1-\ell}(u)
-\phi^{(i)}_\ell(u)\frac{d\psi_{n-1-\ell}^{(i)}}{du}
\Bigr)
=\sum_{j\ge 0}\Phi_{n,j}(u)(\log u)^j,
\end{align}
and the relevant poles are $v=e^{\frac{\pi i}{k}(m-\frac{1}{2})}$ with, $m=1,2,\cdots,2k$.

\begin{figure}
\begin{center}
\includegraphics[width=16cm]{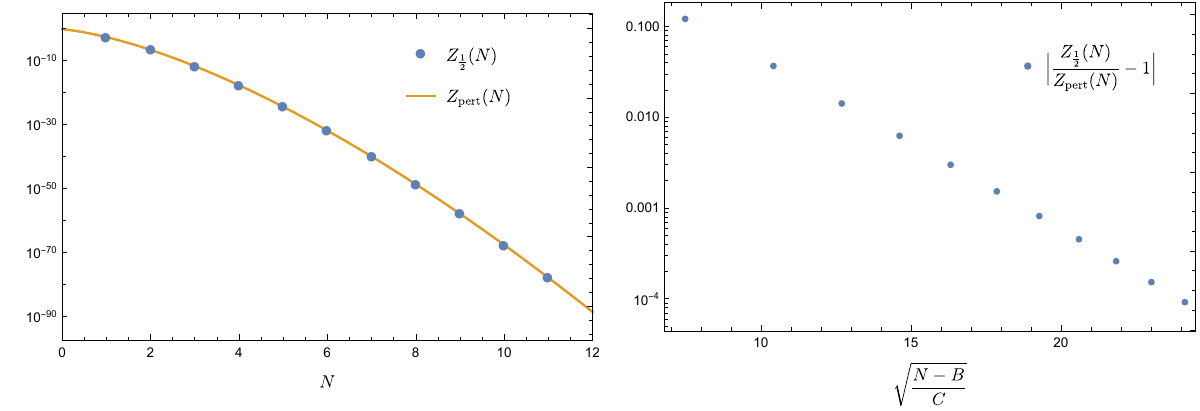}\\
\includegraphics[width=16cm]{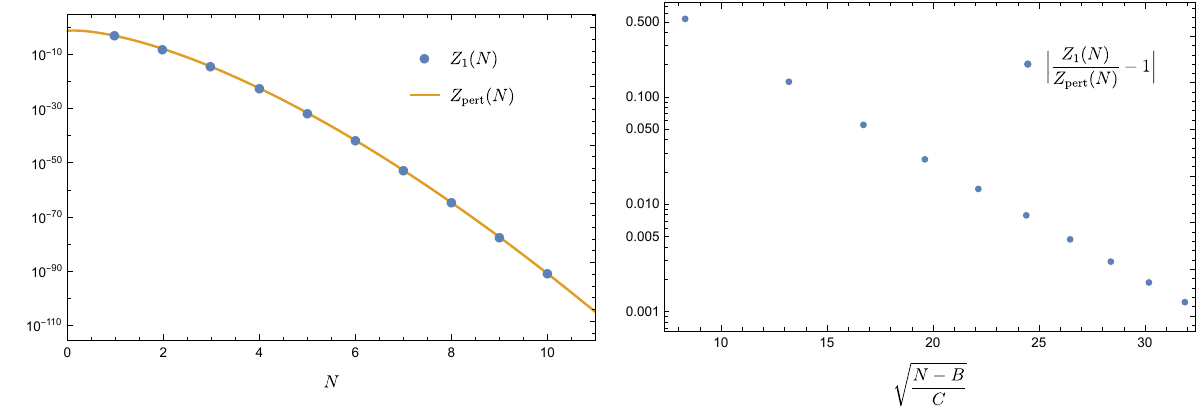}\\
\includegraphics[width=16cm]{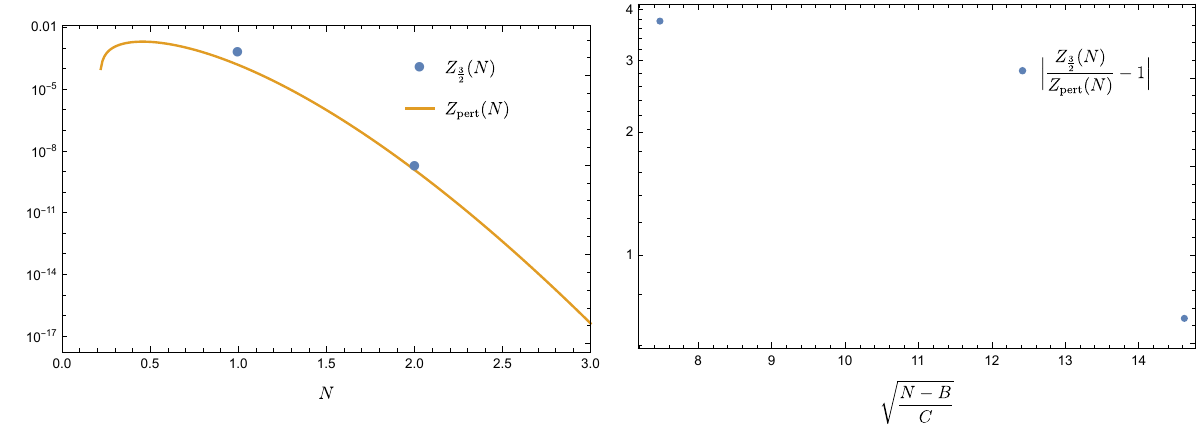}\\
\includegraphics[width=16cm]{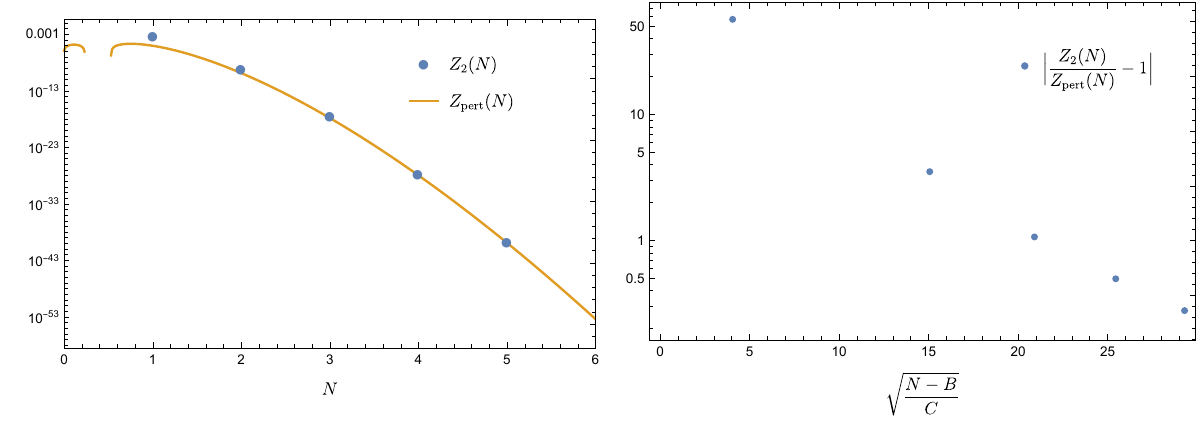}
\caption{
Left: exact partition functions $Z_{k}(N)$ of the affine $D_4$ quiver superconformal Chern-Simons theory with $M_1=M_2=M_3=\zeta_1=\zeta_2=\zeta_3=\zeta_4=0$ and $k=\frac{1}{2},1,\frac{3}{2},2$.
$Z_{\text{pert}}(N)$ is defined in \eqref{Airy}.
Right: deviations of $Z_{k}(N)$ from the perturbative formula $Z_{\text{pert}}(N)$ \eqref{Airy} with $\ell=4$.
}
\label{fig_D4quiver_ZvsAiry_kappa1_M0}
\end{center}
\end{figure}

Note that the above formulas written in $u=e^{\frac{x}{2k}}$ \eqref{D4quiver_M0_TWPY_phifinal},\eqref{D4quiver_M0_TWPY_trrhofinal} contain $u$ and $u^{2k}$, but they do not contain the terms like $u^{k}$.
Hence we can also set $k$ to be an odd half-integer without causing any complications.
In appendix \ref{app_exactvalues} we display the exact values of $Z_{k}(N)$ thus calculated for $k=\frac{1}{2},1,\frac{3}{2},2$.
These results show a good agreement with the all order $1/N$ perturbative expansion of the partition function of the theory on affine $D_\ell$ quiver \eqref{Dr} with $M_1=\cdots=M_{\ell-1}=\zeta_1=\cdots=\zeta_\ell=0$ obtained by the semiclassical (small $k$) expansion \cite{Moriyama:2015jsa}:
\begin{align}
Z_k(N)\approx Z_{\text{pert}}(N)=e^AC^{-\frac{1}{3}}\text{Ai}[C^{-\frac{1}{3}}(N-B)],
\label{Airy}
\end{align}
where
\begin{align}
C&=\frac{1}{2\pi^2k}\frac{\ell-1}{\ell(\ell-2)^2},\quad
B=\frac{\pi^2C}{3}-\frac{1}{12k}\Bigl(\frac{\ell}{\ell-2}+\frac{1}{\ell}\Bigr)+\frac{\ell(\ell-1)}{24}k,\nonumber \\
A&=\frac{1}{2}(A_\text{ABJM}(2\ell k)+\ell^2A_\text{ABJM}(2(\ell-2)k)),
\label{ABC_for_Drquiver}
\end{align}
with (see e.g.~eq(4.8) in \cite{Hatsuda:2015owa}v2)
\begin{align}
A_{\text{ABJM}}(k)=
\frac{2\zeta(3)}{\pi^2k}\Bigl(1-\frac{k^3}{16}\Bigr)+\frac{k^2}{\pi^2}\int_0^\infty\frac{x\log(1-e^{-2x})}{e^{kx}-1}.
\end{align}
Here $\ell=4$.
See figure \ref{fig_D4quiver_ZvsAiry_kappa1_M0} for the comparison between $Z_{\text{pert}}(N)$ and the exact values of $Z_{k}(N)$ for finite $N$.

The partition function for finite $N$ deviates from $Z_{\text{pert}}(N)$ \eqref{Airy} due to the non-perturbative corrections in $1/N$.
To argue the structure of the non-perturbative corrections it is useful to switch to the modified grand potential $J(\mu)$ given by
\begin{align}
\Xi(\kappa=e^{\mu})=\sum_{n\in\mathbb{Z}}e^{J(\mu+2\pi in)}.
\end{align}
The Airy function expression for $Z_{\text{pert}}(N)$ follows from the inversion formula
\begin{align}
Z_k(N)=\int_{-i\infty}^{i\infty}\frac{d\mu}{2\pi i}e^{J(\mu)-\mu N},
\end{align}
together with following perturbative part of the large $\mu$ expansion of $J(\mu)$:
\begin{align}
J(\mu)=J_\text{pert}(\mu)+J_{\text{np}}(\mu),\quad J_\text{pert}(\mu)=\frac{C}{3}\mu^3+B\mu+A.
\end{align}
As already found in \cite{Moriyama:2015jsa}, there are non-perturbative corrections of the following form $J_{\text{np}}(\mu)=\cdots+(\alpha \mu^2+\beta\mu+\gamma)e^{-\omega\mu}+\cdots$, with $\omega\in \mathbb{N}/2$,\footnote{
In \cite{Moriyama:2015jsa} it was found, without fixing $\ell$, that there are three types of the membrane instanton exponents, $\omega=\frac{2n}{\ell}$, $\omega=\frac{2n}{\ell-2}$ and $\omega=\frac{n}{2}$ ($n\in\mathbb{N}$) whose instanton coefficients are at most of order $\mu^1$.
However, when $\ell$ is set to an integer some of the coefficients diverges and the divergences cancel among those non-perturbative effects with the same coefficients.
In this pole cancellation procedure an extra power of $\mu$ can appear due to the $\ell$-dependence of the exponents \cite{Moriyama:2014waa}.
}
which are analogous to the membrane instantons \cite{Drukker:2011zy} in the ABJM theory.
From the results for the circular quiver Chern-Simons theories \cite{Moriyama:2014gxa} it is natural to expect that there are other non-perturbative effects with the exponents $\omega\sim k^{-1}$ and analogous to the worldsheet instantons \cite{Cagnazzo:2009zh} in the ABJM theory, which are not visible in the small $k$ expansion.
In contrast to the membrane instanton, the coefficients of the worldsheet instantons are not guaranteed to be a polynomials in $\mu$ with a definite upper bound on their order.
Nevertheless, here let us assume even after including non-perturbative effects of both kinds that the coefficient of the leading non-perturbative effect is a polynomial of at most second order in $\mu$,
namely,
\begin{align}
J_{\text{np}}(\mu)=(\alpha \mu^2+\beta\mu+\gamma)e^{-\omega\mu}+\cdots,
\label{Jnp}
\end{align}
for each $k\in\mathbb{R}$, with $\omega$ either in $\mathbb{N}/2$ or in $\mathbb{Q}_{>0}/k$.
This is indeed the case for the $\text{U}(N)_k\times \text{U}(N+M)_{-k}$ ABJM theory \cite{Hatsuda:2012dt,Honda:2014npa} as well as for many other circular quiver Chern-Simons theories \cite{Grassi:2014vwa,Hatsuda:2014vsa,Moriyama:2014gxa,Moriyama:2014nca,Hatsuda:2015lpa}.
Substituting \eqref{Jnp} into the inversion formula we obtain
\begin{align}
Z_k(N)&=Z_{\text{pert}}(N)+\int_{-i\infty}^{i\infty}\frac{d\mu}{2\pi i} (\alpha\mu^2+\beta\mu+\gamma)e^{J_{\text{pert}}(\mu)-\mu(N+\omega)}+\cdots\nonumber \\
&=Z_{\text{pert}}(N)+\Bigl[\alpha\Bigl(-\frac{\partial}{\partial N}\Bigr)^2+\beta\Bigl(-\frac{\partial}{\partial N}\Bigr)+\gamma\Bigr]Z_{\text{pert}}(N+\omega)+\cdots.
\end{align}
By using the asymptotics of the Airy function $\text{Ai}(z)\sim (2\sqrt{\pi})^{-1}x^{-\frac{1}{4}} e^{-\frac{2}{3}x^{\frac{3}{2}}}$ at large $z$, we obtain
\begin{align}
\frac{Z_k(N)}{Z_{\text{pert}}(N)}-1\approx \biggl[\alpha x^2+\Bigl(-\frac{\omega^2\alpha}{4C}+\beta\Bigr)x+\Bigl(\frac{3\omega}{4C}+\frac{\omega^4}{32C^2}\Bigr)\alpha-\frac{\omega^2\beta}{4C}+\gamma\biggr]e^{-\omega x},\quad x=\sqrt{\frac{N-B}{C}}.
\label{fittingansatz}
\end{align}
Hence the $1/\mu$ non-perturbative effect in $J(\mu)$ \eqref{Jnp} indeed corresponds to the $1/N$ non-perturbative effect in the partition function.

For $k=1$, by fitting the exact values we find 
\begin{align}
\begin{tabular}{|c|c|c|c|}
\hline
ansatz
&$\alpha\neq 0$
&$\alpha=0,\beta\neq 0$
&$\alpha=\beta=0,\gamma\neq 0$
\\ \hline
$\alpha$
&$0.004143$
&-
&-                      
\\ \hline
$\beta$
&$0.02036$
&$0.09800$
&-
\\ \hline
$\gamma$
&$4.350$
&$2.921$
&$3.383$
\\ \hline
$\omega$
&$0.2793$
&$0.2658$
&$0.2486$
\\ \hline
\end{tabular}.
\end{align}
Here we have used the exact values at $N=7,8,9,10$ for the ansatz $\alpha\neq 0$, $N=8,9,10$ for the ansatz $\alpha=0,\beta\neq 0$ and $N=9,10$ for the ansatz $\alpha=\beta=0,\gamma\neq 0$.
We observe that even if we include $\alpha$ and $\beta$ to the ansatz, these values in the fitting turn out to be significantly small compared with $\gamma$: $\frac{\alpha}{\beta}\sim 10^{-3}$ and $\frac{\beta}{\gamma}\lesssim 0.03$.
This suggests that $\alpha=\beta=0,\gamma\neq 0$ is the correct ansatz, and also supports our working assumption that the leading instanton coefficient is a polynomial of at most second order in $\mu$ \eqref{Jnp}.
We also observe the ansatz $\alpha=\beta=0,\gamma\neq 0$ gives $\omega$ closest to a simple rational number, that is, $\frac{1}{4}$.
Hence we conclude that the exponent of the leading non-perturbative effect for $k=1$ is $\omega=\frac{1}{4}$.
\begin{figure}
\begin{center}
\includegraphics[width=8cm]{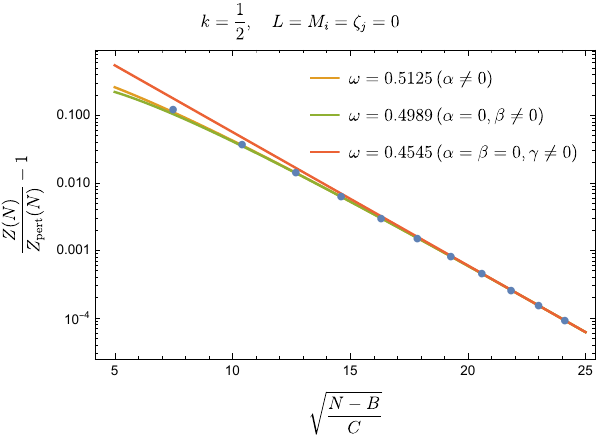}
\caption{
Fitting of the exact values of $Z_{\frac{1}{2}}(N)$ with ansatz \eqref{fittingansatz}.
}
\label{fig_instantonbyfitting_k1/2alpha0only}
\end{center}
\end{figure}
Since the exponent of the leading membrane instanton is $\frac{1}{2}$ \cite{Moriyama:2015jsa}, we conclude that the non-perturbative effect we have identified is the leading worldsheet instanton of the following form:
\begin{align}
J_{\text{np}}(\mu)=\cdots+d_1(k)e^{-\frac{\mu}{4k}},
\label{leadingws}
\end{align}
with the coefficient $d_1(k)$ independent of $\mu$.

For $k=1/2$, by fitting the exact values at $N=8,9,10,11$ with the ansatz \eqref{fittingansatz} we find
\begin{align}
\begin{tabular}{|c|c|c|c|}
\hline
ansatz
&$\alpha\neq 0$
&$\alpha=0,\beta\neq 0$
&$\alpha=\beta=0,\gamma\neq 0$
\\ \hline
$\alpha$
&$0.01427$
&-                     
&-
\\ \hline
$\beta$
& $0.5799$
&$0.6685$
&-
\\ \hline
$\gamma$
&$1.976$
&$1.517$
&$5.305$
\\ \hline
$\omega$
&$0.5125$
&$0.4989$
&$0.45450$
\\ \hline
\end{tabular}.
\end{align}
Here we have used the exact values at $N=8,9,10,11$ for the ansatz $\alpha\neq 0$, $N=9,10,11$ for the ansatz $\alpha=0,\beta\neq 0$ and $N=10,11$ for the ansatz $\alpha=\beta=0,\gamma\neq 0$.
Again we observe that $\alpha$ in the last ansatz is significantly small compared with $\beta,\gamma$, while the magnitude of $\beta$ is not so different from that of $\gamma$.
Taking also into account that the result of the fitting with the ansatz $\alpha=0,\beta\neq 0$ shows better agreement for smaller $N$'s compared with the result with the ansatz $\alpha=\beta=0,\gamma\neq 0$ (see figure \ref{fig_instantonbyfitting_k1/2alpha0only}), we conclude that $\alpha=0,\beta\neq 0$ is the correct ansatz in this case.
The ansatz with $\alpha=0,\beta\neq 0$ also gives $\omega$ closest to a simple rational number, that is, $\frac{1}{2}$.
Hence we conclude that the exponent of the leading non-perturbative effect for $k=\frac{1}{2}$ is $\omega=\frac{1}{2}$.
In this case the non-perturbative effect is understood as the mixture of the leading worldsheet instanton \eqref{leadingws} and the leading membrane instanton, both of which has the exponent $\omega=\frac{1}{2}$.

\section{Discussion\label{sec:Discussion}}

In this paper we revisited the partition function of superconformal Chern-Simons theory on ${\hat D}$-type quiver \cite{Moriyama:2015jsa} and extended the previous analysis in two directions.
When both the FI parameters and the rank differences are turned off, this model was studied in \cite{Moriyama:2015jsa} through the WKB expansion of the Fermi gas formalism.

First, we constructed the Fermi gas formalism with FI parameters and the rank differences both in the open string formalism and the closed string formalism.
The appearance of the quantum curve associated to the (2,2) model with the Chern-Simons level $2k$ in the density matrix \eqref{eq:DM-Closed}, which is $\hat{H}_{2k}$, is interesting, and it would be important to find its physical interpretation.
\begin{figure}
\begin{centering}
\includegraphics[scale=0.6]{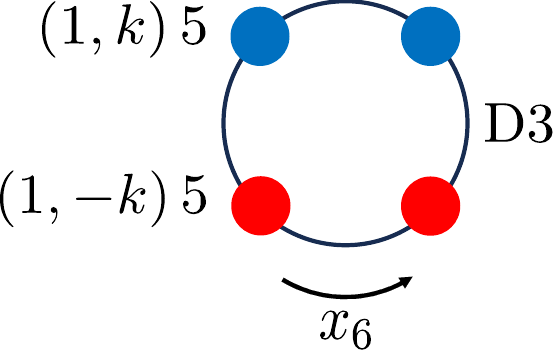}
\par\end{centering}
\caption{The brane configuration whose worldvolume theory is the (2,2) model with the Chern-Simons levels $0$ and $\pm2k$. The big circle represents D3-branes, while the blue circles and the red circles represent $\left(1,\pm k\right)$5-branes, respectively. D3-branes are extended to 0126-directions, where 6-direction is periodic, while the $\left(1,\pm k\right)$5-branes share the 012-directions and are extended to other three directions. Since the Chern-Simons levels between a $\left(1,k_{1}\right)$5-brane and a $\left(1,k_{2}\right)$5-brane is $k_{2}-k_{1}$, this brane configuration provides the correct Chern-Simons levels.\label{fig:BC-22model}}
\end{figure}
For this purpose it seems worthwhile to note that the quantum curve of the (2,2) model is closely related to the brane configuration whose worldvolume theory is the (2,2) model (see figure \ref{fig:BC-22model}).\footnote{
Usually the brane configuration associated to the (2,2) model consists of the $\left(1,2k\right)$5-branes and the NS5-branes.
This brane configuration and the brane configuration in figure \ref{fig:BC-22model} is related by an appropriate ${\rm SL}\left(2,\mathbb{Z}\right)$ transformation (T-transformation), and thus both the worldvolume theories are the same.
}
When the all the ranks are equal, the (2,2) model quantum curve is given by replacing the $\left(1,k\right)5$-brane to $2\cosh\hat{u}/2$ and $\left(1,-k\right)5$-brane to $2\cosh\hat{v}/2$. On the other hand, it is known that the affine D-type quiver theory without Chern-Simons terms can be realized by adding two ${\rm ON^{0}}$ planes, which is a superposition of an ${\rm ON^{-}}$ and an NS5-brane 
\cite{deBoer:1996mp,Kapustin:1998fa,Hanany:1999sj,Porrati:1996xi,Bourget:2023uhe}.
The D3-branes can end on either the NS5-brane or its mirror image, which corresponds to the affine nodes.
The $\hat{D}_{4}$ quiver theory without Chern-Simons terms can be realized by placing the four NS5-branes between the two ${\rm ON^{-}}$-branes.
When an NS5-brane is replaced to a $\left(1,k\right)$5-brane, the mirror image, which was originally an NS5-brane, becomes a $\left(1,-k\right)$5-brane \cite{Gulotta:2012yd}.
\begin{figure}
\begin{centering}
\includegraphics[scale=0.7]{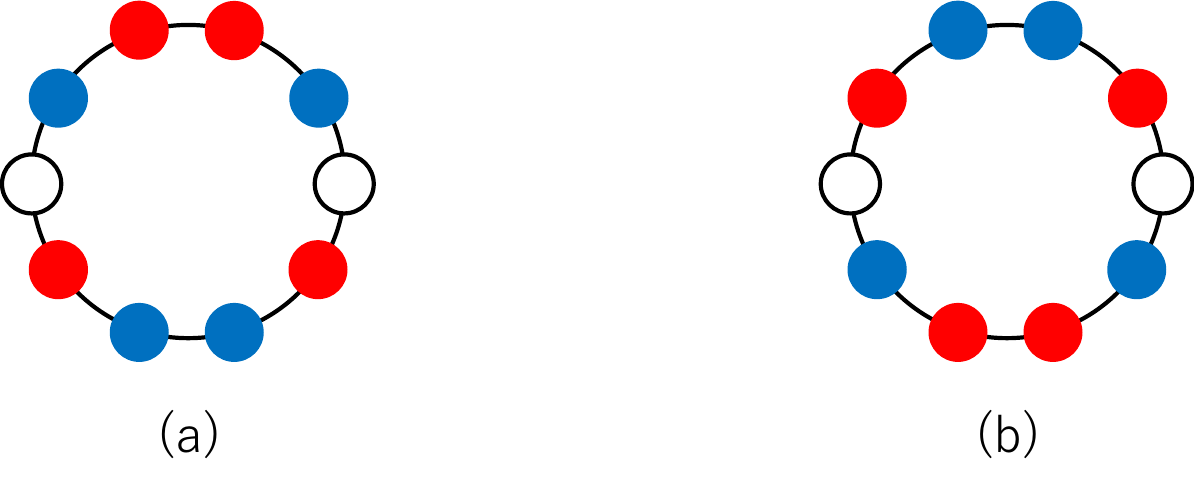}
\par\end{centering}
\caption{(a) The brane configuration associated to the density matrix \eqref{eq:DM-Closed}. The white circle represents ${\rm ON^{-}}$ plane. The $\left(1,\pm k\right)$5-branes on the bottom half are the mirror image of the ones on the top half. We propose that the worldvolume theory of this brane configuration is the $\hat{D_4}$ quiver theory in figure \ref{D4quiver}. (b) The brane configuration after performing the Hanany-Witten transition. The combination of this transition and the rotation explains the symmetry of the density matrix \eqref{eq:DM-Closed}.
\label{fig:BC-D4}}
\end{figure}
Based on these considerations, we expect that the appearance of $\hat{H}_{2k}({\hat u},{\hat v})^{-1}$ in the density matrix \eqref{eq:DM-Closed} implies that the brane configuration which corresponds to our $\hat{D}_{4}$ quiver theory is the one displayed in figure \ref{fig:BC-D4} (a).
${\hat H}_{2k}({\hat u},{\hat v})^{-1}$ in \eqref{eq:DM-Closed} corresponds to the four $\left(1,\pm k\right)$5-branes on the top half while $({\hat H}_{2k}({\hat u},{\hat v})^t)^{-1}$, which is the same as ${\hat H}_{2k}(-{\hat v},-{\hat u})^{-1}$ according to \eqref{eq:QCConj}, corresponds to the ones on the lower half.
Conversely, this prescription implies that the factor 
$2\sinh{\hat p}=2\sinh\frac{\hat{u}+\hat{v}}{2}$
in \eqref{eq:DM-Closed} corresponds to the ${\rm ON^{-}}$ plane.
As a simple check of the proposed brane configuration in figure \ref{fig:BC-D4} (a) for the ${\hat D}_4$ quiver theory, one can easily see that the Chern-Simons levels are consistent with the rule described in figure \ref{fig:BC-22model}. As a more non-trivial check, let us consider the second symmetry discussed in section \ref{subsec:SL-dual}.
As discussed, this symmetry can be generated by the Seiberg-like dualities. It is known that the Seiberg-like duality can be interpreted as the Hanany-Witten transition in the brane configuration \cite{Giveon:2008zn}.
Indeed, one can easily see that, in terms of the brane configuration, the combination of the appropriate Hanany-Witten transitions (see figure \ref{fig:BC-D4} (b)) and the rotation relates the brane configurations with the same order but different number of D3-branes. This rotation appears as the exchange between $\hat{H}_{2k}$ and $\hat{H}_{2k}^{t}$, or in other words the similarity transformation \eqref{eq:SLdualSym}.

Second, by using the Fermi gas formalism we calculated the exact values of the partition function for finite $k$ and $N$.
As was the case for the supersymmetric Chern-Simons theories on circular quivers, such exact values allow us to study the worldsheet instanton corrections, which are non-perturbative both in $k$ and $1/N$ and hence cannot be accessed through the WKB expansion.
In particular, we identified the exponent of the leading worldsheet instanton as $J_{\text{WS}}(\mu)\sim e^{-\frac{\mu}{4k}}$ in terms of the grand potential, or $-\log Z(N)+\log Z_{\text{pert}}(N)\sim e^{-\pi\sqrt{\frac{2N}{k}}}$ in terms of the free energy.
Although we did not find concrete results, by pushing the exact calculation further it would be possible to determine the prefactor of this instanton effect as function of $k$.
As an alternative approach to the worldsheet instantons, it would also be interesting to investigate the 't Hooft expansion \cite{Drukker:2010nc,Grassi:2014vwa} of the partition function of ${\hat D}_4$ quiver theory as well as the theories with other ${\hat D}$-type quivers.

In \cite{Bonelli:2022dse} we conjectured that the grand partition function of the $\text{U}(N+M_1)_k\times \text{U}(N+M)_0\times \text{U}(N+M_2)_{-k}\times \text{U}(N)_0$ circular quiver theory solves the $\mathfrak{q}$-deformed Painlev\'e VI equation in $\tau$-form \cite{1998Nonli..11..823S,2006CMaPh.262..595T,Jimbo:2017ael} with $\mathfrak{q}=e^{\frac{2\pi i}{k}}$ and $(M_1,M_2,M,\zeta_1,\zeta_2)$ being identified appropriately with the Painlev\'e parameters and the time variable $(\theta_0,\theta_1,\theta_t,\theta_\infty,t)$.\footnote{
More precisely, the proposal in \cite{Bonelli:2022dse} is $\tau=F(M_1,M_2,M,\zeta_1,\zeta_2)\det(1+\kappa\Omega(M_1,M_2,M,\zeta_1,\zeta_2){\hat\rho}^{(2,2)}_k(M_1,M_2,M,\zeta_1,\zeta_2))$ with concrete expressions for 
$F(M_1,M_2,M,\zeta_1,\zeta_2)$ and $\Omega(M_1,M_2,M,\zeta_1,\zeta_2)$ were provided only for $M=0$.
However, this problem was solved in \cite{Moriyama:2023mjx}
}
It would be interesting to investigate an analogous set of $\mathfrak{q}$-difference equations for the grand partition function of ${\hat D}_4$ quiver theory.
The results in this paper will be useful in finding such a connection.

There are several further research directions which we hope to address in future.

The theory we consider, as well as the similar supersymmetric Chern-Simons theories on circular quivers, describe $N$ M2-branes in M-theory probing some orbifold background.
Under the AdS/CFT correspondence these theories are dual to the geometries of the form $\text{AdS}_4\times Y_7$, where both of $\text{AdS}_4$ and $Y_7$ have the length scales proportional to $N^{\frac{1}{6}}$.
In this context, the $1/N$ non-perturbative corrections to the free energy can be interpreted as the instanton effects of closed M2-branes in the bulk wrapped on some three-dimensional volume in $Y_7$.
Indeed, in the ABJM theory, the exponent of the worldsheet instanton $e^{-2\pi \sqrt{\frac{2N}{k}}}$ and that of the membrane instantons $e^{-\pi\sqrt{2kN}}$ coincide respectively with the volume of $\mathbb{CP}^1\times S^1/\mathbb{Z}_k\subset S^7/\mathbb{Z}_k$ and $\mathbb{RP}^3\subset S^7/\mathbb{Z}_k$ \cite{Cagnazzo:2009zh,Drukker:2011zy}.
In particular, the gravity dual of these M2-instanton effects was revisited recently in \cite{Gautason:2023igo,Beccaria:2023ujc} where the leading worldsheet instanton of the ABJM theory (and its mass deformation) was reproduced including the prefactor of $e^{-2\pi\sqrt{\frac{2N}{k}}}$ in the large $k$ limit.
It would be interesting if we can study the worldsheet instanton for the ${\hat D}$-type quiver theory in a similar way.
It would also be interesting to work out the exact calculation of the partition function with the FI parameters and the rank differences turned on to identify how the parameter $B$ in $Z_{\text{pert}}(N)$ and the instanton effects depends on these deformations.

The $\mathfrak{q}$-PVI relations satisfied by the grand partition function of $(2,2)$ model are complicated equations involving various shifts of five FI/rank deformation parameters.
On the other hand, it was found in \cite{Bonelli:2022dse} that under a certain limit of the five parameters the partition function of $(2,2)$ model reduces to that of the ABJM theory, where the complicated $\mathfrak{q}$-PVI equations reduce to a simpler, but still non-trivial, bilinear equations of $\mathfrak{q}$-P$\text{III}_3$.
From the viewpoint of the five dimensional Yang-Mills theories associated with the inverse density matrices, this limit corresponds to the mass decoupling limit from $N_\text{f}=4$ to $N_\text{f}=0$.
Also in for the theories with ${\hat D}$-type quiver there may exist a more fundamental matrix model from the viewpoint of $\mathfrak{q}$-IE/M2-MM correspondence which would be obtained from the partition function of ${\hat D}_4$ quiver by taking some limit.
Indeed, since the density matrix for the ${\hat D}_4$-quiver is written in terms of the density matrix of $(2,2)$ model as \eqref{eq:DM-Closed}, we may take the same limit which replace the density matrix of $(2,2)$ model in \eqref{eq:DM-Closed} with that of the ABJM theory.
Although it is not clear whether the matrix model thus obtained has a three dimensional interpretation or not, it can be a useful toy model to understand the relation between ${\hat D}$-type quiver matrix models and discrete integrable systems.

Although in this paper we consider the rank deformation on the non-affine node and the two affine nodes with non-vanishing Chern-Simons levels, it is still possible to turn on another rank deformation on one of the affine nodes with vanishing Chern-Simons level.
To consider this generalization would be important for the following reason.
In \cite{Bonelli:2017gdk,Moriyama:2023mjx,Moriyama:2023pxd} it was observed that the coefficients of the $\mathfrak{q}$-Painlev\'e equations for the grand partition function simplify if we write the equations in terms of the unnormalized grand partition function $Z(0)\text{Det}(1+\kappa{\hat\rho})$ rather than the normalized one $\text{Det}(1+\kappa{\hat\rho})$.
In that viewpoint, one may regard the $\mathfrak{q}$-Painlev\'e equations as the uplift of the same relations satisfied by $Z(0)$ to $N>0$.
This suggests that also in guessing the $\mathfrak{q}$-difference relations for the ${\hat D}$-type quiver it would be useful to study the partition function at $N=0$.
However, if we set $N=0$ in our current deformation, the partition function reduces to that of a four-node linear quiver, and hence it would fail to capture the relation which the partition function with $N>0$ should satisfy.
On the other hand, if we turn on the additional rank deformation, the partition function at $N=0$ is distinctive from that of a linear quiver, hence we may be able to find a non-trivial relation which uplifts to $N>0$ as well.
In \cite{Moriyama:2023pxd} it was also observed in the ABJM theory and $(2,2)$ model that the infinite discrete symmetry of the partition function generated by the duality cascade \cite{Klebanov:2000hb,Elitzur:1997fh,Aharony:2009fc,Evslin:2009pk,Honda:2020uou} coincide with the exceptional affine Weyl group characterizing the $\mathfrak{q}$-Painlev\'e equation \cite{2001CMaPh.220..165S,Kajiwara:2015aaa} corresponding to these theories under $\mathfrak{q}$-IE/M2-MM correspondence.
It would be interesting to investigate the same infinite discrete symmetry for ${\hat D}_4$ quiver theory deformed by the four relative ranks and four FI parameters.

It is also interesting in the following sense to apply the Fermi gas formalism when the rank deformation of one of the affine nodes with vanishing Chern-Simons level is turned on.
In section \ref{symmetries} we studied the symmetries of the density matrix.
However, since all the rank deformations are not turned on, symmetries we can find are restricted in principle.
Because it is known that for several circular quiver theories, symmetries of the density matrix with all the rank deformations include non-trivial symmetries which cannot be explained by the known dualities \cite{Kubo:2019ejc,Kubo:2021enh,Bonelli:2022dse}, finding all the symmetries of the  ${\hat D}_4$ density matrix with all the rank deformations would be also important.

Lastly the $N^{3/2}$ scaling of the free energy, which is characteristic of the theories of M2-branes, is also realized in supersymmetric Chern-Simons theories with the quivers of various other shapes
\cite{Gulotta:2011vp,Amariti:2019pky}.
For these theories the Fermi gas formalism is still not known even in the case without any FI/rank deformations.
It would be interesting if we can extend our analysis to these theories.

\section*{Acknowledgement}
The authors thank Giulio Bonelli, Fran Globlek and Alessandro Tanzini for the collaboration during the early stages of the project.
We are also grateful to Sanefumi Moriyama for valuable discussion.
Preliminary results of this paper were presented by NK in a international school ``The 18th Kavli Asian Winter School on Strings, Particles and Cosmology'' at Yukawa Institute for Theoretical Physics, Kyoto University.
Part of the results in appendix \ref{app_exactvalues} was computed by using the high performance computing facility provided by Yukawa Institute for Theoretical Physics (Sushiki server).
The research activity of NK is partially supported by National key research and development program under grand No.~2022YFE0134300.

\appendix

\section{Notation and formulas for 1d quantum mechanics}
\label{sec_1dqmnotation}
In this appendix we provide our notation for one-dimensional quantum mechanics used in the Fermi gas formalism and formulas for quantum mechanics.

We normalize the canonical position/momentum operator ${\hat x}$,${\hat p}$ as
\begin{align}
[{\hat x},{\hat p}]=2\pi ik.\label{ComRel}
\end{align}
We use the following convention for the position/momentum eigenstates:
\begin{align}
&\ket{a}\text{: eigenstate of }{\hat x}\text{ with }{\hat x}\ket{a}=a\ket{a},\\
&\kket{a}\text{: eigenstate of }{\hat p}\text{ with }{\hat p}\kket{a}=a\kket{a},
\end{align}
normalized as
\begin{align}
\braket{x|y}=2\pi\delta(x-y),\quad
\bbrakket{p|p'}=2\pi\delta(p-p'),\quad
\brakket{x|p}=\frac{1}{\sqrt{k}}e^{\frac{ixp}{2\pi k}},\quad
\bbraket{p|x}=\frac{1}{\sqrt{k}}e^{-\frac{ixp}{2\pi k}}.
\end{align}

We also define ${\hat X}=2{\hat x}$ and ${\hat P}=2{\hat p}$ and use the following convention for the eigenstate of ${\hat X}$ with eigenvalue ${\hat X}=a$ and ${\hat P}$ with eigenvalue ${\hat P}=a$:
\begin{align}
\ket{a}_X=\frac{1}{\sqrt{2}}\ket{a},\quad
\kket{a}_{P}=\frac{1}{\sqrt{2}}\kket{\frac{a}{2}}.
\label{ketXketP}
\end{align}
These states satisfy the followings:
\begin{align}
&{}_X\!\braket{X|X'}_X=2\pi\delta(X-X'),\quad
{}_X\!\brakket{X|p}=\frac{1}{\sqrt{2k}}e^{\frac{iXp}{4\pi k}},\quad
\bbraket{p|X}_X=\frac{1}{\sqrt{2k}}e^{-\frac{iXp}{4\pi k}},\nonumber \\
&{}_P\!\bbrakket{P|P'}_P=2\pi\delta(P-P'),\quad
\brakket{x|P}_P=\frac{1}{\sqrt{2k}}e^{\frac{ixP}{4\pi k}},\quad
{}_P\!\bbraket{P|x}=\frac{1}{\sqrt{2k}}e^{-\frac{ixP}{4\pi k}}.
\label{ketXketPproperties}
\end{align}
The notation of $({\hat x},{\hat P})$ is used in section \ref{sec_closed} while the notation of $({\hat X},{\hat p})$ is used in section \ref{sec_TWPY}.

\subsection{Formulas}
First we list the formulas of similarity transformation:
\begin{align}
&e^{-\frac{i\alpha}{2\pi k}{\hat x}}f({\hat p})e^{\frac{i\alpha}{2\pi k}{\hat x}}=f({\hat p}+\alpha),\quad
e^{-\frac{i\alpha}{4\pi k}{\hat x}^2}f({\hat p})e^{\frac{i\alpha}{4\pi k}{\hat x}^2}=f({\hat p}+\alpha {\hat x}),\quad
e^{-\frac{i\alpha}{4\pi k}{\hat{p}}^{2}}f({\hat{x}})e^{\frac{i\alpha}{4\pi k}{\hat{p}}^{2}}=f({\hat{x}}-\alpha {\hat{p}}),
\label{simtrsfformulaop1} \\
&e^{\frac{i\alpha}{2\pi k}{\hat x}}\kket{p}=\kket{p+\alpha},\quad
\bbra{p}e^{\frac{i\alpha}{2\pi k}{\hat x}}=\bbra{p-\alpha},\label{simtrsfformulastate1}\\
&e^{-\frac{i}{4\pi k}{\hat p}^2}e^{-\frac{i}{4\pi k}{\hat x}^2}\kket{p}=\sqrt{-i}e^{\frac{i}{4\pi k}p^2}\ket{p},\quad
\bbra{p}e^{\frac{i}{4\pi k}{\hat x}^2}e^{\frac{i}{4\pi k}{\hat p}^2}=\sqrt{i}e^{-\frac{i}{4\pi k}p^2}\bra{p}.\label{simtrsfformulastate2}
\end{align}
We will also use the formulas of similarity transformations involving ${\hat P}$ operator and its eigenstates:
\begin{align}
&e^{-\frac{i}{4\pi k}{\hat p}^2}
e^{\frac{i}{4\pi k}{\hat x}^2}
f({\hat p})
e^{-\frac{i}{4\pi k}{\hat x}^2}
e^{\frac{i}{4\pi k}{\hat p}^2}
=e^{\frac{i}{8\pi k}{\hat x}^2}f({\hat P})e^{-\frac{i}{8\pi k}{\hat x}^2},\label{simtrsfformulaPop1}\\
&e^{-\frac{i}{4\pi k}{\hat p}^2}e^{\frac{i}{4\pi k}{\hat x}^2}\kket{p}=
e^{\frac{i}{8\pi k}{\hat x}^2}\kket{p}_P
e^{-\frac{i}{8\pi k}p^2}
,\quad
\bbra{p}e^{-\frac{i}{4\pi k}{\hat x}^2}e^{\frac{i}{4\pi k}{\hat p}^2}=e^{\frac{i}{8\pi k}p^2}{}_P\!\bbra{p}e^{-\frac{i}{8\pi k}{\hat x}^2}.\label{simtrsfformulaPstate1}
\end{align}

We also define the position-basis transpose of a state $\ket{\phi}$ and an operator ${\hat A}$ as
\begin{align}
\left(\bra{\phi}\right)^t=\int_{-\infty}^{\infty}\frac{dx}{2\pi}\left(\braket{\phi|x}\right)\ket{x},\quad
\left(\ket{\phi}\right)^t=\int_{-\infty}^{\infty}\frac{dx}{2\pi}\left(\braket{x|\phi}\right)\bra{x},\quad
{\hat A}^t=\int_{-\infty}^{\infty} \frac{dxdy}{(2\pi)^2}\ket{x}\left(\braket{y|{\hat A}|x}\right)\bra{y}.
\end{align}
Here are useful formulas to calculate the transpose of a given state/operator
\begin{align}
&\left(\kket{p}\right)^t=\bbra{-p},\quad
({\hat A}\ket{\phi})^t=(\ket{\phi})^t{\hat A}^t,\nonumber \\
&({\hat A}{\hat B})^t={\hat B}^t{\hat A}^t,\quad
f({\hat x})^t=f({\hat x}),\quad
f({\hat p})^t=f(-{\hat p}).
\label{transposerules}
\end{align}

For the integration over the position basis, from a trivial identity $1=\int_{-\infty}^\infty \frac{dx}{2\pi}\ket{x}\bra{x}={\hat U}\int_{-\infty}^\infty \frac{dx}{2\pi}\ket{x}\bra{x}{\hat U}^{-1}$ it follows
\begin{align}
\int_{-\infty}^\infty dx\braket{\phi|x}\braket{x|\psi}=\int_{-\infty}^\infty dx\braket{\phi|{\hat U}|x}\braket{x|{\hat U}^{-1}|\psi},\label{completebasisid1}
\end{align}
for any states $\ket{\phi},\ket{\psi}$ and an operator ${\hat U}$.
In the same way, the following identity also holds:
\begin{align}
\int_{-\infty}^\infty dx\braket{x|\phi}\braket{x|\psi}=\int_{-\infty}^\infty dx\braket{x|{\hat U}^t|\phi}\braket{x|{\hat U}^{-1}|\psi}.\label{completebasisid2}
\end{align}

\section{Determinant formulas}
\label{app_detformulas}

In this appendix we list the formulas which we use in sections \ref{sec_open} and \ref{sec_closed} to handle various determinant/Pfaffian arising in the Fermi gas formalism.
Note that the formulas \eqref{CauchyBinet},\eqref{trivialize},\eqref{CBPfaffian},\eqref{FredholmPfaffian2} hold for an arbitrary domain of integration $I$.

\subsection{Cauchy-Vandermonde determinant formula}
The first formula is the Cauchy-Vandermonde determinant formula
\begin{align}
&\frac{\prod_{m<m'}^N2\sinh\frac{\alpha_m-\alpha_{m'}}{2k}\prod_{n<n'}^{N+L}2\sinh\frac{\beta_n-\beta_{n'}}{2k}}{\prod_{m=1}^N\prod_{n=1}^{N+L}2\cosh\frac{\alpha_m-\beta_n}{2k}}
=k^{N+\frac{L}{2}}\det\begin{pmatrix}
\left[\braket{\alpha_m|\frac{1}{2\cosh\frac{{\hat p}-\pi iL}{2}}|\beta_n}\right]_{m,n}^{N\times (N+L)}\\
\left[\bbraket{t_{L,r}|\beta_n}\right]_{r,n}^{L\times (N+L)}
\end{pmatrix}\label{CauchyVdm1}\\
&
=k^{N+\frac{L}{2}}\det
\left(
\begin{array}{cc}
\left[\braket{\beta_m|\frac{1}{2\cosh\frac{{\hat p}+\pi iL}{2}}|\alpha_n}\right]_{m,n}^{(N+L)\times N}&
\left[\brakket{\beta_m|-t_{L,s}}\right]_{m,s}^{(N+L)\times L}
\end{array}
\right),\label{CauchyVdm2}\\
&\frac{\prod_{m<m'}^N2\sinh\frac{\alpha_m-\alpha_{m'}}{2k}\prod_{n<n'}^{N+L}2\sinh\frac{\beta_n-\beta_{n'}}{2k}}{\prod_{m=1}^N\prod_{n=1}^{N+L}2\sinh\frac{\alpha_m-\beta_n}{2k}}=i^{-N^2}k^{N+\frac{L}{2}}\det\begin{pmatrix}
\left[\braket{\alpha_m|\frac{\tanh\frac{{\hat p}-\pi iL}{2}}{2}|\beta_n}\right]_{m,n}^{N\times (N+L)}\\
\left[\bbraket{t_{L,r}|\beta_n}\right]_{r,n}^{L\times (N+L)}
\end{pmatrix}\label{CauchyVdm3}\\
&=
i^{N^2}
k^{N+\frac{L}{2}}\det\left(\begin{array}{cc}
\left[\braket{\beta_m|\frac{\tanh\frac{{\hat p}+\pi iL}{2}}{2}|\alpha_n}\right]_{m,n}^{(N+L)\times N}&
\left[\brakket{\beta_m|-t_{L,s}}\right]_{m,s}^{(N+L)\times L}
\end{array}\right),\label{CauchyVdm4}
\end{align}
where on the right-hand sides we have used the notation for 1d quantum mechanics introduced in appendix \ref{sec_1dqmnotation}, with
\begin{align}
t_{L,r}=2\pi i\left(\frac{L+1}{2}-r\right),\quad (r=1,2,\cdots,L).
\label{t_nr}
\end{align}
Note that the above four formula \eqref{CauchyVdm1},\eqref{CauchyVdm2},\eqref{CauchyVdm3},\eqref{CauchyVdm4} follows from the following Cauchy determinant formula
\begin{align}
\frac{
\prod_{m<m'}^N(\alpha_m-\alpha_{m'})
\prod_{n<n'}^N(\beta_n-\beta_{n'})
}{\prod_{m,n=1}^N(\alpha_m+\beta_n)}=\det\left(\left[\frac{1}{\alpha_m+\beta_n}\right]_{m,n}\right),
\end{align}
with appropriate redefinitions and limits of the arguments (see for example \cite{Matsumoto:2013nya}).

\subsection{Cauchy-Binet formula and determinant trivialization formula}
The second formula is the Cauchy-Binet formula
\begin{align}
&\frac{1}{N!}
\int_I
d^N\alpha
\det\left(
\begin{array}{cc}
\left[f_m(\alpha_n)\right]_{m,n}^{(N+L_1)\times N}
&\left[v_{ms}\right]_{m,s}^{(N+L_1)\times L_1}
\end{array}
\right)
\det\left(
\begin{array}{cc}
\left[g_m(\alpha_n)\right]_{m,n}^{(N+L_2)\times N}
&\left[w_{ms}\right]_{m,s}^{(N+L_2)\times L_2}
\end{array}
\right)\nonumber \\
&=(-1)^{L_1L_2}\det\begin{pmatrix}
\left[\int_I d\alpha f_m(\alpha)g_n(\alpha)\right]_{m,n}^{(N+L_1)\times (N+L_2)}
&\left[v_{ms}\right]_{m,s}^{(N+L_1)\times L_1}\\
\left[w_{nr}\right]_{r,n}^{L_2\times (N+L_2)}
&\left[0\right]^{L_1\times L_2}
\end{pmatrix}.
\label{CauchyBinet}
\end{align}
In particular, for $L_1,L_2=0$, this formula can be easily obtained from the following determinant trivialization formula
\begin{align}
\frac{1}{N!}
\int_I
d^N\alpha\det\left(\left[f_n(\alpha_m)\right]_{m,n}\right)\det\left(\left[g_n(\alpha_m)\right]_{m,n}\right)=\int_I d^N\alpha\prod_{m=1}^Nf_m(\alpha_m)\det\left(\left[g_n(\alpha_m)\right]_{m,n}\right).
\label{trivialize}
\end{align}

\subsection{Cauchy-Binet-like formula for Pfaffian}
The third formula is a Cauchy-Binet-like formula for Pfaffian
\begin{align}
&\frac{1}{N!}
\int_I
d^N\alpha
\det\left(
\begin{array}{ccc}
\left[f_m(\alpha_n)\right]_{m,n}^{(2N+L)\times N}
&\left[g_{m}(\alpha_n)\right]_{m,n}^{(2N+L)\times N}
&\left[v_{ms}\right]_{m,s}^{(2N+L)\times L}
\end{array}
\right)\nonumber \\
&=(-1)^{\frac{N(N-1)}{2}+\frac{L(L-1)}{2}}\text{pf}\begin{pmatrix}
\left[\int_I d\alpha(f_m(\alpha)g_n(\alpha)-g_m(\alpha)f_n(\alpha))\right]_{m,n}^{(2N+L)\times (2N+L)}
&\left[v_{ms}\right]_{m,s}^{(2N+L)\times L}\\
\left[-v_{nr}\right]_{r,n}^{L\times (2N+L)}
&\left[0\right]^{L\times L}
\end{pmatrix},
\label{CBPfaffian}
\end{align}
where $\text{pf}(A)$ is Pfaffian which is defined for an $2n\times 2n$ anti-symmetric matrix $A$ as
\begin{align}
\text{pf}(A)=(-1)^{\frac{n(n-1)}{2}}\frac{1}{2^nn!}\sum_{\sigma\in S_{2n}}(-1)^\sigma \prod_{i=1}^nA_{\sigma(i),\sigma(n+i)}.
\end{align}
We have checked that \eqref{CBPfaffian} is satisfied for a generic set of $\{f_m(\alpha),g_m(\alpha),v_{ms}\}$ for $N\le 3$ and $L\le 4$, although we do not have an explicit proof for general $N$ and $L$.

\subsection{Fredholm Pfaffian formula}
The fourth formula is the Fredholm Pfaffian formula. For $N_\infty\ge 0$ and $L\ge 0$, let $A,B,C,D$ to be any $N_\infty\times N_\infty$ matrices, $v,w$ to be any $N_\infty\times 2L$ matrices and $\alpha$ to be any $2L\times 2L$ matrix, satisfying the following properties
\begin{align}
A_{ji}=-A_{ij},\quad
C_{ji}=-B_{ij},\quad
D_{ji}=-D_{ij},\quad
\alpha_{sr}=-\alpha_{rs}.
\end{align}
Then the following identity holds:
\begin{align}
(-1)^{\frac{L(L-1)}{2}}
\sqrt{\text{det}\left[
\begin{pmatrix}
\bar{\Omega}&0\\
0&0
\end{pmatrix}
+\begin{pmatrix}
\kappa\bar{P}&\kappa V\\
-V^t&\alpha
\end{pmatrix}
\right]}
=
\sum_{k=0}^{N_\infty}
(-1)^{\frac{k(k-1)}{2}}\kappa^k
\sum_{\substack{S\subset \{1,2,\cdots,N_\infty\}\\
(|S|=k)}
}\text{pf}\begin{pmatrix}
\bar{P}_S&V_S\\
-V_S^t&\alpha
\end{pmatrix}.
\label{FredholmPfaffian}
\end{align}
Here the symbols on the left-hand side are
\begin{align}
\bar{\Omega}=\begin{pmatrix}
[0]^{N_\infty\times N_\infty}&[\delta_{ij}]_{i,j}^{N_\infty\times N_\infty}\\
[-\delta_{ij}]_{i,j}^{N_\infty\times N_\infty}&[0]^{N_\infty\times N_\infty}
\end{pmatrix},\quad
\bar{P}=\begin{pmatrix}
[A_{ij}]_{i,j}^{N_\infty\times N_\infty}&[B_{ij}]_{i,j}^{N_\infty\times N_\infty}\\
[C_{ij}]_{i,j}^{N_\infty\times N_\infty}&[D_{ij}]_{i,j}^{N_\infty\times N_\infty}
\end{pmatrix},\nonumber \\
V=
\begin{pmatrix}
[v_{is}]_{i,s}^{N_\infty\times 2L}\\
[w_{is}]_{i,s}^{N_\infty\times 2L}
\end{pmatrix},\quad
V^t=
\left(\begin{array}{cc}
[v_{jr}]_{r,j}^{2L\times N_\infty}&
[w_{jr}]_{r,j}^{2L\times N_\infty}
\end{array}\right),\quad
\alpha=[\alpha_{rs}]_{r,s}^{2L\times 2L},
\end{align}
while the symbols in the summand under $\sum_{S\subset \{1,2,\cdots,N_\infty\}\,(|S|=k)}$ are
\begin{align}
\bar{P}_S&=\begin{pmatrix}
[A_{n_a,n_b}]_{a,b}^{k\times k}&[B_{n_a,n_b}]_{a,b}^{k\times k}\\
[C_{n_a,n_b}]_{a,b}^{k\times k}&[D_{n_a,n_b}]_{a,b}^{k\times k}\\
\end{pmatrix},\nonumber \\
V_S&=
\begin{pmatrix}
[v_{n_a,s}]_{a,s}^{k\times 2L}\\
[w_{n_a,s}]_{a,s}^{k\times 2L}
\end{pmatrix},\quad
V^t_S=
\begin{pmatrix}
[v_{n_b,r}]_{r,b}^{2L\times k}&
[w_{n_b,r}]_{r,b}^{2L\times k}
\end{pmatrix},
\end{align}
for $S=\{n_1,n_2,\cdots,n_k\}$ $(1\le n_1<n_2<\cdots<n_k\le N_\infty)$.
For $L=0$, the identity \eqref{FredholmPfaffian} is proved in \cite{Matsumoto:2005aaa}.
For $L>0$, we have checked for $N_\infty\le 10$ and $L\le 5$ that \eqref{FredholmPfaffian} is satisfied although we do not have an explicit proof.
By taking the continuum limit $N_\infty\rightarrow\infty$ where
\begin{align}
&A_{ij}\text{ with }1\le i,j\le N_\infty\rightarrow A(x,y)\text{ with }x,y\in I,\nonumber \\
&v_{is}\text{ with }1\le i\le N_\infty\rightarrow v_s(x)\text{ with }x\in I,
\end{align}
and so on, with $I$ an arbitrary domain, the identity \eqref{FredholmPfaffian} reduces to the following Fredholm Pfaffian formula
\begin{align}
&(-1)^{\frac{L(L-1)}{2}}
\sqrt{\text{Det}\left[
\begin{pmatrix}
\kappa A&1+\kappa B&\left[\kappa v_s\right]_s^{2L}\\
-1+\kappa C&\kappa D&\left[\kappa w_s\right]_s^{2L}\\
\left[-v_r\right]_r^{2L}&\left[-w_r\right]_r^{2L}&\left[\alpha_{rs}\right]_{r,s}^{2L\times 2L}
\end{pmatrix}
\right]}\nonumber \\
&=
\sum_{N=0}^{\infty}
(-1)^{\frac{N(N-1)}{2}}
\kappa^N
\frac{
1
}{N!}
\int_I d^Nx
\text{pf}\begin{pmatrix}
\left[A(x_i,x_j)\right]_{i,j}^{N\times N}&\left[B(x_i,x_j)\right]_{i,j}^{N\times N}&\left[v_s(x_i)\right]_{i,s}^{N\times 2L}\\
\left[C(x_i,x_j)\right]_{i,j}^{N\times N}&\left[D(x_i,x_j)\right]_{i,j}^{N\times N}&\left[w_s(x_i)\right]_{i,s}^{N\times 2L}\\
\left[-v_r(x_j)\right]_{r,j}^{2L\times N}&\left[-w_r(x_j)\right]_{r,j}^{2L\times N}&\left[\alpha_{rs}\right]_{r,s}^{2L\times 2L}
\end{pmatrix},
\label{FredholmPfaffian2}
\end{align}
which we have used in the main text.

\section{$D_{4}^{\left(1\right)}$ quantum curve conjecture\label{sec:YkQC}}

In this appendix we study the factor ${\cal Y}_{k}$ defined in \eqref{eq:Ik}. We conjecture that this is equal to 
\begin{align}
 & {\cal Y}_{k}(\alpha_{m},\beta_{n};N;L_{1},L_{2},L;\zeta_{a})\nonumber \\
 & =k^{N}{\cal Y}_{k}^{\left(0\right)}(L_{1},L_{2},L;\zeta_{a})\det\left(\left[\braket{\alpha_{m}|\frac{1}{\hat{H}_{k}\left(\hat{x},\hat{p};L_{1},L_{2},L;\zeta_{a}\right)}|\beta_{n}}\right]_{m,n}^{N\times N}\right),\label{eq:YkConj}
\end{align}
where ${\cal Y}_{k}^{\left(0\right)}$ is $N=0$ part of ${\cal Y}_{k}$,
\begin{equation}
{\cal Y}_{k}^{\left(0\right)}(L_{1},L_{2},L;\zeta_{a})={\cal Y}_{k}(\cdot,\cdot;0;L_{1},L_{2},L;\zeta_{a}).\label{eq:Yk0Def}
\end{equation}
$\hat{H}_{k}$ is called the quantum curve because this is a Laurent series of the exponential of the position and the momentum operators. The explicit form of the quantum curve in our case is
\begin{align}
\hat{H}_{k}\left(\hat{x},\hat{p};L_{1},L_{2},L;\zeta_{a}\right) 
 &=e^{\frac{\pi i\left(-L_{1}+L_{2}\right)}{2}+\frac{1}{2}\left(\zeta_{1}+\zeta_{2}+\zeta_{3}+\zeta_{4}\right)}e^{-\hat{x}+\hat{p}}\nonumber \\
 & \quad+\left[e^{\frac{\pi i\left(-L_{1}-L_{2}\right)}{2}+\frac{1}{2}\left(\zeta_{1}+\zeta_{2}+\zeta_{3}-\zeta_{4}\right)+i\pi k}+e^{\frac{\pi i\left(L_{1}+L_{2}\right)}{2}+\frac{1}{2}\left(-\zeta_{1}+\zeta_{2}+\zeta_{3}+\zeta_{4}\right)-i\pi k}\right]e^{\hat{p}}\nonumber \\
 & \quad+e^{\frac{\pi i\left(L_{1}-L_{2}\right)}{2}+\frac{1}{2}\left(-\zeta_{1}+\zeta_{2}+\zeta_{3}-\zeta_{4}\right)}e^{\hat{x}+\hat{p}}\nonumber \\
 & \quad+\left[e^{\frac{\pi i\left(-L_{1}-L_{2}+2L\right)}{2}+\frac{1}{2}\left(\zeta_{1}+\zeta_{2}-\zeta_{3}+\zeta_{4}\right)}+e^{\frac{\pi i\left(L_{1}+L_{2}-2L\right)}{2}+\frac{1}{2}\left(\zeta_{1}-\zeta_{2}+\zeta_{3}+\zeta_{4}\right)}\right]e^{-\hat{x}}+E\nonumber \\
 & \quad+\left[e^{\frac{\pi i\left(-L_{1}-L_{2}+2L\right)}{2}+\frac{1}{2}\left(-\zeta_{1}-\zeta_{2}+\zeta_{3}-\zeta_{4}\right)}+e^{\frac{\pi i\left(L_{1}+L_{2}-2L\right)}{2}+\frac{1}{2}\left(-\zeta_{1}+\zeta_{2}-\zeta_{3}-\zeta_{4}\right)}\right]e^{\hat{x}}\nonumber \\
 & \quad+e^{\frac{\pi i\left(L_{1}-L_{2}\right)}{2}+\frac{1}{2}\left(\zeta_{1}-\zeta_{2}-\zeta_{3}+\zeta_{4}\right)}e^{-\hat{x}-\hat{p}}\nonumber \\
 & \quad+\left[e^{\frac{\pi i\left(-L_{1}-L_{2}\right)}{2}+\frac{1}{2}\left(-\zeta_{1}-\zeta_{2}-\zeta_{3}+\zeta_{4}\right)+i\pi k}+e^{\frac{\pi i\left(L_{1}+L_{2}\right)}{2}+\frac{1}{2}\left(\zeta_{1}-\zeta_{2}-\zeta_{3}-\zeta_{4}\right)-i\pi k}\right]e^{-\hat{p}}\nonumber \\
 & \quad+e^{\frac{\pi i\left(-L_{1}+L_{2}\right)}{2}+\frac{1}{2}\left(-\zeta_{1}-\zeta_{2}-\zeta_{3}-\zeta_{4}\right)}e^{\hat{x}-\hat{p}},\label{eq:QCConj}
\end{align}
where $E$ is a constant.
In this expression, $\hat{x}$ and $\hat{p}$ must satisfy the commutation relation $\left[ \hat{x},\hat{p} \right] =2\pi ik$ (as \eqref{ComRel}).

In the following sections we show two evidences for this conjecture.

\subsection{$L=0$ case\label{subsec:YkL0}}

The important point is that when $L=0$, the conjectured relation \eqref{eq:YkConj} and the conjectured quantum curve \eqref{eq:QCConj} can be analytically obtained. In this section we perform this computation.

First, by using the Cauchy-Vandermonde determinant formula \eqref{CauchyVdm1} and \eqref{CauchyVdm2} the factors at the center of the third and fourth line of \eqref{eq:Ik} can be written by the determinants. The remaining factors at the third and fourth line of ${\cal Y}_{k}$ can be put into the determinants. The phase factor and the FI factors can also be put into the determinants as 
\begin{align}
 & {\cal Y}_{k}(\alpha_{m},\beta_{n};N;L_{1},L_{2},0;\zeta_{a})\nonumber \\
 & =\frac{k^{N}}{N!}\int_{-\infty}^{\infty}\frac{d^{N}\gamma}{(2\pi)^{N}}\nonumber \\
 & \quad\times\det\left(\left[\braket{\alpha_{m}|i^{-L_{1}}\frac{\prod_{r=1}^{L_{1}}2\sinh\frac{\hat{x}-\zeta_{1}-t_{L_{1},r}}{2k}}{2\cosh\frac{\hat{x}-\zeta_{1}-\pi iL_{1}}{2}}\frac{1}{2\cosh\frac{\hat{p}+\zeta_{2}}{2}}\frac{1}{\prod_{r=1}^{L_{1}}2\cosh\frac{\hat{x}-\zeta_{1}-t_{L_{1},r}}{2k}}|\gamma_{n}}\right]_{m,n}^{N\times N}\right)\nonumber \\
 & \quad\times\det\left(\left[\braket{\gamma_{m}|i^{L_{2}}\frac{1}{\prod_{r=1}^{L_{2}}2\cosh\frac{\hat{x}-\zeta_{4}-t_{L_{2},r}}{2k}}\frac{1}{2\cosh\frac{\hat{p}+\zeta_{3}}{2}}\frac{\prod_{r=1}^{L_{2}}2\sinh\frac{\hat{x}-\zeta_{4}-t_{L_{2},r}}{2k}}{2\cosh\frac{\hat{x}-\zeta_{4}+\pi iL_{2}}{2}}|\beta_{n}}\right]_{m,n}^{N\times N}\right).\label{eq:YkL01}
\end{align}
It is known that the inverse of the operators inside the bras and kets can be written by the quantum curves as \cite{Kashaev:2015wia,Kubo:2019ejc,Bonelli:2022dse} 
\begin{align}
 & i^{L}\frac{1}{\prod_{j}^{L}2\cosh\frac{\hat{x}-t_{L,j}}{2k}}\frac{1}{2\cosh\frac{\hat{p}}{2}}\frac{\prod_{j}^{L}2\sinh\frac{\hat{x}-t_{L,j}}{2k}}{2\cosh\frac{\hat{x}+\pi iL}{2}}\nonumber \\
 & =\left[\left(e^{-\frac{1}{2}i\pi L}e^{\frac{1}{2}\hat{x}}+e^{\frac{1}{2}i\pi L}e^{-\frac{1}{2}\hat{x}}\right)e^{\frac{1}{2}\hat{p}}+\left(e^{\frac{1}{2}i\pi L}e^{\frac{1}{2}\hat{x}}+e^{-\frac{1}{2}i\pi L}e^{-\frac{1}{2}\hat{x}}\right)e^{-\frac{1}{2}\hat{p}}\right]^{-1}.
 \end{align}
This formula and the formula which is the Hermitian conjugate to this formula reads 
\begin{align}
 & i^{-L_{1}}\frac{\prod_{r=1}^{L_{1}}2\sinh\frac{\hat{x}-\zeta_{1}-t_{L_{1},r}}{2k}}{2\cosh\frac{\hat{x}-\zeta_{1}-\pi iL_{1}}{2}}\frac{1}{2\cosh\frac{\hat{p}+\zeta_{2}}{2}}\frac{1}{\prod_{r=1}^{L_{1}}2\cosh\frac{\hat{x}-\zeta_{1}-t_{L_{1},r}}{2k}}\nonumber \\
 & =\left.\left[e^{\frac{1}{2}\hat{p}}\left(e^{\frac{1}{2}i\pi L_{1}}e^{\frac{1}{2}\hat{x}}+e^{-\frac{1}{2}i\pi L_{1}}e^{-\frac{1}{2}\hat{x}}\right)+e^{-\frac{1}{2}\hat{p}}\left(e^{-\frac{1}{2}i\pi L_{1}}e^{\frac{1}{2}\hat{x}}+e^{\frac{1}{2}i\pi L_{1}}e^{-\frac{1}{2}\hat{x}}\right)\right]^{-1}\right|_{\hat{x}\rightarrow\hat{x}-\zeta_{1},\quad\hat{p}\rightarrow\hat{p}+\zeta_{2}}\nonumber \\
 & i^{L_{2}}\frac{1}{\prod_{r=1}^{L_{2}}2\cosh\frac{\hat{x}-\zeta_{4}-t_{L_{2},r}}{2k}}\frac{1}{2\cosh\frac{\hat{p}+\zeta_{3}}{2}}\frac{\prod_{r=1}^{L_{2}}2\sinh\frac{\hat{x}-\zeta_{4}-t_{L_{2},r}}{2k}}{2\cosh\frac{\hat{x}-\zeta_{4}+\pi iL_{2}}{2}}\nonumber \\
 & =\left.\left[\left(e^{-\frac{1}{2}i\pi L_{2}}e^{\frac{1}{2}\hat{x}}+e^{\frac{1}{2}i\pi L_{2}}e^{-\frac{1}{2}\hat{x}}\right)e^{\frac{1}{2}\hat{p}}+\left(e^{\frac{1}{2}i\pi L_{2}}e^{\frac{1}{2}\hat{x}}+e^{-\frac{1}{2}i\pi L_{2}}e^{-\frac{1}{2}\hat{x}}\right)e^{-\frac{1}{2}\hat{p}}\right]^{-1}\right|_{\hat{x}\rightarrow\hat{x}-\zeta_{4},\quad\hat{p}\rightarrow\hat{p}+\zeta_{3}}.\label{eq:QCInv}
\end{align}
The two determinants in \eqref{eq:YkL01} can be glued by using the Cauchy-Binet formula \eqref{CauchyBinet}. At this step the integrations over $\gamma_{n}$ become trivial with the kets and bras $\ket{\gamma_{n}}\bra{\gamma_{n}}$, and the inverse of the quantum curves inside the bras and kets are multiplied. Note that when inverses of two operators are multiplied, the order is reversed. Therefore, we finally arrive at 
\begin{align}
{\cal Y}_{k}(\alpha_{m},\beta_{n};N;L_{1},L_{2},0;\zeta_{a}) & =k^{N}\det\left(\left[\braket{\alpha_{m}|\frac{1}{\hat{H}_{k}^{\left(L=0\right)}\left(\hat{x},\hat{p}\right)}|\beta_{n}}\right]_{m,n}^{N\times N}\right),
\end{align}
where $\hat{H}_{k}^{\left(L=0\right)}$ is the product of two quantum curves in \eqref{eq:QCInv} with reversed order. Explicitly, 
\begin{align}
\hat{H}_{k}^{\left(L=0\right)}\left(\hat{x},\hat{p}\right) & =e^{\frac{\pi i\left(-L_{1}+L_{2}\right)}{2}+\frac{1}{2}\left(\zeta_{1}+\zeta_{2}+\zeta_{3}+\zeta_{4}\right)}e^{-\hat{x}+\hat{p}}\nonumber \\
 & \quad+\left[e^{\frac{\pi i\left(-L_{1}-L_{2}\right)}{2}+\frac{1}{2}\left(\zeta_{1}+\zeta_{2}+\zeta_{3}-\zeta_{4}\right)+i\pi k}+e^{\frac{\pi i\left(L_{1}+L_{2}\right)}{2}+\frac{1}{2}\left(-\zeta_{1}+\zeta_{2}+\zeta_{3}+\zeta_{4}\right)-i\pi k}\right]e^{\hat{p}}\nonumber \\
 & \quad+e^{\frac{\pi i\left(L_{1}-L_{2}\right)}{2}+\frac{1}{2}\left(-\zeta_{1}+\zeta_{2}+\zeta_{3}-\zeta_{4}\right)}e^{\hat{x}+\hat{p}}\nonumber \\
 & \quad+\left[e^{\frac{\pi i\left(-L_{1}-L_{2}\right)}{2}+\frac{1}{2}\left(\zeta_{1}+\zeta_{2}-\zeta_{3}+\zeta_{4}\right)}+e^{\frac{\pi i\left(L_{1}+L_{2}\right)}{2}+\frac{1}{2}\left(\zeta_{1}-\zeta_{2}+\zeta_{3}+\zeta_{4}\right)}\right]e^{-\hat{x}}+E^{\left(L=0\right)}\nonumber \\
 & \quad+\left[e^{\frac{\pi i\left(-L_{1}-L_{2}\right)}{2}+\frac{1}{2}\left(-\zeta_{1}-\zeta_{2}+\zeta_{3}-\zeta_{4}\right)}+e^{\frac{\pi i\left(L_{1}+L_{2}\right)}{2}+\frac{1}{2}\left(-\zeta_{1}+\zeta_{2}-\zeta_{3}-\zeta_{4}\right)}\right]e^{\hat{x}}\nonumber \\
 & \quad+e^{\frac{\pi i\left(L_{1}-L_{2}\right)}{2}+\frac{1}{2}\left(\zeta_{1}-\zeta_{2}-\zeta_{3}+\zeta_{4}\right)}e^{-\hat{x}-\hat{p}}\nonumber \\
 & \quad+\left[e^{\frac{\pi i\left(-L_{1}-L_{2}\right)}{2}+\frac{1}{2}\left(-\zeta_{1}-\zeta_{2}-\zeta_{3}+\zeta_{4}\right)+i\pi k}+e^{\frac{\pi i\left(L_{1}+L_{2}\right)}{2}+\frac{1}{2}\left(\zeta_{1}-\zeta_{2}-\zeta_{3}-\zeta_{4}\right)-i\pi k}\right]e^{-\hat{p}}\nonumber \\
 & \quad+e^{\frac{\pi i\left(-L_{1}+L_{2}\right)}{2}+\frac{1}{2}\left(-\zeta_{1}-\zeta_{2}-\zeta_{3}-\zeta_{4}\right)}e^{\hat{x}-\hat{p}}.\label{eq:QCL0}
\end{align}
Note that in this case the constant term is 
\begin{align}
E^{\left(L=0\right)}= & e^{\frac{\pi i\left(L_{1}-L_{2}\right)}{2}+\frac{1}{2}\left(\zeta_{1}-\zeta_{2}+\zeta_{3}-\zeta_{4}\right)}+e^{\frac{\pi i\left(-L_{1}+L_{2}\right)}{2}+\frac{1}{2}\left(-\zeta_{1}-\zeta_{2}+\zeta_{3}+\zeta_{4}\right)}\nonumber \\
 & +e^{\frac{\pi i\left(-L_{1}+L_{2}\right)}{2}+\frac{1}{2}\left(\zeta_{1}+\zeta_{2}-\zeta_{3}-\zeta_{4}\right)}+e^{\frac{\pi i\left(L_{1}-L_{2}\right)}{2}+\frac{1}{2}\left(-\zeta_{1}+\zeta_{2}-\zeta_{3}+\zeta_{4}\right)}.\label{eq:E0}
\end{align}
This expression is consistent with \eqref{eq:QCConj}.

\subsection{$\zeta_{1}=\zeta_{3}=0$ case\label{subsec:22model}}

In the last section we conjectured that the factor ${\cal Y}_{k}$ defined in \eqref{eq:Ik} can be written in the quantum curve when $L=0$. In this section, we anticipate how this parameter changes the quantum curve.

Interestingly, a factor similar to ${\cal Y}_{k}$ appears in the partition function of so called (2,2) model.
The (2,2) model is a 3d ${\cal N}=4$ superconformal Chern-Simons gauge theory where the gauge group is affine $A$-type quiver with four nodes.
The quiver diagram of the relevant (2,2) model with specific ranks of the gauge group, the Chern-Simons levels and the FI parameters are shown in figure \ref{fig:Quiver-22}.
The partition function of this theory can be deformed as\footnote{We ignored an overall factor which is independent of $N$ because this factor does not play any role for the quantum curve. The phase factor of the partition function is determined so that this expression is obtained.} \cite{Bonelli:2022dse}\footnote{In \cite{Bonelli:2022dse}, by substituting (B.14) to (B.10), we obtain this expression.} 
\begin{align}
 & Z_{k}^{(2,2)}(N;L_{1},L_{2},L;\zeta_{1},\zeta_{2},\zeta_{3},\zeta_{4})\nonumber \\
 & =\frac{i^{-(L_{1}-L_{2})N}}{N!(N+L)!}\int_{-\infty}^{\infty}\frac{d^{N}\alpha}{(2\pi k)^{N}}\int_{-\infty}^{\infty}\frac{d^{N+L}\gamma}{(2\pi k)^{N+L}}e^{\frac{i}{2\pi k}((\zeta_{3}-\zeta_{2})\sum_{m=1}^{N}\alpha_{m}+(\zeta_{2}-\zeta_{3})\sum_{m=1}^{N+L}\gamma_{m})}\nonumber \\
 & \times\left(\prod_{m=1}^{N}\frac{\prod_{r=1}^{L_{1}}2\sinh\frac{\alpha_{m}-\zeta_{1}-t_{L_{1},r}}{2k}}{2\cosh\frac{\alpha_{m}-\zeta_{1}-\pi iL_{1}}{2}}\right)\frac{\prod_{m<m'}^{N}2\sinh\frac{\alpha_{m}-\alpha_{m'}}{2k}\prod_{m<m'}^{N+L}2\sinh\frac{\gamma_{m}-\gamma_{m'}}{2k}}{\prod_{m=1}^{N}\prod_{n=1}^{N+L}2\cosh\frac{\alpha_{m}-\gamma_{n}}{2k}}\left(\prod_{m=1}^{N+L}\frac{1}{\prod_{r=1}^{L_{1}}2\cosh\frac{\gamma_{m}-\zeta_{1}-t_{L_{1},r}}{2k}}\right)\nonumber \\
 & \times\left(\prod_{m=1}^{N+L}\frac{1}{\prod_{r=1}^{L_{2}}2\cosh\frac{\gamma_{m}-\zeta_{4}-t_{L_{2},r}}{2k}}\right)\frac{\prod_{m<m'}^{N+L}2\sinh\frac{\gamma_{m}-\gamma_{m'}}{2k}\prod_{m<m'}^{N}2\sinh\frac{\alpha_{m}-\alpha_{m'}}{2k}}{\prod_{m=1}^{N+L}\prod_{n=1}^{N}2\cosh\frac{\gamma_{m}-\alpha_{n}}{2k}}\left(\prod_{m=1}^{N}\frac{\prod_{r=1}^{L_{2}}2\sinh\frac{\alpha_{m}-\zeta_{4}-t_{L_{2},r}}{2k}}{2\cosh\frac{\alpha_{m}-\zeta_{4}+\pi iL_{2}}{2}}\right),
\end{align}
by partially applying the Fermi gas formalism. When $\zeta_{1}=\zeta_{3}=0$, it was conjectured in \cite{Kubo:2019ejc,Bonelli:2022dse} that this partition function can be written as 
\begin{align}
 & Z_{k}^{(2,2)}(N;L_{1},L_{2},L;0,\zeta_{2},0,\zeta_{4})\nonumber \\
 & =Z_{k}^{(2,2)}(0;L_{1},L_{2},L;0,\zeta_{2},0,\zeta_{4})\int_{-\infty}^{\infty}\frac{d^{N}\alpha}{(2\pi)^{N}}\det\left(\left[\braket{\alpha_{m}|\frac{1}{\hat{H}_{k}^{(2,2)}\left(\hat{x},\hat{p}\right)}|\alpha_{n}}\right]^{N\times N}\right),\label{eq:22PFRes}
\end{align}
where 
\begin{align}
\hat{H}_{k}^{(2,2)}\left(\hat{x},\hat{p}\right) & =e^{\frac{\pi i\left(-M_{1}+M_{2}\right)}{2}+\frac{1}{2}\left(\zeta_{2}+\zeta_{4}\right)}e^{-\hat{x}+\hat{p}}\nonumber \\
 & \quad+\left[e^{\frac{\pi i\left(-M_{1}-M_{2}\right)}{2}+\frac{1}{2}\left(\zeta_{2}-\zeta_{4}\right)+i\pi k}+e^{\frac{\pi i\left(M_{1}+M_{2}\right)}{2}+\frac{1}{2}\left(\zeta_{2}+\zeta_{4}\right)-i\pi k}\right]e^{\hat{p}}\nonumber \\
 & \quad+e^{\frac{\pi i\left(M_{1}-M_{2}\right)}{2}+\frac{1}{2}\left(\zeta_{2}-\zeta_{4}\right)}e^{\hat{x}+\hat{p}}\nonumber \\
 & \quad+\left[e^{\frac{\pi i\left(-M_{1}-M_{2}+2L\right)}{2}+\frac{1}{2}\left(\zeta_{2}+\zeta_{4}\right)}+e^{\frac{\pi i\left(M_{1}+M_{2}-2L\right)}{2}+\frac{1}{2}\left(-\zeta_{2}+\zeta_{4}\right)}\right]e^{-\hat{x}}+E\nonumber \\
 & \quad+\left[e^{\frac{\pi i\left(-M_{1}-M_{2}+2L\right)}{2}+\frac{1}{2}\left(-\zeta_{2}-\zeta_{4}\right)}+e^{\frac{\pi i\left(M_{1}+M_{2}-2L\right)}{2}+\frac{1}{2}\left(\zeta_{2}-\zeta_{4}\right)}\right]e^{\hat{x}}\nonumber \\
 & \quad+e^{\frac{\pi i\left(M_{1}-M_{2}\right)}{2}+\frac{1}{2}\left(-\zeta_{2}+\zeta_{4}\right)}e^{-\hat{x}-\hat{p}}\nonumber \\
 & \quad+\left[e^{\frac{\pi i\left(-M_{1}-M_{2}\right)}{2}+\frac{1}{2}\left(-\zeta_{2}+\zeta_{4}\right)+i\pi k}+e^{\frac{\pi i\left(M_{1}+M_{2}\right)}{2}+\frac{1}{2}\left(-\zeta_{2}-\zeta_{4}\right)-i\pi k}\right]e^{-\hat{p}}\nonumber \\
 & \quad+e^{\frac{\pi i\left(-M_{1}+M_{2}\right)}{2}+\frac{1}{2}\left(-\zeta_{2}-\zeta_{4}\right)}e^{\hat{x}-\hat{p}},\label{eq:QC22}
\end{align}
with a constant $E$. It is worth noting that the matrix appeared in \eqref{eq:22PFRes} is a $N\times N$ matrix although the number of $\gamma_{n}$ was $N+L$. This conjecture was used for proposing a relation between the (2,2) model and the $\mathfrak{q}$-Painlev\'e equations, and the proposed relation was checked in many ways.

The difference between ${\cal Y}_{k}$ \eqref{eq:Ik} and the partition function of the (2,2) model $Z_{k}^{(2,2)}$ is whether $\alpha_{m}$ and $\beta_{m}$ are ``glued'' by the integration or not. Here we claim that this conjecture holds even if it is not glued. Under this assumption, \eqref{eq:QCL0} and \eqref{eq:QC22} lead the conjectured quantum curve \eqref{eq:QCConj}.

\section{Exact values of partition function with $\ell=4$, $M_1=M_2=M_3=\zeta_1=\zeta_2=\zeta_3=\zeta_4=0$}
\label{app_exactvalues}
In this appendix we display the exact values of the partition function of ${\hat D}_4$ quiver super Chern-Simons theory without deformations, $M_1=M_2=M_3=\zeta_1=\zeta_2=\zeta_3=\zeta_4=0$, for finite $N$.
These exact values can be calculated by using the techniques explained in section \ref{sec_TWPY}.
We denote the partition function as $Z_{k}(N)$.

For $k=\frac{1}{2}$ we find
\begin{align}
&Z_{\frac{1}{2}}(1)=
\frac{1}{48 \pi^2},\nonumber \\
&Z_{\frac{1}{2}}(2)=
\frac{1}{302400 \pi^2}
-\frac{1}{7680 \pi^6},\nonumber \\
&Z_{\frac{1}{2}}(3)=
-\frac{1763}{5588352000 \pi^2}
+\frac{337}{99532800 \pi^4}
-\frac{13}{5529600 \pi^6}
-\frac{17}{5160960 \pi^8},\nonumber \\
&Z_{\frac{1}{2}}(4)=
\frac{33161}{9323233920000 \pi^2}
-\frac{10879}{243855360000 \pi^4}
+\frac{110671}{1170505728000 \pi^6}
-\frac{151}{65028096000 \pi^8}
-\frac{19}{412876800 \pi^{10}}\nonumber \\
&\quad -\frac{1}{9083289600 \pi^{12}},\nonumber \\
&Z_{\frac{1}{2}}(5)=
-\frac{100577927}{2959608397794048000 \pi^2}
+\frac{225860389}{483316446412800000 \pi^4}
-\frac{17004137}{12051526975488000 \pi^6}
+\frac{348757}{337983528960000 \pi^8}\nonumber \\
&\quad +\frac{731}{1498247331840 \pi^{10}}
-\frac{1583}{3269984256000 \pi^{12}}
-\frac{1}{991895224320 \pi^{14}},\nonumber \\
&Z_{\frac{1}{2}}(6)=
\frac{7167457201}{23144137670749455360000 \pi^2}
-\frac{814993194619}{179445730224144384000000 \pi^4}
+\frac{215989200687971}{12920092576138395648000000 \pi^6}\nonumber \\
&\quad -\frac{369357430831}{17816977480561459200000 \pi^8}
+\frac{11111333}{3460951336550400000 \pi^{10}}
+\frac{273731}{30764011880448000 \pi^{12}}\nonumber \\
&\quad -\frac{606803}{142832912302080000 \pi^{14}}
+\frac{89}{12696258871296000 \pi^{16}}
+\frac{23}{323754601218048000 \pi^{18}},\nonumber \\
&Z_{\frac{1}{2}}(7)=
-\frac{255567041}{92135709965459736576000 \pi^2}
+\frac{9532006332801721}{223149737820234748723200000000 \pi^4}\nonumber \\
&\quad -\frac{661077767954033}{3682226384199442759680000000 \pi^6}
+\frac{4084729947556969}{13862499328750843330560000000 \pi^8}\nonumber \\
&\quad -\frac{14535520549283}{89797566502029754368000000 \pi^{10}}
-\frac{273553817}{4399253698904064000000 \pi^{12}}
+\frac{1570580677}{15837313316054630400000 \pi^{14}}\nonumber \\
&\quad -\frac{91615637}{2879511512009932800000 \pi^{16}}
+\frac{3277}{17266912064962560000 \pi^{18}}
+\frac{4973}{2952641963108597760000 \pi^{20}},\nonumber \\
&Z_{\frac{1}{2}}(8)=
\frac{105894309952369}{4290739121319225264349051392000 \pi^2}
-\frac{9817458311361607}{24845180055594746889000960000000 \pi^4}\nonumber \\
&\quad +\frac{9716302689296534892019}{5308014613186554797173034188800000000 \pi^6}
-\frac{41906686480383245447}{11610148162814063807897272320000000 \pi^8}\nonumber \\
&\quad +\frac{365906521658898091}{118773894248737225656238080000000 \pi^{10}}
-\frac{20644076149708721}{55011066809941451882889216000000 \pi^{12}}\nonumber \\
&\quad -\frac{87228726371}{73051318100140116620083200 \pi^{14}}
+\frac{13393301521}{15964011822583067443200000 \pi^{16}}
-\frac{832492987}{4111942439150184038400000 \pi^{18}}\nonumber \\
&\quad +\frac{4657}{2735535936409436160000 \pi^{20}}
+\frac{44951}{2425103265699861626880000 \pi^{22}}
+\frac{31}{13386570026663236180377600 \pi^{24}},\nonumber \\
&Z_{\frac{1}{2}}(9)=
-\frac{1222112355705161}{5577960857714992843653766809600000 \pi^2}
+\frac{450240072271594833733}{124298056703929134461424250060800000000 \pi^4}\nonumber \\
&\quad -\frac{72422488788288487332997801}{4002243018342662317068467778355200000000000 \pi^6}
+\frac{166092010073368557801499}{4076555222927274084228890256998400000000 \pi^8}\nonumber \\
&\quad -\frac{178072162981682949906389}{3961025317824071984675844784128000000000 \pi^{10}}
+\frac{690805087039015992607}{36122466910079954964380374794240000000 \pi^{12}}\nonumber \\
&\quad +\frac{88780310614562126821}{11858385601793924609518809907200000000 \pi^{14}}
-\frac{3131080097471}{253566558694701231243264000000 \pi^{16}}\nonumber \\
&\quad +\frac{6918725990689}{1209615981528294138839040000000 \pi^{18}}
-\frac{716452345649}{660016104792874340371660800000 \pi^{20}}\nonumber \\
&\quad +\frac{285250811}{31731543499657420210176000000 \pi^{22}}
+\frac{533413}{4381059281453422749941760000 \pi^{24}}\nonumber \\
&\quad +\frac{45791}{474614755490787464577024000000 \pi^{26}},\nonumber \\
&Z_{\frac{1}{2}}(10)=
+\frac{3140650919827501}{1615041015961574356525212069427200000 \pi^2}\nonumber \\
&\quad -\frac{917339511187428149583839}{27771069582573299794601682506184130560000000 \pi^4}\nonumber \\
&\quad +\frac{9562896985170392458894206396253}{54532282089416661414854212921234292736000000000000 \pi^6}\nonumber \\
&\quad -\frac{27939971960672603601912392051}{63920623694554328398363912581666570240000000000 \pi^8}\nonumber \\
&\quad +\frac{6495142520737412438266438933}{11300211077954403761482483792399564800000000000 \pi^{10}}\nonumber \\
&\quad -\frac{14774500755288226174610501}{39448009581222645858266125238922117120000000 \pi^{12}}\nonumber \\
&\quad +\frac{3049956620974601051928271}{81175168113319255426586407298138112000000000 \pi^{14}}\nonumber \\
&\quad +\frac{1686824782932342149}{13980412498957047960695860101120000000 \pi^{16}}
-\frac{220404541617135463933}{2376670124822698153318296217190400000000 \pi^{18}}\nonumber \\
&\quad +\frac{3373372604237267}{105708179343626754353925193728000000 \pi^{20}}
-\frac{26957213883811}{5544135280260144459121950720000000 \pi^{22}}\nonumber \\
&\quad +\frac{144467193323}{5100604457839332902392194662400000 \pi^{24}}
+\frac{1165432561}{2392058367673568821468200960000000 \pi^{26}}\nonumber \\
&\quad +\frac{40769}{21262741045987278413050675200000 \pi^{28}}
+\frac{9103}{33914071968349709068815826944000000 \pi^{30}},\nonumber \\
&Z_{\frac{1}{2}}(11)=
-\frac{3318063015235826625571}{192100733571784480361246147328717272064000000 \pi^2}\nonumber \\
&\quad +\frac{27383101232602032285535901873}{91182974446029521079599272307154850598092800000000 \pi^4}\nonumber \\
&\quad -\frac{36840165408746460842212932722387549}{21972692422288655383887208012352933572116480000000000000 \pi^6}\nonumber \\
&\quad +\frac{11744476360220468507581891060730693}{2591374044889079750433872198017053590814720000000000000 \pi^8}\nonumber \\
&\quad -\frac{3710468893622337813847902146348519}{546751446833739903388245562658543175204864000000000000 \pi^{10}}\nonumber \\
&\quad +\frac{208360751883394766009606371291807}{36807951539999320284201664807307254431744000000000000 \pi^{12}}\nonumber \\
&\quad -\frac{151559967939496216113666770107}{73846673936048793046674186447262203248640000000000 \pi^{14}}\nonumber \\
&\quad -\frac{609536524418481656244559153}{1017936608141023463049393547518651924480000000000 \pi^{16}}\nonumber \\
&\quad +\frac{2251489499699878645493339}{2145847922299917708668023288576868352000000000 \pi^{18}}\nonumber \\
&\quad -\frac{1218428140249441917680011}{2265061695761024248038469026831138816000000000 \pi^{20}}\nonumber \\
&\quad +\frac{17727409310383517}{122319464669053815752399152742400000000 \pi^{22}}
-\frac{69512699592013}{3875134555631181490778485555200000000 \pi^{24}}\nonumber \\
&\quad +\frac{195547428018511}{8752637249652295260505006040678400000000 \pi^{26}}
+\frac{57443236369}{75780409087898660264112606412800000000 \pi^{28}}\nonumber \\
&\quad +\frac{585868009}{24418131817211790529547395399680000000 \pi^{30}}
+\frac{1033327}{504641390889043670943979504926720000000 \pi^{32}}.
\end{align}
For $k=1$ we find
\begin{align}
&Z_{1}(1)=
\frac{1}{96 \pi^{2}},\nonumber \\
&Z_{1}(2)=
\frac{3}{1048576}
-\frac{36277}{1238630400 \pi^{2}}
+\frac{7}{589824 \pi^{4}}
-\frac{1}{245760 \pi^{6}},\nonumber \\
&Z_{1}(3)=
-\frac{7}{268435456}
+\frac{180619357}{732476473344000 \pi^{2}}
+\frac{191887}{1426902220800 \pi^{4}}
-\frac{1541}{5662310400 \pi^{6}}
-\frac{31}{660602880 \pi^{8}},\nonumber \\
&Z_{1}(4)=
\frac{243}{1099511627776}
-\frac{8380522497631}{3754029823064801280000 \pi^{2}}
+\frac{828980993}{16876257945845760000 \pi^{4}}
+\frac{59868161}{12785043898368000 \pi^{6}}\nonumber \\
&\quad -\frac{3834259}{2130840649728000 \pi^{8}}
-\frac{293}{1268357529600 \pi^{10}}
+\frac{19}{4650644275200 \pi^{12}},\nonumber \\
&Z_{1}(5)=
-\frac{525}{281474976710656}
+\frac{20535314431317819373}{1016913702199391576276336640000 \pi^{2}}
-\frac{484825215577667}{32434817671279883059200000 \pi^{4}}\nonumber \\
&\quad -\frac{226751681927}{7658752696878366720000 \pi^{6}}
+\frac{92575189433}{2835211334902087680000 \pi^{8}}
-\frac{16850329}{2454728428486656000 \pi^{10}}\nonumber \\
&\quad -\frac{3779}{5952824672256000 \pi^{12}}
+\frac{8443}{162512113552588800 \pi^{14}},\nonumber \\
&Z_{1}(6)=
\frac{18259}{1152921504606846976}
-\frac{3023122814444987222623}{16618611254751069260237745684480000 \pi^{2}}\nonumber \\
&\quad +\frac{4922913325177801163971}{19524743082228318264505663488000000 \pi^{4}}
+\frac{44875365636933892397}{1734105471118962477439647744000000 \pi^{6}}\nonumber \\
&\quad -\frac{4323711039919111}{20380859976716959245926400000 \pi^{8}}
+\frac{14440823119}{109971833596202188800000 \pi^{10}}
-\frac{375619789}{14516282034698688921600 \pi^{12}}\nonumber \\
&\quad -\frac{45351133}{37442790962516459520000 \pi^{14}}
+\frac{434663}{1248093032083881984000 \pi^{16}}
+\frac{53}{482217762396045312000 \pi^{18}},\nonumber \\
&Z_{1}(7)=
-\frac{39987}{295147905179352825856}
+\frac{47487406181331788129032013}{29045964434757239516858837940012318720000 \pi^{2}}\nonumber \\
&\quad -\frac{61295601520701228218606900453}{19348051967231384875539193035698995200000000 \pi^{4}}
+\frac{2347460954980793837250137}{1012162681382716018831973595217920000000 \pi^{6}}\nonumber \\
&\quad -\frac{63874819589031109281163}{126520335172839502353996699402240000000 \pi^{8}}
-\frac{236870598633828277}{430443762708262179273965568000000 \pi^{10}}\nonumber \\
&\quad +\frac{1420570653182033}{2989192796585154022735872000000 \pi^{12}}
-\frac{10594384618007}{99639759886171800757862400000 \pi^{14}}\nonumber \\
&\quad -\frac{198986999}{53917618986023701708800000 \pi^{16}}
+\frac{66932053}{40737756567217907957760000 \pi^{18}}\nonumber \\
&\quad +\frac{183461}{193504343694285062799360000 \pi^{20}},\nonumber \\
&Z_{1}(8)=
\frac{1410951}{1208925819614629174706176}
-\frac{215184170590039571823658114471811}{14657438232968360641379481167913943744307527680000 \pi^{2}}\nonumber \\
&\quad +\frac{358595498584834693869842771919637}{10144942090056530560532320044983982583971840000000 \pi^{4}}\nonumber \\
&\quad -\frac{3314788326228301048257288240263489}{74703664481325361400283447603972962663792640000000 \pi^{6}}\nonumber \\
&\quad +\frac{1537215964672513180189335900353}{42551643017387423900822196174511826534400000000 \pi^{8}}\nonumber \\
&\quad -\frac{8346193067065646427089671}{583005704476444426847216790845521920000000 \pi^{10}}\nonumber \\
&\quad -\frac{729576824639802483987497}{1181353664333847917558834023555399680000000 \pi^{12}}
+\frac{27944615298660887647}{13787974607071054126503665074176000000 \pi^{14}}\nonumber \\
&\quad -\frac{2677575061727459}{7142177988640794678323576832000000 \pi^{16}}
-\frac{3085400875153}{137973892962378988103978188800000 \pi^{18}}\nonumber \\
&\quad +\frac{382282874269}{62416761102028589856561561600000 \pi^{20}}
+\frac{40862741}{11443072868705241473702952960000 \pi^{22}}\nonumber \\
&\quad -\frac{16669}{43865112663370092315861319680000 \pi^{24}},\nonumber \\
&Z_{1}(9)=
-\frac{3130751}{309485009821345068724781056}
+\frac{26610166940528058837625882794885793}{201848087335111879508321020662693537025057043251200000 \pi^{2}}\nonumber \\
&\quad -\frac{326022198339635174306732134182548958214391}{879611629881369521590803630622410693943887449712230400000000 \pi^{4}}\nonumber \\
&\quad +\frac{138802741029004473340322180062520038033}{229478590077078721279241079417537686049443020800000000000 \pi^{6}}\nonumber \\
&\quad -\frac{4170394522854968903094393458794603}{6290834903690556749497553482439828434845696000000000 \pi^{8}}\nonumber \\
&\quad +\frac{23397444802608203170324525015575169}{56617514133215010745477981341958455913611264000000000 \pi^{10}}\nonumber \\
&\quad -\frac{42486308444752830197004583999}{387862035074088948292916386613708822937600000000 \pi^{12}}\nonumber \\
&\quad -\frac{3574933121001522330827665417}{560245161773684036423101447330912744243200000000 \pi^{14}}\nonumber \\
&\quad +\frac{6288209417295162763573}{741241514876139869840837034387701760000000 \pi^{16}}
-\frac{1229676424867533397}{1298815270345740856245540052008960000000 \pi^{18}}\nonumber \\
&\quad -\frac{1100873764104559}{9227693961323906724394061266944000000 \pi^{20}}
+\frac{530861692444781}{27683081883971720173182183800832000000 \pi^{22}}\nonumber \\
&\quad +\frac{1519619701}{252663048941011731739361201356800000 \pi^{24}}
-\frac{2938183}{288124529494136185316815405056000000 \pi^{26}},\nonumber \\
&Z_{1}(10)=
\frac{55870047}{633825300114114700748351602688}\nonumber \\
&\quad -\frac{699330967749490343133483508719309800266891}{590425350661215118702414882156977145722550949235721009889280000 \pi^{2}}\nonumber \\
&\quad +\frac{92101779960515063032984101042662289588753923}{24589719035659661249975233654775615431340162434195343278080000000 \pi^{4}}\nonumber \\
&\quad -\frac{9965898626323168693267695815873394146131420212707}{1383171695755855945311106893081128368012884136923488059392000000000000 \pi^{6}}\nonumber \\
&\quad +\frac{31749093463398874261538561141395335373824830491}{3419930016978764699945044515859932778053834404481151795200000000000 \pi^{8}}\nonumber \\
&\quad -\frac{62586043161390022151715910777121269081}{8810877146794218093275226064836987497124515020800000000000 \pi^{10}}\nonumber \\
&\quad +\frac{81641928257594884647192024388096579}{28535227123140365415720902596347061780460077056000000000 \pi^{12}}\nonumber \\
&\quad -\frac{9548499437368284584566846479701663}{22313260306816526340112585488872890414795849728000000000 \pi^{14}}\nonumber \\
&\quad -\frac{8047015474711735692855395507}{120475119587813015192423735234039476522057728000000 \pi^{16}}\nonumber \\
&\quad +\frac{3182305255955068457268457801}{112284925054430990036798437444006091056742400000000 \pi^{18}}\nonumber \\
&\quad -\frac{38154507391808944120183}{30645889191039126778699566349725141565440000000 \pi^{20}}\nonumber \\
&\quad -\frac{365751446009319589}{744121241041159838255137100566364160000000 \pi^{22}}
+\frac{6631611383969123}{126776211436642046517541876392787968000000 \pi^{24}}\nonumber \\
&\quad -\frac{3357873156997}{441452879109735697695011891010600960000000 \pi^{26}}
-\frac{66326527}{774478735280238066131599808790528000000 \pi^{28}}\nonumber \\
&\quad -\frac{56633}{3743313887187817319636065742487552000000 \pi^{30}}.
\end{align}
For $k=\frac{3}{2}$ we find
\begin{align}
&Z_{\frac{3}{2}}(1)=
\frac{1}{144 \pi^{2}},\nonumber \\
&Z_{\frac{3}{2}}(2)=
 \frac{50}{14348907}
-\frac{37}{4251528 \sqrt{3} \pi}
-\frac{112789}{5952139200 \pi^{2}}
-\frac{1}{236196 \sqrt{3} \pi^{3}}
+\frac{19}{1679616 \pi^{4}}
-\frac{1}{1866240 \pi^{6}}.
\end{align}
For $k=2$ we find
\begin{align}
&Z_{2}(1)=
\frac{1}{192 \pi^{2}},\nonumber \\
&Z_{2}(2)=
\frac{457}{134217728}
-\frac{7}{1048576 \pi}
-\frac{509557}{39636172800 \pi^{2}}
-\frac{1}{524288 \pi^{3}}
+\frac{13}{1572864 \pi^{4}}
-\frac{1}{7864320 \pi^{6}},\nonumber \\
&Z_{2}(3)=
-\frac{16349}{549755813888}
+\frac{637561}{9019431321600 \pi}
+\frac{15815844251}{187513977176064000 \pi^{2}}
-\frac{19909}{579820584960 \pi^{3}}\nonumber \\
&\quad -\frac{11114843}{730573937049600 \pi^{4}}
-\frac{193}{16106127360 \pi^{5}}
-\frac{13861}{2899102924800 \pi^{6}}
-\frac{59}{84557168640 \pi^{8}},\nonumber \\
&Z_{2}(4)=
\frac{34871}{140737488355328}
-\frac{22439575817}{34135877800584806400 \pi}
-\frac{758352829008529}{1281375512939452170240000 \pi^{2}}\nonumber \\
&\quad +\frac{140758399}{218198082532147200 \pi^{3}}
+\frac{299580676621}{4608343503078948864000 \pi^{4}}
-\frac{56603}{974098582732800 \pi^{5}}
+\frac{25650277139}{157102619423145984000 \pi^{6}}\nonumber \\
&\quad -\frac{2291}{57724360458240 \pi^{7}}
-\frac{10626851}{290930776709529600 \pi^{8}}
-\frac{11}{115448720916480 \pi^{9}}
-\frac{433}{288621802291200 \pi^{10}}\nonumber \\
&\quad +\frac{23}{3809807790243840 \pi^{12}},\nonumber \\
&Z_{2}(5)=
-\frac{19127015}{9223372036854775808}
+\frac{71146579910239}{11996603659428562489835520 \pi}
+\frac{806101730071123295741}{189330842009486722564903403520000 \pi^{2}}\nonumber \\
&\quad -\frac{898762473394111}{115024253836850563645440000 \pi^{3}}
-\frac{43755859532907383}{850259284361999366467092480000 \pi^{4}}
+\frac{195481980961}{73733496049263181824000 \pi^{5}}\nonumber \\
&\quad -\frac{1275515856625181}{828174627625324058247168000 \pi^{6}}
+\frac{16858367}{239394467692412928000 \pi^{7}}
+\frac{1948443158159}{3378334728075331239936000 \pi^{8}}\nonumber \\
&\quad -\frac{177203}{1994953897436774400 \pi^{9}}
-\frac{507029783}{8043654114465074380800 \pi^{10}}
-\frac{5239}{9753107943024230400 \pi^{11}}\nonumber \\
&\quad -\frac{12839}{10639754119662796800 \pi^{12}}
+\frac{4639}{133129923422280744960 \pi^{14}}.
\end{align}

\bibliography{bunken_240722Nosaka.bib}
% \bibliography{bunken_210824Nosaka.bib}

\end{document}